\documentclass[pre,twocolumn,longbibliography,superscriptaddress,showpacs,pdfoutput=1]{revtex4-1}
\usepackage{graphicx}
\usepackage{epsfig}
\usepackage{amssymb,amsmath,amsfonts,hyperref}
\usepackage{wasysym}
\usepackage{bm}
\usepackage{lpic}
\usepackage{latexsym}
\usepackage{eucal}
\usepackage{float}
\usepackage{verbatim}
\usepackage[normalem]{ulem}
\usepackage{epstopdf}
\usepackage{xcolor}
\usepackage{braket}
\usepackage{lipsum}

\newcommand{\be}{\begin{equation}}
\newcommand{\ee}{\end{equation}}

\newcommand{\eea}{\end{eqnarray}}

\begin{document}

\title{Quantum half-orphans in kagom\'e antiferromagnets}

\author{Pranay Patil}
\affiliation{Laboratoire de Physique Th\'eorique, Universit\'e de Toulouse, CNRS, UPS, France}

\author{Fabien Alet}
\affiliation{Laboratoire de Physique Th\'eorique, Universit\'e de Toulouse, CNRS, UPS, France}

\author{Sylvain Capponi}
\affiliation{Laboratoire de Physique Th\'eorique, Universit\'e de Toulouse, CNRS, UPS, France}

\author{Kedar Damle}
\affiliation{Department of Theoretical Physics, Tata Institute of Fundamental Research, Mumbai 400005, India}

\begin{abstract}
We numerically study the effects of non-magnetic impurities (vacancies) in the spin-$S$ Heisenberg
antiferromagnet on the kagom\'e lattice. For a range of low but nonzero temperatures, and spin values that extend down to $S=2$, we find that the magnetization response to an external magnetic field is consistent with the response of emergent ``half-orphan'' degrees of freedom that are expected
to dominate the response of the corresponding classical magnet in a similar temperature range whenever there are two vacancies on the same triangle. Specifically, for all spin values we have considered (from $S=1/2$ to $S=4$), there is a large enhancement of the local susceptibility of the lone spin on such a triangle with two vacancies; in the presence of a uniform magnetic field $h$, this lone-spin behaves effectively as an almost free spin $S$ in an effective field $h/2$. Quite remarkably, in the zero temperature limit, the ground state in the presence of a half-orphan has a non-zero total spin value $S_{GS}$ that shows a trend similar to $S/2$ when $S\ge 2$. These qualitative aspects of the response differ strikingly from the more conventional response of diluted samples without such half-orphan degrees of freedom. We discuss how these findings could be checked experimentally. 
\end{abstract}

\maketitle

\section{Introduction}

The low-temperature physics of frustrated quantum magnets is controlled by the interplay between the geometric frustration, quantum fluctuations, entropic effects, and quenched disorder. In many interesting cases~\cite{ExpQSL}, the geometric frustration renders the leading exchange interactions unable to drive magnetic ordering even at temperatures much lower than the scale of these interactions. Instead, in a semiclassical picture, they confine the system to a manifold of minimally frustrated configurations, which can admit a description in terms of emergent degrees of freedom as in the case of classical spin ice~\cite{castelnovo2008magnetic}. The low temperature physics is then controlled by how these emergent degrees of freedom behave in the presence of subdominant interactions as well as quantum and thermal fluctuations. In the extreme quantum limit, frustrated $S=1/2$ magnets can provide an arena for the physics of quantum number fractionalization; the Majorana fermion excitations of Kitaev's honeycomb model being a celebrated example~\cite{kitaev2006anyons}.

Interestingly, quenched disorder, {\em i.e.} non-magnetic substitutional impurities or lattice imperfections that affect the bond strengths in their vicinity, can provide a powerful probe of these emergent excitations and their unusual quantum numbers. For instance, the emergent magnetic monopole excitations of classical spin ice can be nucleated in the ground state of a disordered sample. Similarly, vacancies in Kitaev's honeycomb model nucleate a Majorana fermion-Z$_2$ flux complex in their vicinity. More generally, a wide variety of features can be probed or created by non-magnetic impurities in magnetic systems, depending on the nature of the underlying state. If the clean system is a spin liquid, then non-magnetic impurities can lead to spinon deconfinement~\cite{Poilblanc2006} or even spin-charge separation if they are mobile~\cite{Laeuchli2004}. More conventional ordering patterns can also be characterized by the response to non-magnetic impurities~\cite{Alloul2009}. An enhancement of local correlations is often observed near spinless impurities in antiferromagnets~\cite{Martins}. In valence bond crystals, that break lattice symmetries, or valence bond solids that are non-degenerate, non-magnetic impurities generically induce a localized spin texture around them \cite{Poilblanc2006}. In systems with competing phases at low temperatures, non-magnetic impurities can furthermore reveal or seed competing orders~\cite{Kaul}. Quantum critical systems have also been shown~\cite{sachdev,hoglund} to host responses characteristic of their criticality to the inclusion of spinless defects. All these different physical behaviors and the corresponding local textures or responses have been detected in magnetic compounds using local probe techniques, such as nuclear magnetic resonance (NMR). Let us for instance mention the spin ``polaron'' observed in the two-dimensional doped material,   SrCu$_2$(BO$_3$)$_2$ material~\cite{Haravifard2006,Yoshida2015}, the imaging of spin-1/2 edge excitations in spin 1 Haldane compound Y$_2$BaNi$_{1-x}$Mg$_x$O$_5$~\cite{Tedoldi}, the magnetic response around an impurity observed in a $S=1/2$ kagom\'e compound (herbertsmithite)~\cite{Olariu2008} and the localized bound states generated by impurities in topological magnets~\cite{yin2020spin}.

Coming back to frustrated magnets, the  SrCr$_{9p}$Ga$_{12-9p}$O$_{19}$ (SCGO) compound, in which Cr$^{3+}$ $S=3/2$ moments lie on the vertices of the pyrochlore-slab (bilayer kagom\'e) lattice, provides another striking example of the interplay between impurities and the underlying physical state~\cite{SCGO1988,Ramirez1990,Limot2002}. Even in the best samples of SCGO, nonmagnetic Ga atoms disrupt the corner-sharing network of triangles and tetrahedra formed by the Cr ions; these Ga impurities can be modeled as static vacancies in the Heisenberg antiferromagnet on this lattice.
At a classical level, the Heisenberg model on the pyrochlore slab lattice remains disordered down to $T=0$, providing an interesting example of a classical spin liquid with full SU(2) symmetry of interactions~\cite{Moessner_Berlinsky,moessner1998properties}. 

Following an earlier phenomenological analysis~\cite{schiffer_two_1997} of experimental results on SCGO, Henley~\cite{henley_effective_2001} described the $T=0$ liquid state in terms of an emergent fluctuating polarization field and noted that a triangle with two vacancies led to a lone or ``orphan'' spin on that triangle which behaves as a source (charge) for this polarization field. The polarization field of this charge leads, in this description, to an oscillating spin texture with a power-law envelope. Even without a detailed computation of this texture, this effective theory predicts that the classical spin-$S$ Heisenberg model on the pyrochlore slab lattice will have a net spin polarization of $S/2$ in the direction of an infinitesimal external magnetic field at $T=0$. 

Motivated by this factor of two, which is suggestive of {\em spin fractionalization}, Henley dubbed this combination of the lone spin and the resulting spin-texture a ``half-orphan'', modifying the terminology of ``orphan spins'' introduced in the earlier phenomenological studies~\cite{schiffer_two_1997}. In closely related work~\cite{Moessner_Berlinsky} that studied the classical Heisenberg antiferromagnet on the pyrochlore and pyrochlore-slab lattices, Moessner and Berlinsky also recognized the special role of these half-orphan degrees of freedom in dominating the low temperature magnetic response, and used this insight to  model the experimental results on SCGO within a single-unit approximation that used as input properties of individual simplices (triangles and tetrahedra) with various degrees of dilution.

This striking $T=0$ prediction of a saturation magnetization of $S/2$ for the system with two vacancies on the same triangle
led immediately to the questions: Does the low-temperature susceptibility show signatures of this ``spin fractionalization''~? In other words, does the impurity susceptibility of this orphan-texture complex correspond to that of a classical spin $S/2$ object? If yes, how is this response resolved spatially~?
Motivated by these questions, Ref.~\onlinecite{sen_fractional_2011} used a hybrid large-N field-theory
as well as direct Monte-Carlo simulations to study the low but nonzero temperature behaviour. Perhaps surprisingly, the answer turns out to be in the affirmative, with an interesting spatial structure that will be important in our present study: At low temperatures $T$ in the spin-liquid regime, each lone spin on a triangle with two vacancies ``sees'' an effective magnetic field that has magnitude $h/2$ (where $h$ is the applied external field) and responds to it as a free spin $S$ in this field at temperature $T$. Exactly half of this paramagnetic response is cancelled off by the net diamagnetic response of the surrounding spin texture for which this orphan spin serves as a source, giving rise to a net susceptibility that equals the susceptibility of a spin $S/2$ at temperature $T$. 

In conjunction with subsequent work~\cite{sen_vacancy_2012} that also characterized the entropic interactions between these half-orphan degrees of freedom, this approach provides a fairly detailed picture of both the orphan-texture complex and its susceptibility, as well as interactions between orphans, including a theory for low temperature glassy behaviour in the multi-orphan case. Importantly, it provides a reasonably satisfactory fit to NMR data on SCGO~\cite{Limot2002,Bono2004}, that could not previously be accounted for by more conventional ideas. We provide more details on orphan physics in the classical case in the Appendix for completeness, along with classical Monte Carlo simulations for kagom\'e systems, as starting from now we will focus on kagom\'e planes (and not bilayers).

While one expects that the classical physics of half-orphans would be reflected in some form in a semiclassical treatment of the corresponding quantum magnet with large but finite spin $S$, this expectation has thus far not been shored up by an actual controlled calculation of $1/S$ corrections to the classical picture. Nor has the fairly impressive success of classical descriptions of $S=3/2$ frustrated magnets such as SCGO been quantitatively examined from this point of view. 

Spin $S=1/2$ kagom\'e magnets such as Herbertsmithite~\cite{Mendels_2011,Khuntia2020} also feature some degree of dilution by nonmagnetic vacancies. Since it is well known that the classical kagom\'e magnet has a fairly broad low-temperature spin liquid regime, extending upwards from a lower cutoff of $T^{*} \sim 10^{-3} JS^2$ (that marks a subtle crossover to an ordered regime below this temperature~\cite{Chern_Moessner,Zhitomirsky,Rutenberg_Huse}), one expects that half-orphans seeded by vacancy-pairs should be fairly well-defined at the classical level in this regime. These observations provide the central motivation for the present study, in which we ask:
{\it Do these emergent half-orphan degrees of freedom nucleated by pairs of vacancies on the same triangle survive quantum effects in the Heisenberg antiferromagnet on the kagom\'e lattice?}

The presence of nonmagnetic impurities in some of the best-known experimental realizations of frustrated quantum magnets on the kagom\'e lattice has motivated several studies of vacancy effects, which we build on here with our work. For instance, a single non-magnetic impurity is known to induce local dimer order in the spin-1/2 case~\cite{Dommange2003} and does not generate a localized spin in the spin-3/2 case either~\cite{Laeuchli2007}, while a finite concentration could lead to valence bond glass in the spin-1/2 case~\cite{Singh2010}. Using a combination of series expansion and variational wavefunction studies, Gregor and Motrunich~\cite{gregor_nonmagnetic_2008} have also modeled inhomogeneous Knight shifts in compounds such as Herbertsmithite~\cite{Mendels_2011} by studying the response of the $S=1/2$ kagom\'e antiferromagnet to nonmagnetic impurities. Again, there is no evidence that single vacancies induce the kind of dramatic enhancement that pairs of vacancies do in the classical picture. 

In this manuscript, we study the finite-temperature as well as ground-state signatures of orphan physics in the spin-$S$ quantum Heisenberg model on the kagom\'e lattice for a large set of spin values ranging from $S=1/2$ up to $S=6$, allowing us to discuss the quantum to classical crossover. One striking result of our study is that the ground state of the quantum system with two vacancies on the same triangle has nonzero spin quantum number $S_{GS}$.
This tracks $[S/2]$, which is the total quantum spin value (allowed to exist in the system) which is the closest to $S/2$, i.e. we find $S_{GS} \simeq [S/2]$ down to $S=2$. This provides a dramatic quantum manifestation of half-orphan physics at $T=0$. In some way, this result is reminiscent of the $S=1/2$ end-spins in cut $S=1$ Haldane gap chains, which provide the clearest and experimentally relevant signature of the underlying symmetry-protected topological order in these systems. However, this is qualitatively different: The presence of quantum half-orphans in the ground state of samples with two vacancies on the same triangle is blind to the integer or half-integer nature of elementary spin $S$, and is perhaps best thought of as a semiclassical effect that survives to surprisingly low values of $S$. Indeed, there is no obvious underlying topological order at the heart of this effect.

In our finite temperature studies, we find that the local susceptibility offers a very clear signature of orphan physics down to $S=2$. We attribute this to presence of quantum half-orphans in the ground state of these systems. From other probes such as the magnetization curve, we find that orphan physics can be detected to even lower values of quantum spin (down to $S=1/2$), with specific signatures not present for other types of impurity patterns. We also present an analysis of the histogram of spin magnetization due to a distribution of impurities, akin to a NMR Knight shift experiment, specific to the experimentally relevant $S=1/2$ situation.  Our results are obtained with a finite-temperature method working in Krylov bases generated by an appropriate number of randomly distributed initial states (see details below), as well as Lanczos exact diagonalization (ED) and Density Matrix Renormalization Group (DMRG) for the study of ground-state properties.

The layout of this manuscript is as follows:
In Sec.~\ref{Sec4} we first study the $T=0$ ground state and present our results on the ground-state polarization, as well as spin textures obtained through ED and DMRG computations.
 We begin Section~\ref{Sec3} with a brief description of the finite-temperature random sampling technique used to simulate
thermodynamics of the quantum system, and then present finite-temperature results for orphan physics, including the local susceptibility response, the magnetization curve, as well as experimental predictions for the NMR Knight shift.  Finally, Sec.~\ref{Sec5} summarizes our results along with a discussion of some outstanding issues. For completeness, in the Appendix, we also present the results of finite-temperature Monte Carlo simulations of vacancy effects in the classical Heisenberg model on the kagom\'e lattice, and identify the temperature regime in which orphan physics is clearly visible in the classical case.

\section{Ground State Physics in the Quantum Case}\label{Sec4}
\begin{figure}[!ht]
\begin{minipage}[t]{0.21\textwidth}
\includegraphics[width=\textwidth]{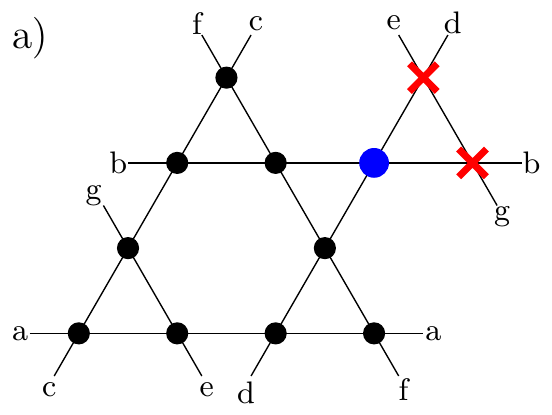} \label{figk12o}
\end{minipage}
\begin{minipage}[t]{0.21\textwidth}
\includegraphics[width=\textwidth]{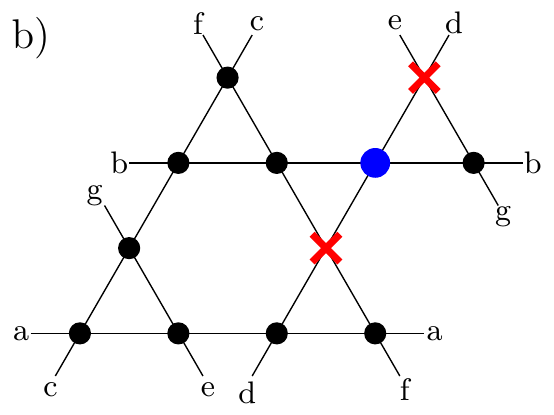}\label{figk12n}
\end{minipage}
\vspace{5mm}
\begin{minipage}[t]{0.26\textwidth}
\includegraphics[width=\textwidth]{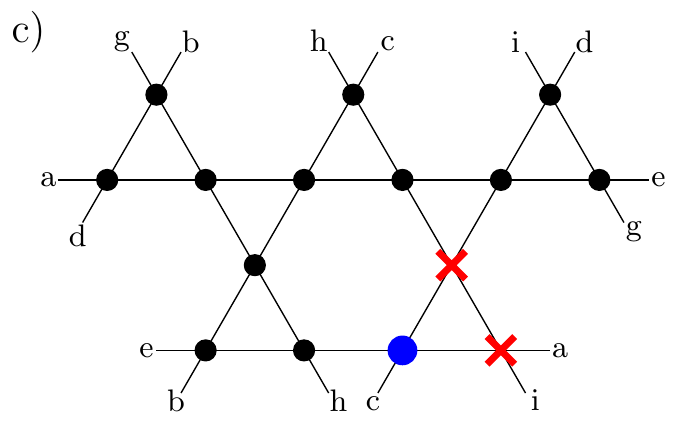}\label{figk15o}
\end{minipage}
\vspace{5mm}
\begin{minipage}[t]{0.22\textwidth}
\includegraphics[width=\textwidth]{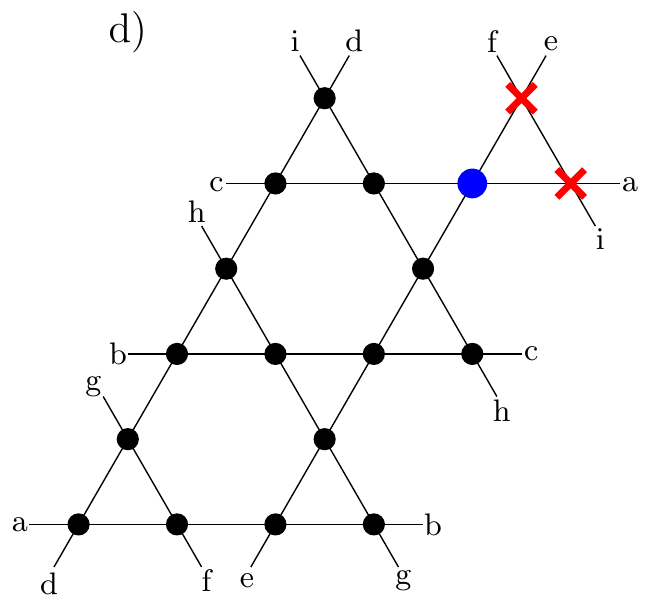}\label{figk18o}
\end{minipage}
\begin{minipage}[t]{0.22\textwidth}
\includegraphics[width=\textwidth]{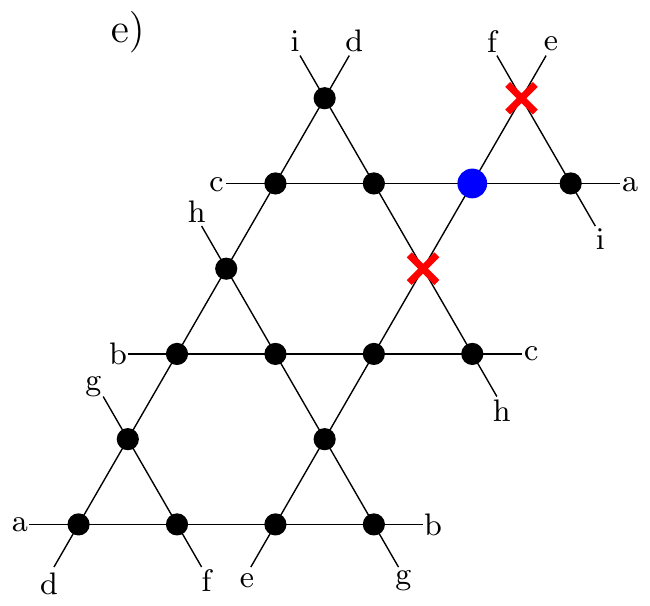}\label{figk18n}
\end{minipage}
\begin{minipage}[t]{0.30\textwidth}
\includegraphics[width=\textwidth]{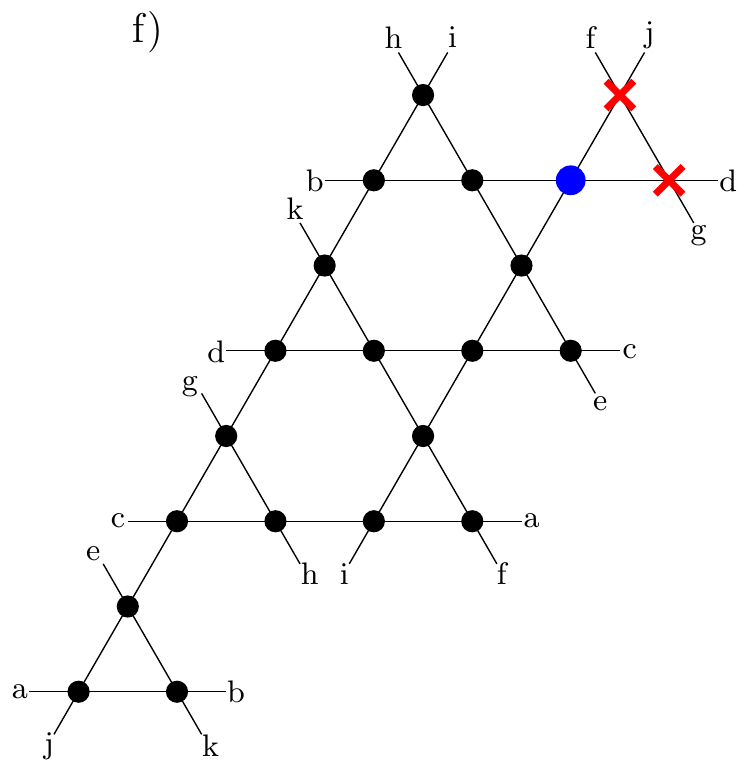}\label{figk21o}
\end{minipage}
\caption{Kagom\'e lattices with $N$ sites and two non-magnetic impurities (denoted as red crosses). Periodic boundary conditions are shown through system spanning bonds denoted by alphabetical indices. (a), (c), (d) and (f): Samples with respectively $N=12$, $N=15$, $N=18$ and $N=21$ sites, for which the position of the impurities create an orphan spin (denoted as blue dot). (b) and (e): Samples with $N=12$ and $N=18$ where the impurities are next-nearest neighbors and do {\it not} create an orphan spin. The blue dot in those cases denotes the spin which we monitor the magnetization or local suceptibility in Sec.~\ref{Sec3}.}
\label{figlat}
\end{figure}

We begin with perhaps the simplest question that gets to the heart of the intriguing conundrum posed by
Henley, Moessner and Berlinksy's {\em classical} $T=0$ argument for a saturation magnetization of $S/2$ that reflects the presence of a half-orphan nucleated by two vacancies on a triangle: What is the ground state spin quantum number of a spin-$S$ quantum antiferromagnet on a kagom\'e lattice with two vacancies on one triangle of the lattice? 

\begin{figure}[h]
\includegraphics[width=\hsize]{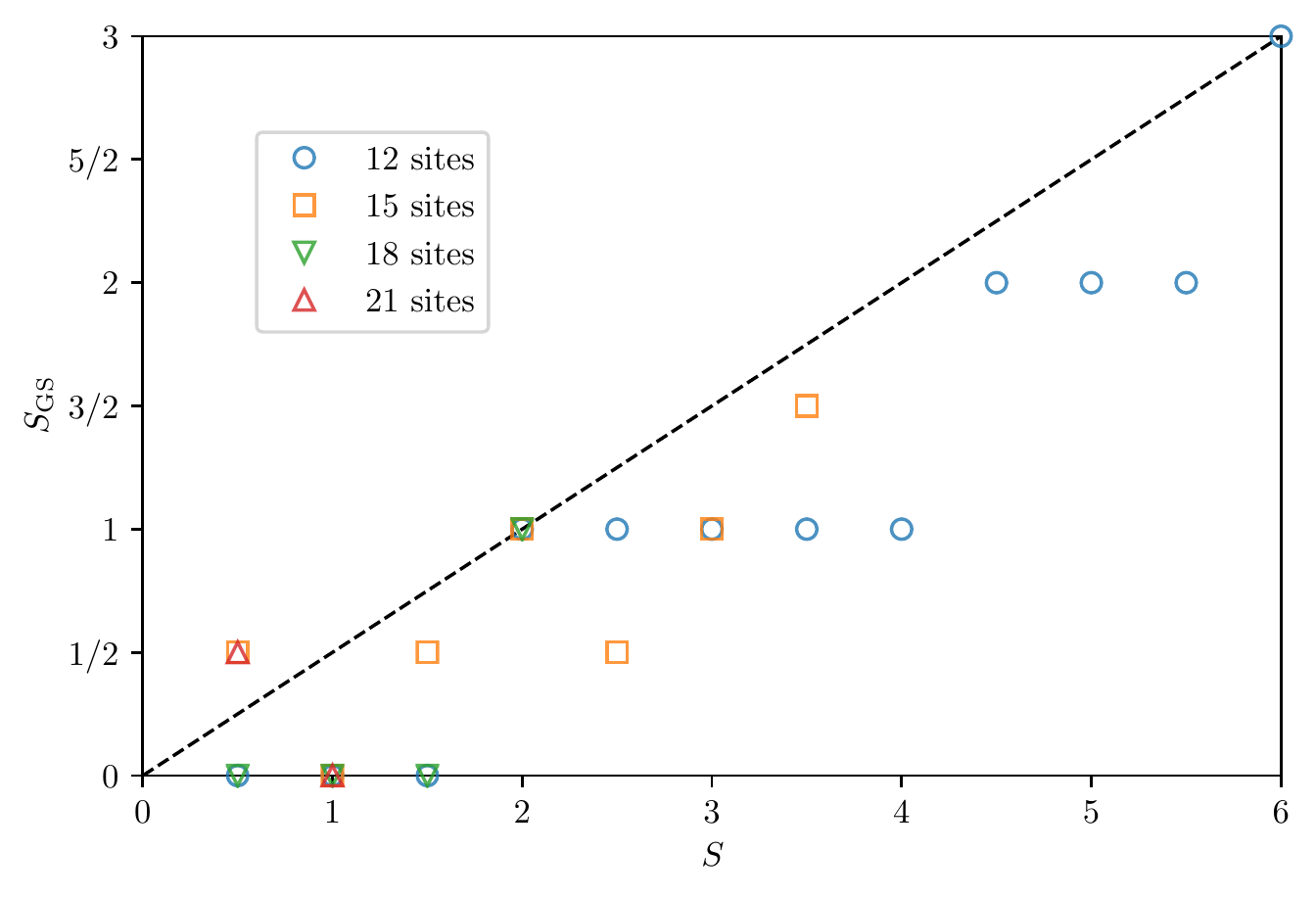}
\caption{Ground state global spin $S_{\rm GS}$ as a function of local spin-$S$ for kagom\'e lattices of varying sizes with an orphan spin. The straight line is the classical prediction $S_{GS}=S/2$.}
\label{fig:ground_state_spin}
\end{figure}

To answer this question for the kagom\'e magnet with Hamiltonian
\begin{equation}
H=\sum_{\langle i,j\rangle} \vec{S_i}\cdot\vec{S_j} \; ,
\label{eq:H}
\end{equation}
we first compute the total spin of the ground-state for some of the lattices shown in Fig.~\ref{figlat}.
This is done using Lanczos diagonalization for kagom\'e samples with an orphan spin, up to $S=6$. This is represented in Fig.~\ref{fig:ground_state_spin}, where we find that starting from $S\geq 2$, the ground-state is no longer a singlet. We compare the total spin $S_{\rm GS}$ of the ground state to $S/2$ (dashed line in Fig.~\ref{fig:ground_state_spin}) and find an overall trend towards this classical prediction. In particular, the ground state spin $S_{\rm GS}$ tracks in most cases the (half-)integer closest to $S/2$ which is allowed in the sample.
The results in Fig.~\ref{fig:ground_state_spin} are mostly for the $12$-sites samples with a pair impurity, but we find the same for
the few larger clusters which we are able to simulate. In particular, we find 
$S_{\rm GS}=1$ for spin-2 lattices of sizes 15 and 18 with an orphan impurity. We were able to check also a combination of spin and lattice size where the 
global total spin is half integer $\in \{ 1/2,3/2,5/2,...\}$. Note for instance the case of $S=7/2$ spins on a $15$-sites sample with a pair impurity (13 sites effective) where we find that $S_{GS}=3/2$, which is indeed the allowed value closest to the classical prediction.

These results should be contrasted with all other cases we have tested, such as the pure case and different arrangements of the impurities (for one or two non-magnetic impurities). In these more conventional configurations, we always find the ground state to be of the lowest possible spin (a global singlet $S_{\rm GS}=0$ when the total number of spins is even). This is consistent with what is expected for conventional antiferromagnets. 

\begin{figure}[h]
\includegraphics[width=0.8\hsize]{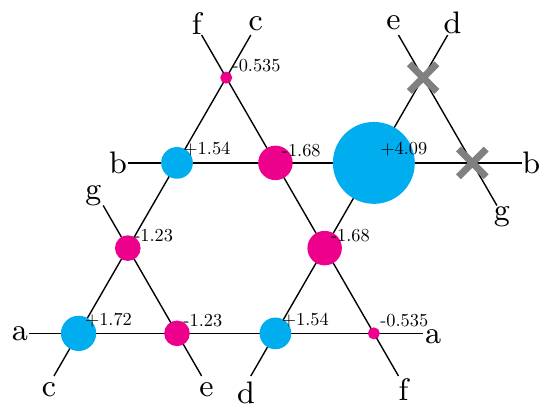}
\caption{Ground state texture $\braket{S^z_i}$ for $S=11/2$ measured on the highest polarized state of the ground-state multiplet, with $S_{\rm GS}=2$, obtained from ED for a 12-site lattice (periodic boundary conditions are indicated with identical letters) with 2 impurities shown as crosses. Radii of the circles are proportional to the $\braket{S^z_i}$ value; blue and red correspond to positive and negative values respectively.}
\label{fig:ED1}
\end{figure}

In addition to this, we also study the ground state texture for cases where it is not a global singlet. This can be done by calculating $\braket{S^z_i}$
for all sites on the lattice in the highest polarized state in the ground
state multiplet. This is shown explicitly in Fig.~\ref{fig:ED1} for $S=11/2$.
Since the ground-state has total spin $S_{\rm GS}=2$ (see Fig.~\ref{fig:ground_state_spin}), we have measured the distribution of magnetization in the $S^z_{\rm total}=\sum_{i=1}^N S_i^z=2$ sector. There is clearly a larger effect at the orphan site which tapers off with increasing distance, but due to the relative small size of the lattice, other sites are affected as well. 

\begin{figure*}[t]
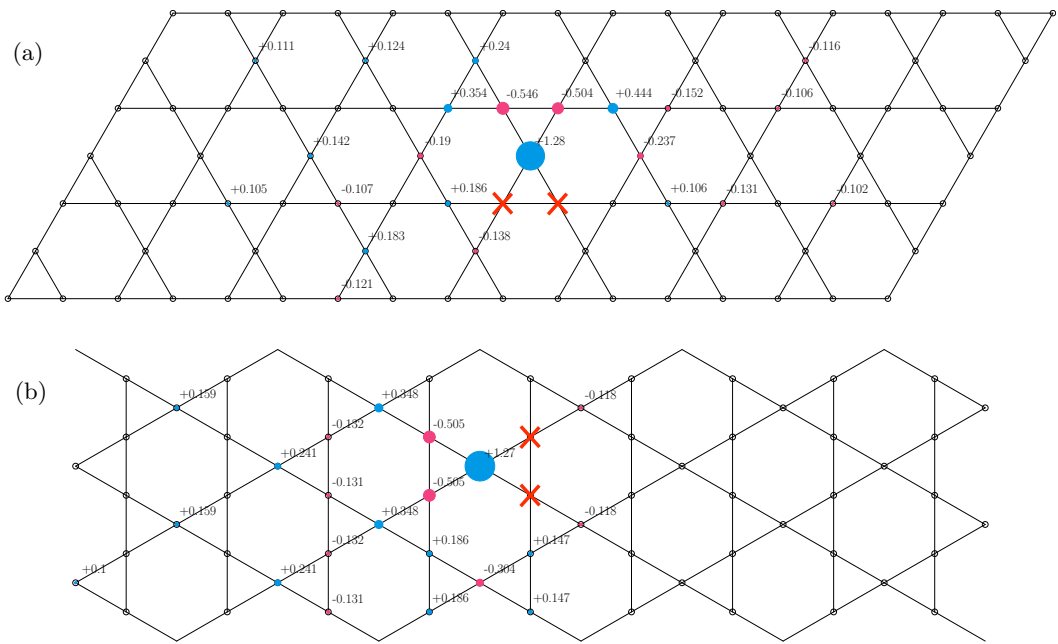

\begin{lpic}[clean]{sz_XC(0.11)}
\lbl{28,300; \text{(a)}}
\end{lpic}\\ 
\vspace{6mm}
\begin{lpic}[clean]{sz_YC(0.11)}
\lbl{-50,300; \text{(b)}}
\end{lpic}\\ 
\caption{Ground-state texture $\braket{S^z_i}$ for $S=2$ measured in the $S^z_{\rm total}=1$ sector, achieved using 
DMRG for (a) a $6\times 17$ lattice; (b) a $4\times 19$ lattice with a different orientation. Open/Periodic boundary conditions are used in the long/short direction. The two crosses denote the positions of the two impurities, whereas radius of circles and closeby numbers denote $\braket{S^z_i}$ (values are only shown when the absolute value is larger than 0.1).}
\label{fig:DMRG1}
\end{figure*}

In Fig.\ref{fig:DMRG1}, we present other magnetization profiles for much larger system
sizes using DMRG simulations~\cite{White1992,Schollwoeck2011,itensor}. We have chosen a spin $S=2$ so that the ground-state is polarized and we measure the local $\braket{S^z_i}$ values in the $S^z_{\rm total}=1$ ground-state. Typically, we have kept up to $m=8000$ states (using only U(1) quantum number corresponding to the conservation of $S^z$) to achieve a discarded weight below $5.10^{-5}$. As often used in such simulations, we have chosen cylinder geometries, i.e. periodic boundary conditions in the short direction and open ones in the other. From these plots, it is clear that the spin texture is well localized around the orphan spin. Moreover, the numerical values found in these larger clusters are within $10\%$ of the ones obtained by ED on a much smaller 12-site cluster with 2 impurities.

In order to make contact with experiments, we now move to the finite-temperature regime to see if signatures of orphan spins remain and how to characterize them. 

\section{Thermodynamics for the Quantum Version}\label{Sec3}

In this section, we study the thermodynamics of the quantum version of the Heisenberg model in Eq.~\ref{eq:H}, for various values of the spin $S$ ranging from $S=1/2$ up to $S=4$, on samples of the kagom\'e lattice containing one or several non-magnetic impurities, some of which are depicted in Fig.~\ref{figlat}. 

While a full computation of the thermodynamic properties does require the complete set of eigenvalues (to get the partition function), or even eigenstates (to compute observables), it has been known for a long-time that finite-temperature properties can be approximated using only a few well-chosen pure states in the correct energy window, as done in the finite-temperature Lanczos method~\cite{Prelovsek1994}. This is rooted at the foundations of statistical mechanics since a pure state with energy $E$ of a large system has the same local properties as a thermodynamic mixed state at the related temperature. These ideas have been put on more rigorous grounds over the years, and are known as quantum typicality~\cite{Hams2000,goldstein_canonical_2006,Popescu2006,Reimann2007}. 

In most of our results, we thus use a typicality scheme based on the exact application of the Hamiltonian to a random initial vector $|v_0 \rangle$.  Rather than using a power-method with repeated applications of the Hamiltonian~\cite{sugiura_canonical_2013,sugiura_thermal_2012} which has a slower convergence, we use the formulation in the Krylov basis ${\rm Span} \{ |v_0 \rangle, H |v_0 \rangle, H^2 |v_0 \rangle \ldots, H^m | v_0 \rangle \}$, and choose $m$ large enough (typically between 100 and 500) such that at least two of the lowest energy states are converged. We found that this criterion implies large-enough Krylov spaces to ensure convergence at the lower temperature. Several previous works show that these approaches are particularly successful to study thermodynamic properties of frustrated magnets on lattices larger than those available with full diagonalization methods~\cite{Shimokawa2016,Schnack2018,wietek_thermodynamic_2019,Prelovsek2020}, and we refer to them for details about this method. Despite using exact (machine precision) application of $H$, all the data presented using this technique present error bars, which results from averaging over a finite number of initial random vectors $|v_0 \rangle$. In our simulations, we use between $100$ and $800$ initial random vectors, depending on the system size, which is quite a high number (see discussion in Ref.~\cite{Schnack2020} which averages over $100$ initial states). The error bars are computed using standard jackknife techniques to correctly take into account the important correlations between the numerator and denominators in all thermal expectations values~\cite{Aichhorn2003,Schnack2020}.

Specific aspects of our computations are: (i) the presence of impurities imply the absence of translation symmetries (for some impurity patterns, a reflection symmetry is still present), (ii) we measure expectation values of observables (such as the orphan magnetization $S^z_{\rm orphan}$) which do not commute with $H$, hence requiring to store all Krylov vectors~\cite{wietek_thermodynamic_2019}. Even though we use total magnetization $S^z_{\rm total}$ conservation and spin inversion $S_i^z \rightarrow -S_i^z$  symmetry, the first point implies very large Hilbert spaces, especially for large spin $S$ values. The second point implies a much larger memory requirement than for most applications of the typicality method which deal with observables which commute with the Hamiltonian. We mitigate this by storing all Krylov vectors in parallel for the largest samples. Both computationally demanding points limit the typicality method to kagom\'e samples with a relatively small number of spins. We can nevertheless reach samples with up to $N=25$ spins for $S=1/2$ and up to $N=10$ spins for $S=4$. 

In the following, we will present results for the total susceptibility of the sample $\chi_{\rm tot}= \langle (S^z_{\rm total})^2\rangle / T$, the local susceptibility, defined as (see {\it e.g.} ~\cite{gregor_nonmagnetic_2008}) $\chi_{loc}=\langle S^z_{\rm orphan}S^z_{\rm total}\rangle/T$, or the magnetization curve $S^z_{\rm orphan}(h)$. Here $S^z_{\rm orphan}$ is the magnetization at an orphan site.

\subsection{Local Susceptibility}

\begin{figure}[h]
\includegraphics[width=\hsize]{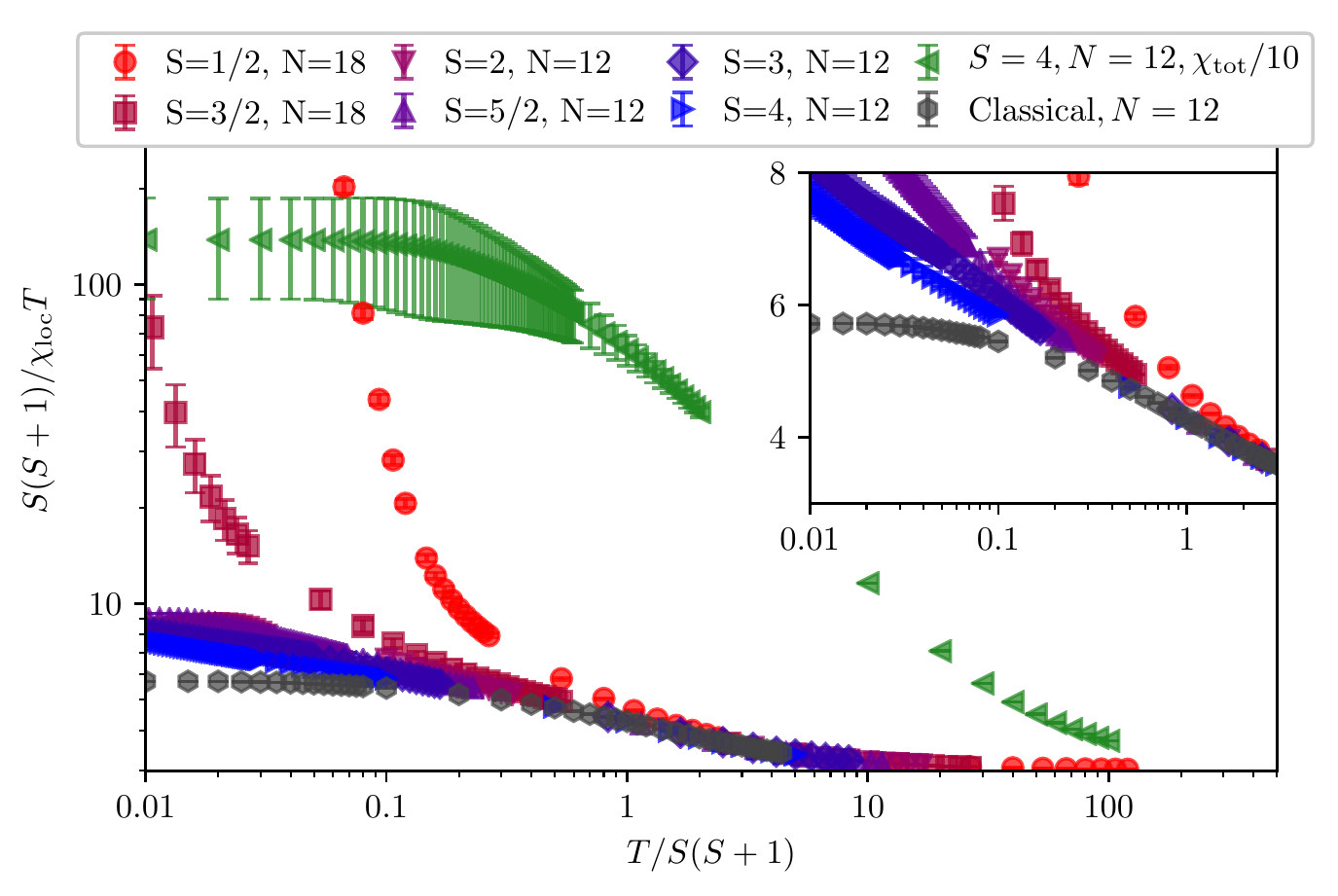}
\caption{Scaled inverse local susceptibility at the orphan spin site for different values of the spin-$S$. 
A comparison to the average susceptibility per site for $S=4$ for the 12-site lattice with a pair impurity is shown. This is calculated by normalizing the total susceptibility by number of spins.
Inset: magnified version where variations between different spin values can be clearly seen.}
\label{fig:chi_local_S4}
\end{figure}

We compute the magnetic response of the spin closest to non-magnetic impurities: it is in most cases the orphan spin (such as in Figs.~\ref{figlat}a,~\ref{figlat}c,~\ref{figlat}d, and~\ref{figlat}f), but we also consider the closest spin for other impurity patterns (such as in Fig.~\ref{figlat}b and~\ref{figlat}e). Motivated by the classical expectations outlined in the Introduction, we study the local susceptibility as a sensitive probe of orphan physics. 

We compare in particular in Fig.~\ref{fig:chi_local_S4} the local susceptibility of the orphan spin for the $12$-sites kagom\'e sample with a pair of impurities (Fig.~\ref{figlat}a) for various values of spin $S$ to the classical result. To provide a simple comparison, we scale $T\to T/S(S+1)$ and $1/(\chi_{loc}T)\to  S(S+1)/(\chi_{loc}T)$, in analogy with the classical case. We find an agreement between the classical version for a range of temperatures lying between a (spin-dependent) minimum temperature and arbitrarily large temperatures. The inset of Fig.~\ref{fig:chi_local_S4} allows us to identify the dependence on the spin $S$ of the typical temperature scale above which the local susceptibility is almost indistinguishable from the classical response. As expected, this temperature decreases with increasing spin $S$, but note that it does so, even when rescaled by $S(S+1)$. More strikingly, we find that for $S\geq 2$, the inverse local susceptibility $S(S+1)/(\chi_{loc}T)$ appears to converge to a finite value (around $
\sim 7-10$, depending on $S$), relatively close to the one of the classical case ($6$ in theory, around $\sim 5.5$ for the $12$-site samples, as shown in the Appendix). The lower spin values ($S=1/2$ and $S=3/2$), for which we can use a larger 18-sites sample, depart earlier from the classical case (as expected); the major difference lying at low temperatures where the inverse response $S(S+1)/(\chi_{loc}T)$ diverges. Clearly, this difference has its origins in the fact that the ground state for these cases does not host a half-orphan, {\em i.e.} is a spin singlet and does not lie in the total spin sector $S_{\rm GS} \simeq [S/2]$.

A clear signature of the orphan spin is found by comparing its local (inverse) susceptibility to the averaged (inverse) susceptibility per site, also represented for $S=4$ in Fig.~\ref{fig:chi_local_S4}: as is readily seen in this figure, the orphan spin is approximately an order of magnitude more sensitive to magnetic field at low temperatures.
This is justified by assuming that a large portion of the magnetic response of the entire system is provided by the orphan spin,
implying that the averaged susceptibility is $1/N$ times the orphan susceptibility, where $N$ is the number of spins ($10$ in this case).
It is important to understand whether this signature is unique to the orphan spin, or could also be found for spins close to the non-magnetic impurities for other impurity patterns. To answer this question, we study two other kinds of impurities (i) a single site impurity and (ii) two impurities on neighboring plaquettes (this case corresponds to Figs.~\ref{figlat}b and~\ref{figlat}e). 

\begin{figure}[h]
\includegraphics[width=\hsize]{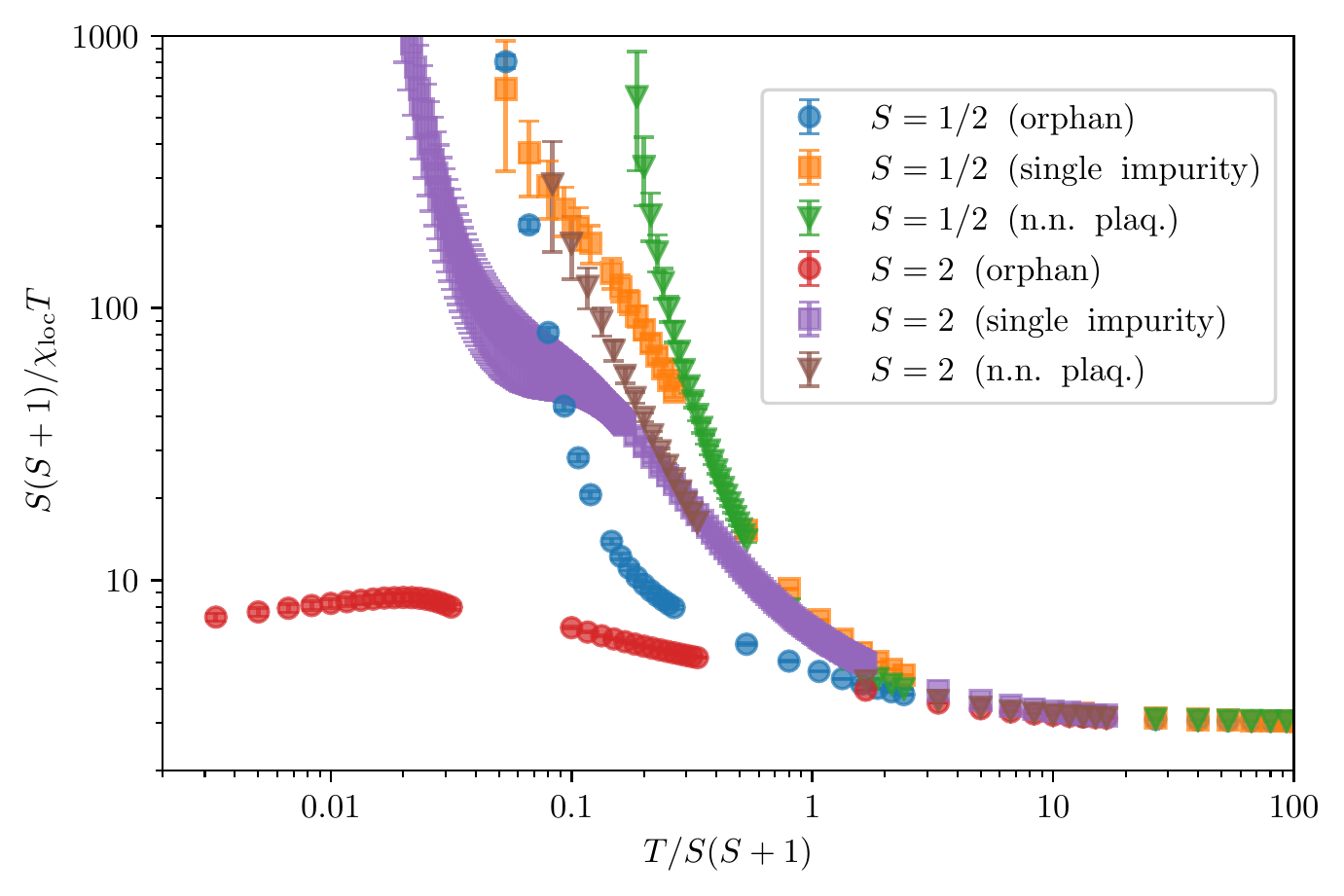}
\caption{Comparison between the scaled inverse local susceptibility for the orphan spin situation and other impurity patterns, for $S=2$ on a $12$-site sample and $S=1/2$ on a 
$18$-site sample.}
\label{fig:chi_compare_other_imp}
\end{figure}

For the former case, we consider the local susceptibility of any spin close to the impurity, while for the latter, we consider the spin in between the two impurities (blue dot in Figs.~\ref{figlat}b and~\ref{figlat}e). The corresponding three scaled inverse local susceptibilities are represented in Fig.~\ref{fig:chi_compare_other_imp} for $S=2$ (representative of the generic $S\geq 2$ case) as well as for $S=1/2$. In both cases, we find the magnetic response to be significantly stronger for the orphan spin for low to medium ($T \approx S(S+1)$) temperatures.

\begin{figure}[h]
\includegraphics[width=\hsize]{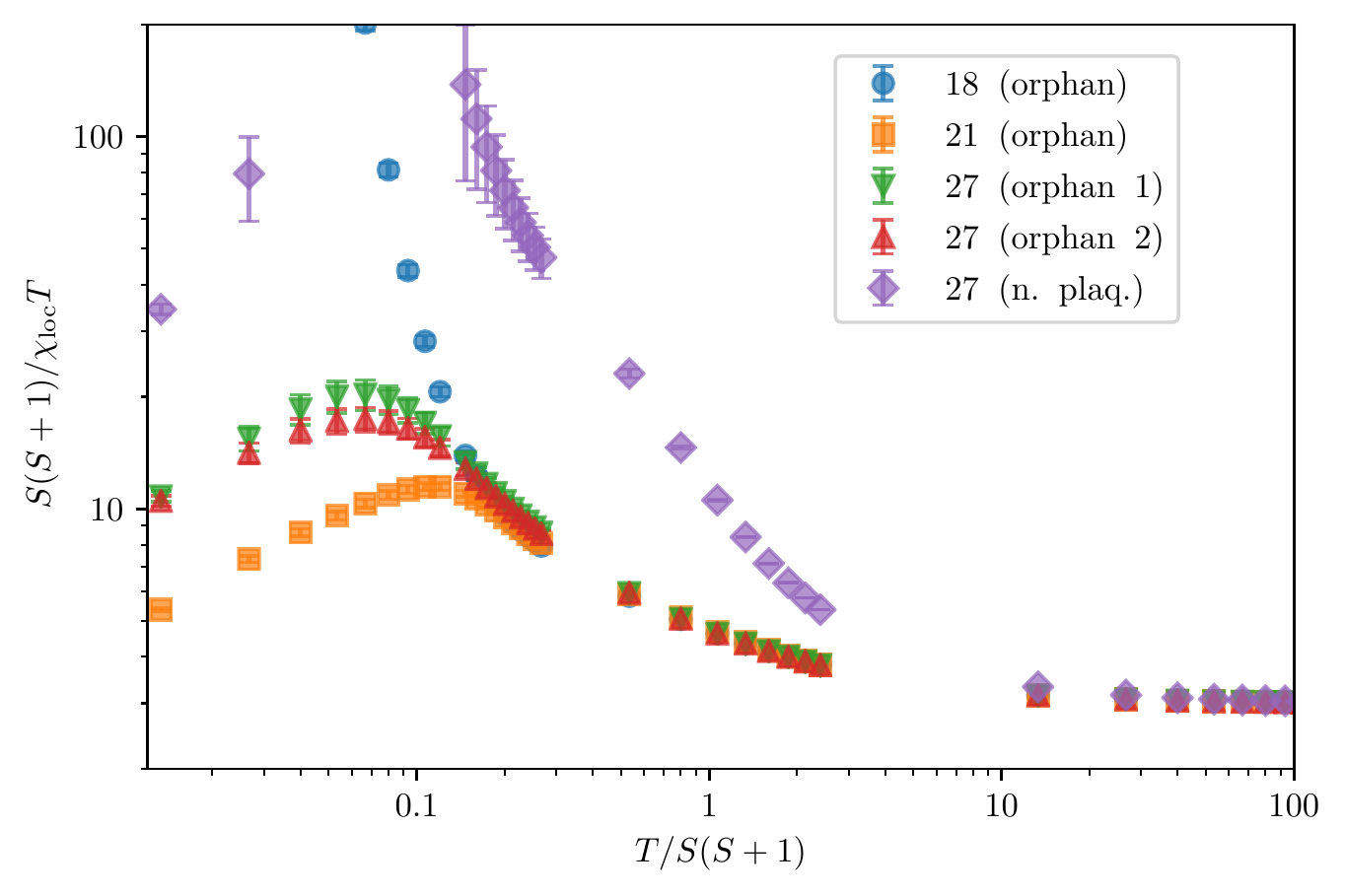}
\caption{Scaled inverse local susceptibility comparisons of the orphan spin for
$S=1/2$ on various samples of large sizes. For the $27$-site sample, we consider two situations: one with two orphan spins (for which we distinguish the response) and one with impurities on neighboring plaquettes.}
\label{fig:chi_local_S0.5}
\end{figure}

The spin-$1/2$ case is of special importance for several compounds~\cite{Hiroi2001,Mendels2007,Mendels_2011,Arh2020}, for which local magnetic responses can be probed by NMR. Motivated by this, we study in more details the $S=1/2$ case for a
larger variety of sample sizes and impurity patterns. In Fig.~\ref{fig:chi_local_S0.5}, we study the orphan
spin configuration for an $18$-site kagom\'e sample (with effectively $16$ spins), a $21$-site sample
($19$ spins effective) and a $27$-site sample with two orphan spins which are maximally
separated ($23$ spins effective), as well as the same $27$-site sample with inpurities on neighboring plaquettes ($25$ sites effective). All of these configurations are shown in Figs.~\ref{figlat} and ~\ref{fig:impurity_pattern_27}.  When the number of spins is even, the ground-state can be and is a global singlet.
This feature is absent in all other cases with an odd number of spins, leading to a divergence in $\chi_{\rm loc}$ at low temperatures. Besides the lattice independent behavior in the high temperature range, we observe a clear distinction in the medium to high-temperature range in the local response between the orphan spin and other spins close to impurities, before the low-temperature signal (dominated by the ground-state nature) eventually appears. For
the case of the lattice with two orphan impurities, both orphans are chosen
to be equivalent from the perspective of lattice geometry and we expect that
they should have identical behavior, which is what we observe within error bars.

\subsection{Orphan magnetization curve and effects of the screening cloud}
\begin{figure}[h]
\includegraphics[width=\hsize]{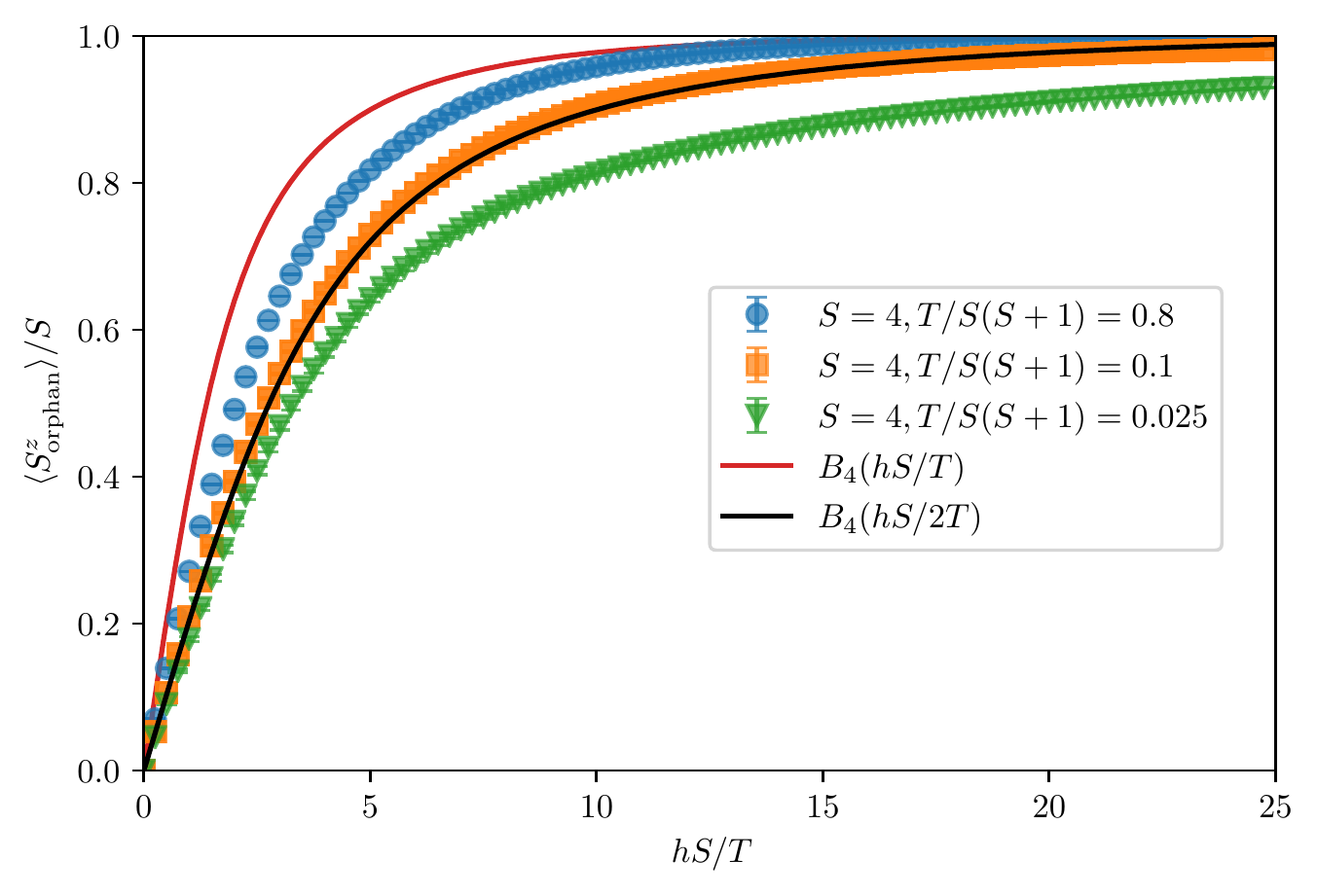}
\caption{Scaled orphan magnetization curve $\braket{S^z_{\rm orphan}}/S$ as a function of scaled applied field $hS/T$ at
different temperatures, for a spin $S=4$ $12$-sites kagom\'e sample. The continuous curves are Brillouin functions at fields $h/2$ and $h$.}
\label{fig:Brillouin_S4}
\end{figure}

Another important feature of orphan physics which is important to test for in the quantum case is the effect
of the screening cloud on the orphan spin itself. As discussed in the Introduction, in the classical case the cloud screens half of the applied magnetic field, implying that the orphan spin responds like a free spin $S$ in a magnetic field $h/2$. This
description should work only in the temperature range where quantum and thermal
effects are not strong, i.e, in an intermediate temperature range with respect to $JS^2$, the coupling value.

Fig.~\ref{fig:Brillouin_S4} displays a comparison between the scaled magnetization curve of the orphan spin $\langle S^z_{\rm orphan} (h,T) \rangle /S$ for the spin $S=4$ kagom\'e $12$-sites sample for three different temperatures, and compares it to
the expected  Brillouin functions for the full field $B_S(hS/T)$ or half-field $B_S(hS/2T)$, where
\begin{equation*}
B_{S}(x=hS/T)={\frac {2S+1}{2S}}\mathrm {coth} \left({\frac {2S+1}{2S}}x\right)-{\frac {1}{2S}}\mathrm {coth} \left({\frac {x}{2S}}\right)
\end{equation*}
is the response to a magnetic field $h$ of a free spin-$S$ at temperature $T$. We find good agreement between the typicality data and the half-field response for a temperature $T/(S(S+1)) \approx 0.1$, and expected deviations at low and high temperature. We discuss below how to quantify the temperature scale for which the agreement is the best. We find similar behavior for lower spins, all the way down to spin-1/2. 

\begin{figure}[h]
\includegraphics[width=\hsize]{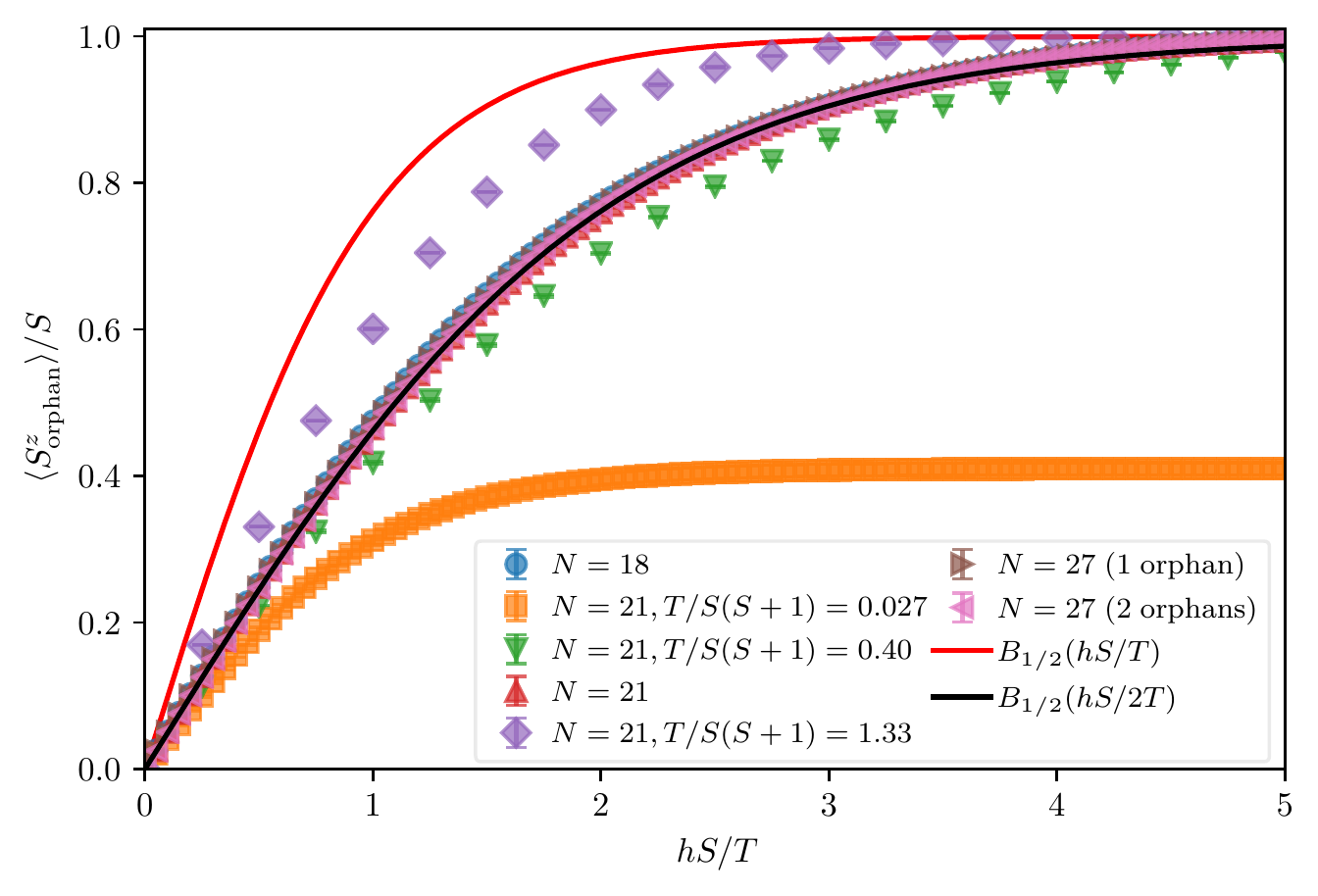}
\caption{Scaled orphan magnetization curve $\braket{S^z_{\rm orphan}}/S$ as a function of scaled applied field $hS/T$, for various $S=1/2$ kagom\'e samples. For the $27$-sites kagom\'e sample, we consider the case with two equivalent orphan spins. The temperature is $T/S(S+1)=0.53$, unless notified in the legend. Also shown are the expected Brillouin functions for spin $S=1/2$
at temperature $T$ in field $h$ and $h/2$.}
\label{fig:Brillouin_S0.5}
\end{figure}

In particular, motivated by the specific signature of the orphan local susceptibility in Fig.~\ref{fig:chi_local_S0.5}, we also study the field dependent response of the orphan impuritiy complexes at different temperatures for the low-spin case $S=1/2$. Once again, we
find a temperature range where the screening cloud of the orphan provides a net cancellation of
half the magnetic field, making the response of the orphan spin consistent with
a free spin-$S$ in a magnetic field $h/2$. This can be observed in Fig.~\ref{fig:Brillouin_S0.5}, for a different set of lattice sizes where the adherence to the half-field Brillouin function is consistent for different samples with the same temperature scale. An interesting feature to note here is the behavior of the orphan impurity complexes in the $27$-sites sample with two such complexes.
The screening cloud appears to be highly local as each of the two orphan 
spins respond exactly as a lattice with only one orphan spin. Furthermore, we find a plateau at low temperatures in the orphan response $\langle S^z_{\rm orphan} \rangle$ as well as an approach to its saturation value through a series of plateaus at higher
fields (not shown in Fig.~\ref{fig:Brillouin_S0.5}). This is due to the discrete nature of the finite-size spectrum on such a small lattice (and at such low temperature) and appears without any  relationship to other well-known quantized plateaus that exist in the clean case~\cite{Schulenburg2002,Cabra2005,Nishimoto2013,Capponi2013b}.

The field dependence discussed above only appears for a particular temperature
range which is controlled by thermal and quantum correlations. At high
temperatures, the thermal correlation length is too short for the texture
to survive, and at low temperatures the quantum correlation length is too long
for the classical orphan description to work. It was noted in Ref.~\onlinecite{sen_fractional_2011},
that in the classical case, the expectation that the orphan texture in the SCGO
bilayer provides exactly $h/2$ shielding, works at temperatures below 
$T/S^2\approx 0.2$. In the case of the kagom\'e lattice, the classical results presented in the Appendix
indicate that the orphan physics is slightly less stable to temperature (see Fig.~\ref{fig:chi_imp_classical}) and we must thus expect a lower $T/S^2$ bound. This sets an approximate upper bound of 
temperature congruent with orphan physics for the quantum case as well for large spin-$S$. 

To make a more quantitative prediction for this temperature range, we define the integrated square of the difference between the numerical data obtained with the typicality method and the expected form of the field response of the orphan. Considering the analytical Brillouin function for spin-$S$ in magnetic field $h/2$, we compute the integral
\be\label{isd}
I=\int dh (\langle S^z_{\rm orphan}\rangle (h,T)-B_S(h/2,T))^2,
\ee
where $T$ is fixed. This quantity approaches zero only when the data fits the
function to high accuracy in the window defined above. For numerical convenience, we define the integral to run from $h=0$ to an $h$ satisfying $B_{1/2}(hS/2T)=0.9$. We have checked that this does not affect conclusions.

\begin{figure}[h]
\includegraphics[width=\hsize]{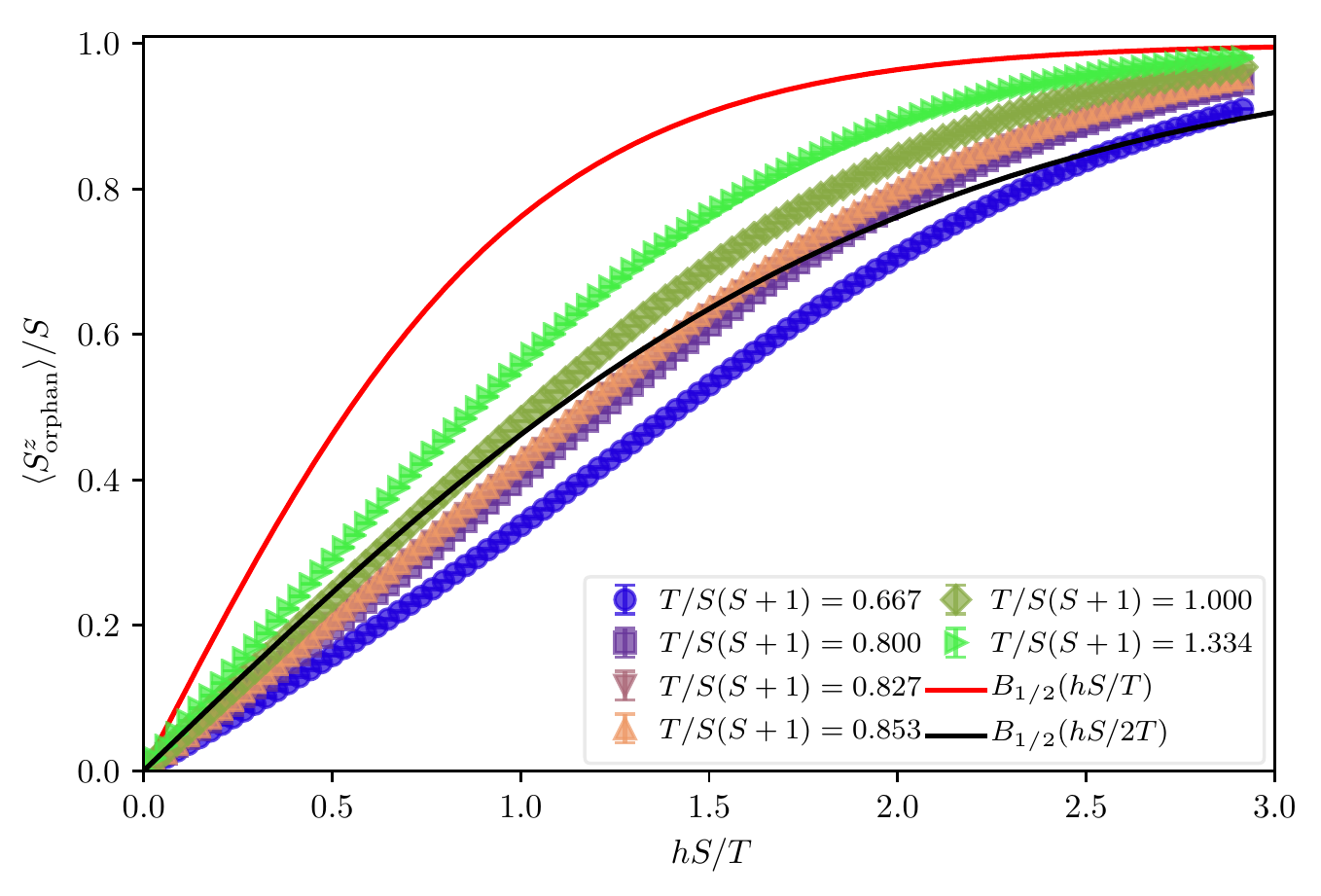}
\caption{Scaled orphan magnetization curve for S=1/2 27-site lattice with neighboring
plaquette impurities (Fig.~\ref{fig:impurity_pattern_27}c), where the points are shown for the range of $hS/T$
used for the integrated difference.}
\label{fig:Brillouin_nplaq}
\end{figure}

\begin{figure}
\includegraphics[width=\hsize]{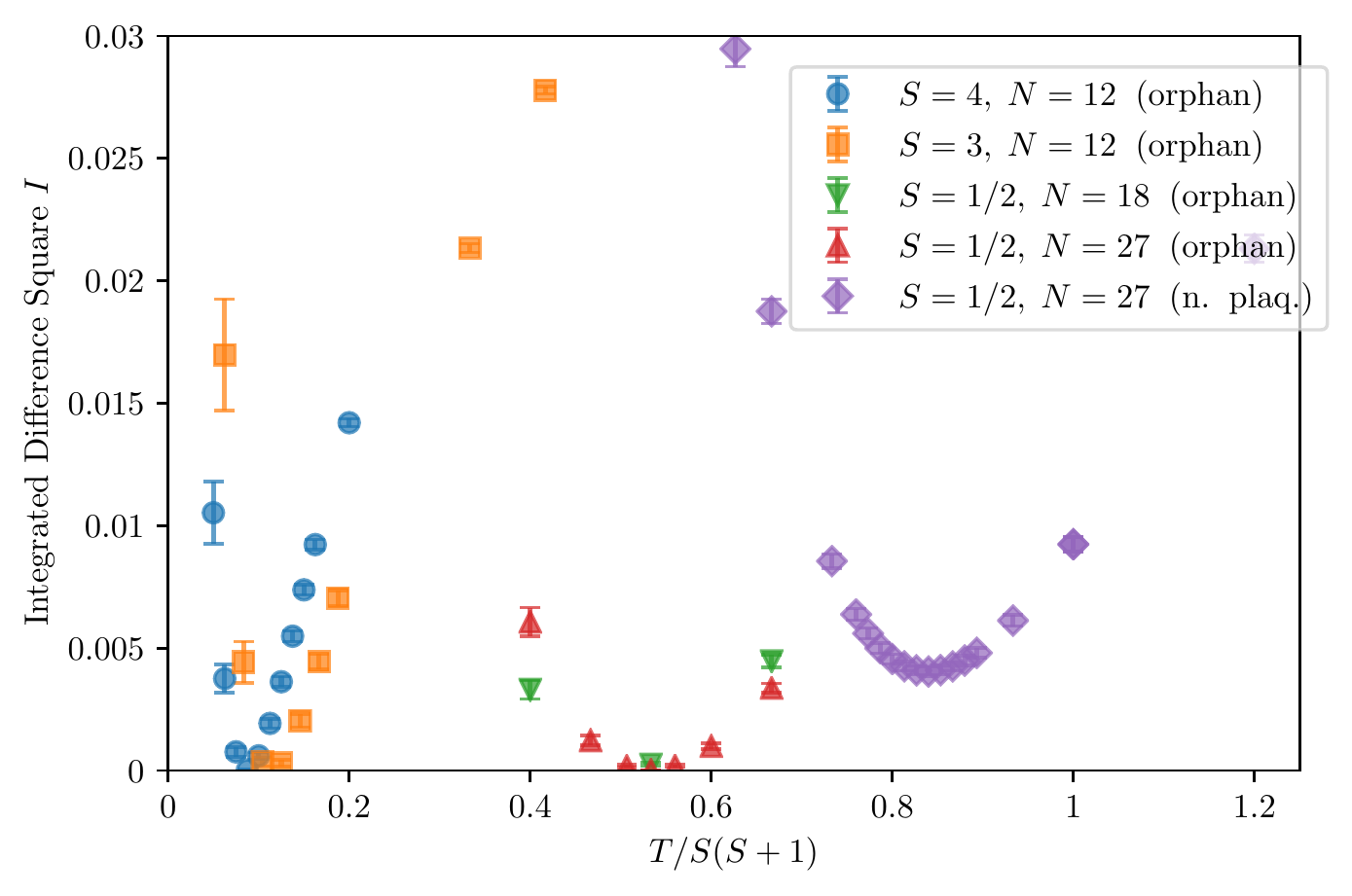}
\caption{Integrated difference square between the orphan magnetization curve and the half-field Brillouin function, as defined in Eq.~\eqref{isd}, for
different values of $S$ and as a function of scaled temperature $T/S(S+1)$.}
\label{fig:integrated_diff}
\end{figure}

We show results for the quantum case in Fig.~\ref{fig:integrated_diff}, for $S=3$ and $4$ on the
$12$-site lattice with an orphan spin, along with $S=1/2$ results on larger
lattices. For all spin values, we find a temperature range where the data are extremely close to the half-field Brillouin function. This region can also be estimated directly from Fig.~\ref{fig:chi_local_S4} by comparing the susceptibility with the expected value for the orphan defect.
The deviations from orphan behavior at high temperatures arise from the same thermal fluctuations which melt orphan defects for the classical case, whereas the deviations at low temperatures arise from strong quantum fluctuations expected to render the spin liquid description inapplicable.
We chose in Figs.~\ref{fig:Brillouin_S4} and~\ref{fig:Brillouin_S0.5} the temperature close to the minima of $I$ for $S=4$ and $S=1/2$ respectively, but let us emphasize that displaying data at temperatures corresponding to values of $I \lesssim 0.01$ (respectively $I \lesssim 0.002$) would result in curves which would not be distinguishable for $S=4$ (respectively $S=1/2$) on the scale of Figs.~\ref{fig:Brillouin_S4} and~\ref{fig:Brillouin_S0.5}.
Based on the numerical evidence provided in Fig.~\ref{fig:integrated_diff}, this leaves a fair range of temperatures where the orphan physics could be probed experimentally with local probes for instance.

As can be furthermore noted in Fig.~\ref{fig:integrated_diff}, there is no significant
difference between $18$ and $27$-sites samples for $S=1/2$, implying that the
orphan shielding is fairly local (as also expected from Refs.~\cite{sen_fractional_2011,sen_vacancy_2012}).
We also compare to the case where the two impurities are on neighboring
plaquettes to show that a strong $h/2$ shielding does not occur for any temperature
range in this case. Although there is a minimum in the integrated difference square measure, the values at the minimum are significantly larger than corresponding values in cases where orphan spins are present. The corresponding magnetization curves are also quite distinct from the half-field Brillouin function (as seen in Fig.~\ref{fig:Brillouin_nplaq}).

\subsection{Local susceptibility distribution for doped $S=1/2$ kagom\'e systems}

In order to make connections with possible experimental detection of orphan physics using NMR, we  develop in this section a crude estimate for the distribution of 
local susceptibilities on large kagom\'e samples for the case $S=1/2$. 

\begin{figure}
\begin{minipage}[t]{0.23\textwidth}
\includegraphics[width=\textwidth]{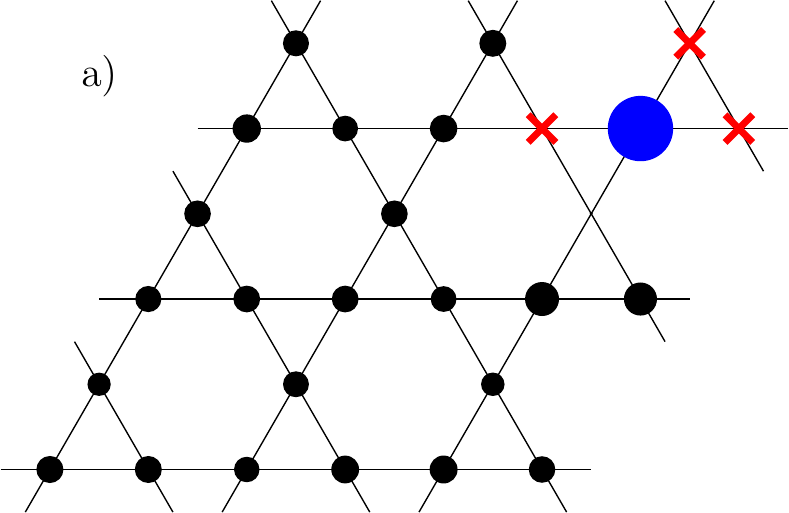}
\end{minipage}
\begin{minipage}[t]{0.23\textwidth}
\includegraphics[width=\textwidth]{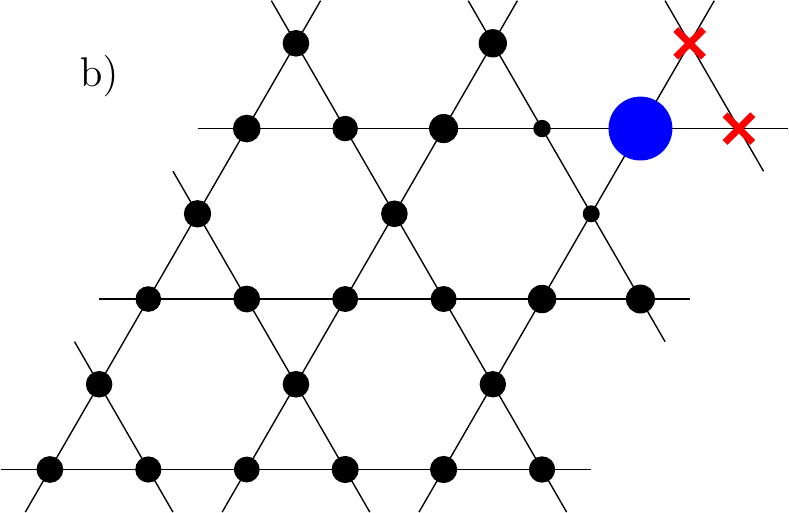}
\end{minipage}
\begin{minipage}[t]{0.23\textwidth}
\includegraphics[width=\textwidth]{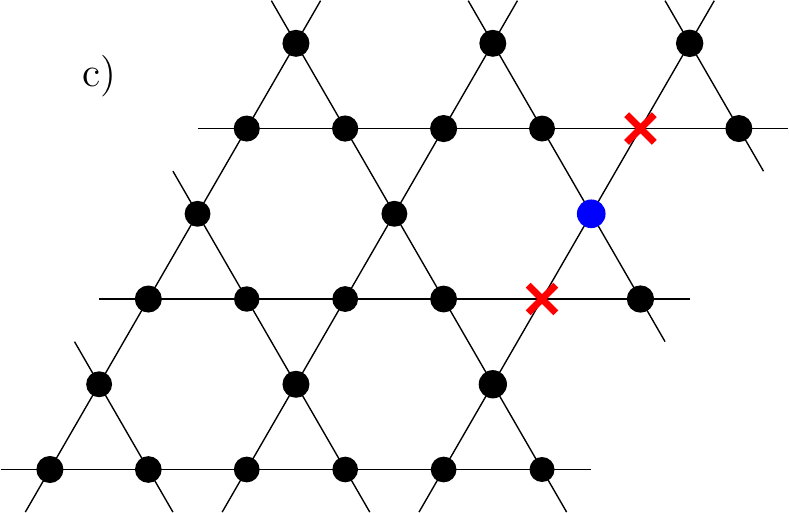}
\end{minipage}
\begin{minipage}[t]{0.23\textwidth}
\includegraphics[width=\textwidth]{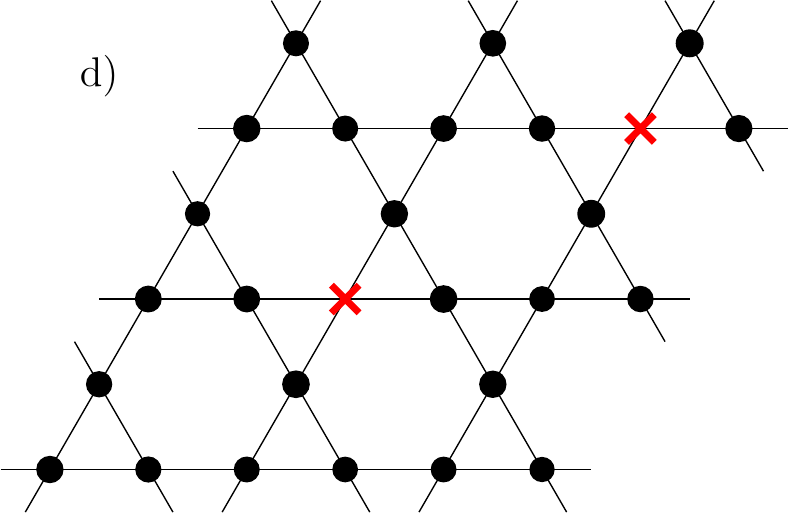}
\end{minipage}
\caption{Kagom\'e $27$-site samples showing the various non-magnetic impurities (represented as red crosses) pattern considered for $S=1/2$ experimental
predictions (see text): (a) triple defect, (b) orphan spin, (c) impurities on neighboring plaquettes, (d) two single impurities sufficiently separated. The case of two maximally separated orphans is the same as (d) with orphans in the locations of the crosses and the impurities in the triangle right of the orphan. The cases with all or no neighbours being impurities are not represented for simplicity.  The radii of the circles is proportional to the value of the local
susceptibility $\chi_{\rm loc}$ at that site, at $T=0.4$. Some sites may appear empty as the radius is too small to be visible.}
\label{fig:impurity_pattern_27}
\end{figure}

The procedure we consider is as follows. We work with a large lattice with $3\times300\times300=270,000$ sites to be representative of the thermodynamic limit. On this lattice, we randomly pick sites to host a non-magnetic impurity with probability $0.1$. This non-magnetic impurity doping  fraction of $10\%$ is chosen to ensure that we get a sizeable number of orphans. 
The choice of temperature is guided by Fig.~\ref{fig:Brillouin_S0.5}, where we see
that $T/S(S+1)=0.53$ (or $T=0.4$) yields a good agreement with the orphan
picture. At this temperature, we study a range of impurity patterns on the $27$-sites
sample and find that only certain impurity patterns produce a local 
susceptibility which differs significantly from that of the pure system. Using
this information, we find in particular that there are only six patterns which are relevant
for sites which host a magnetic ion, some of which are presented pictorially in Fig.~\ref{fig:impurity_pattern_27}. The motivation to create this simple division is driven by the observation that for a temperature $T=0.4$, quantum correlations are quite weak and the
impurities do not influence significantly the sites that are separated by more
than one bond, confirming that the physics is very local. In increasing order of probabilities of occurrence and together with a description of their magnetic response, these six patterns are:

\begin{itemize}
    \item[P1.] All four neighbors of the ion are non-magnetic impurities. This implies that
the spin is completely decoupled from the lattice, and acts as a free spin $S=1/2$.
\item[P2.] Three neighbors are non-magnetic impurities: this is the situation where an orphan spin has an additional non-magnetic impurity next to it (see Fig.~\ref{fig:impurity_pattern_27}a). We find that the
response of this spin is only slightly stronger than that of the orphan spin.
\item[P3.] Two neighbors are non-magnetic impurities  (Fig.~\ref{fig:impurity_pattern_27}b): this is the orphan spin which response has been detailed earlier in this section.
\item[P4.] Two neighbors are non-magnetic impurities on neighboring plaquettes but do not form an orphan (see Fig.~\ref{fig:impurity_pattern_27}c). This has a 
response which is slightly stronger than the pure case.
\item[P5.] Only one neighbor is a non-magnetic impurity (Fig.~\ref{fig:impurity_pattern_27}d which displays two of these cases). We find also that this situation behaves similarly to the
pure case with a slightly stronger response.
\item[P6. ]No impurities are in the nearest neighbors. Response is well approximated by
the pure case.
\end{itemize}

\begin{figure}[!ht]
\includegraphics[width=\hsize]{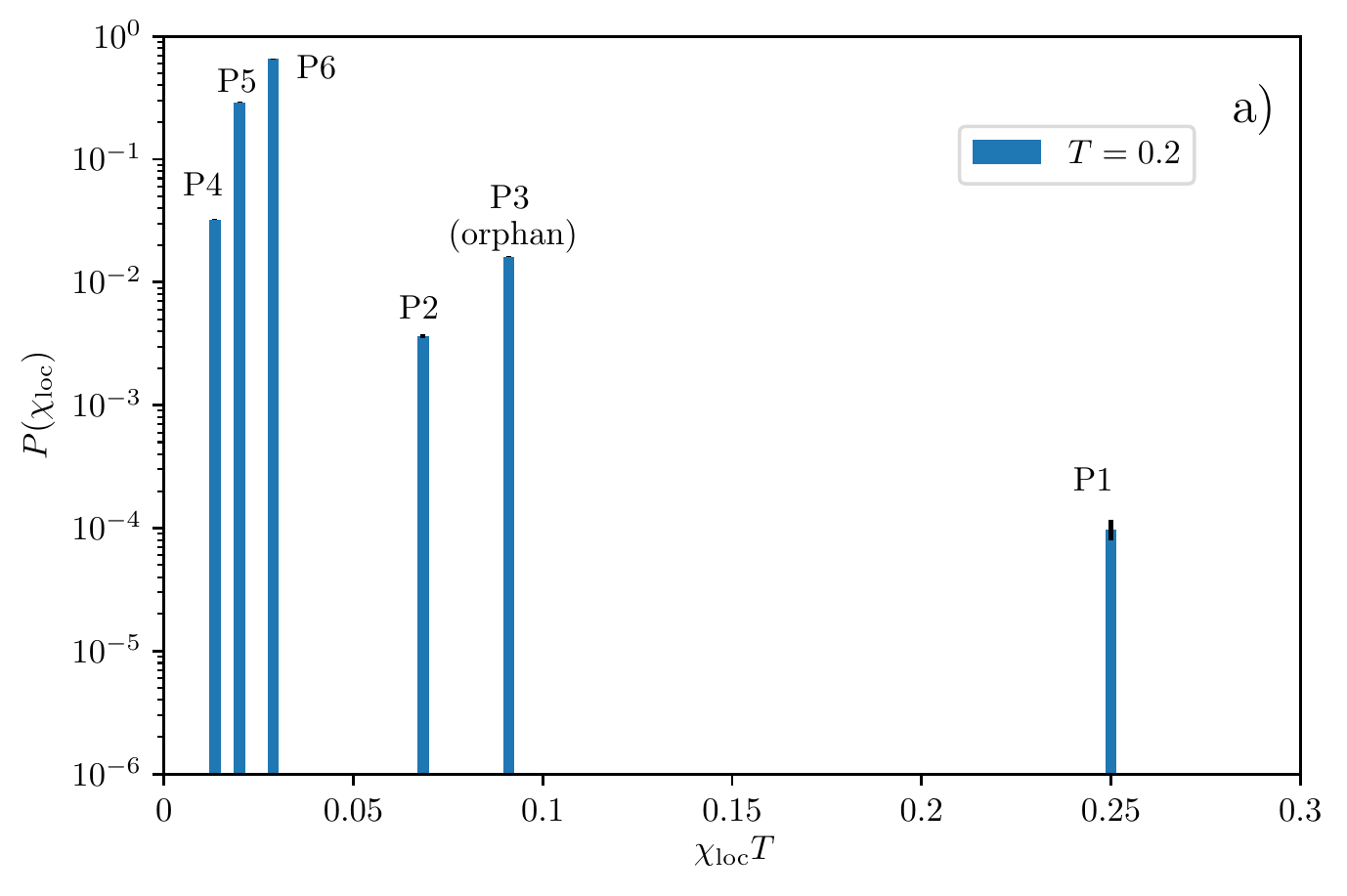}
\includegraphics[width=\hsize]{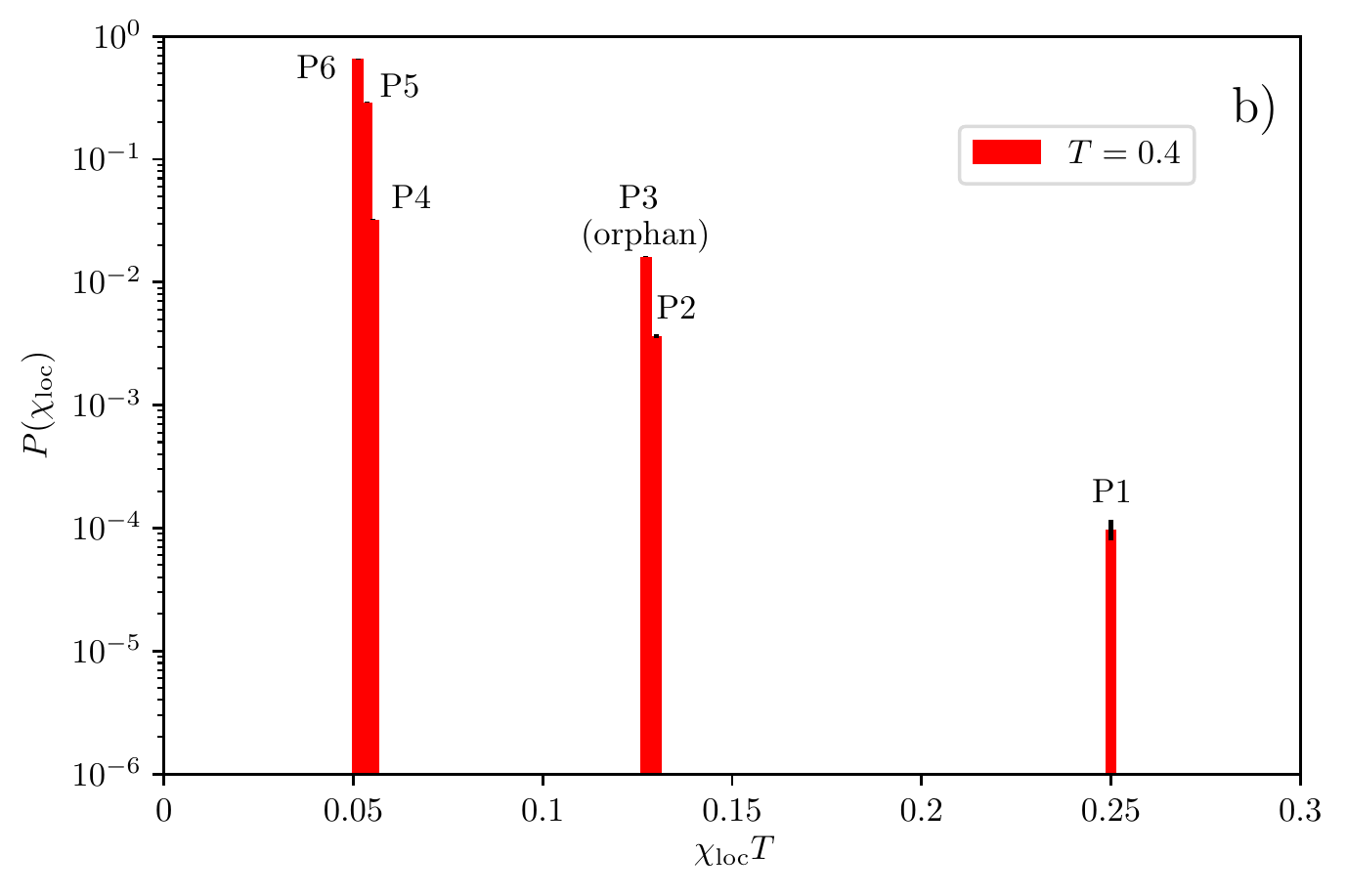}
\includegraphics[width=\hsize]{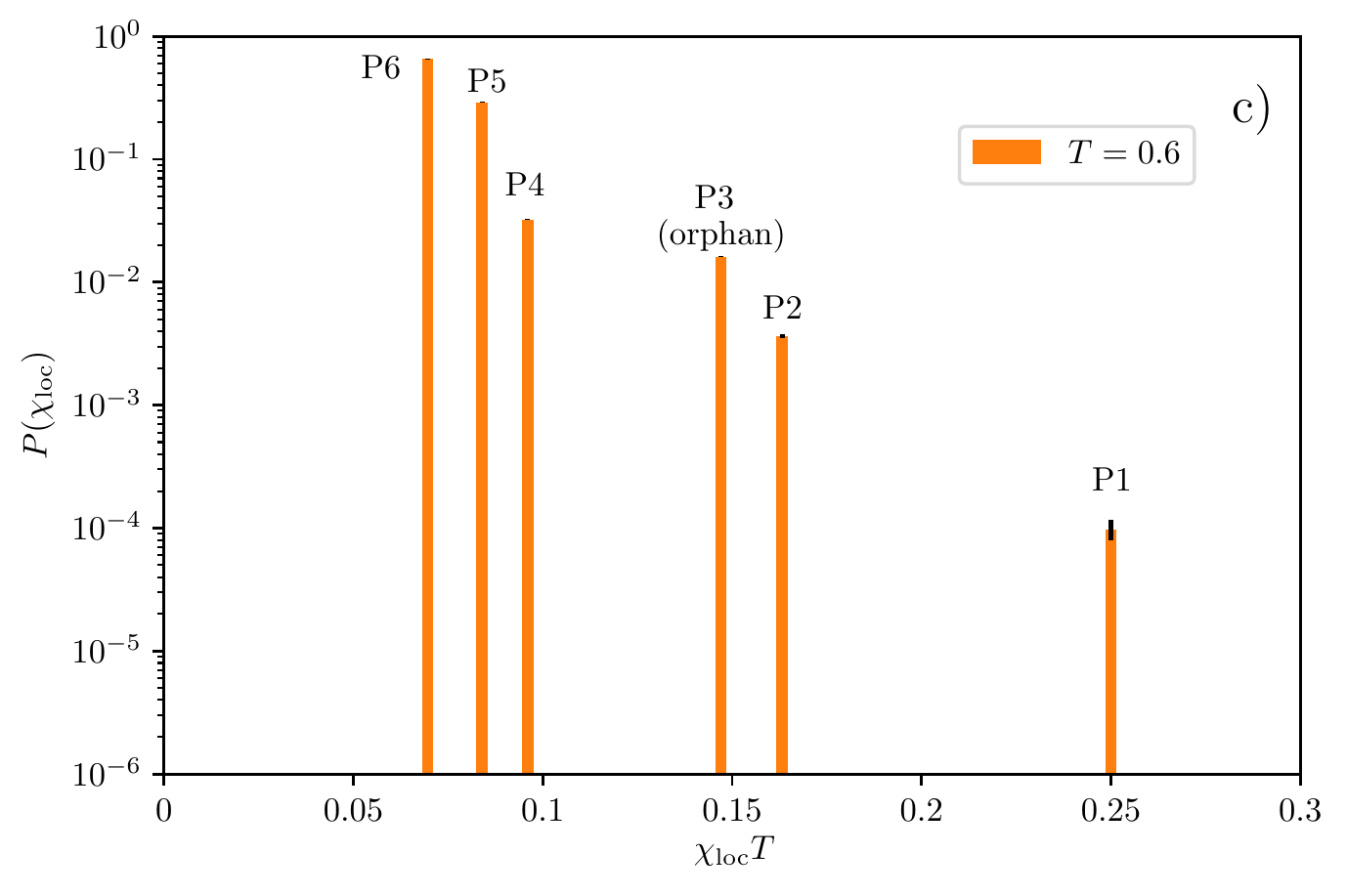}
\caption{The probabilities of occurrence of local susceptibilities in a 270,000-site lattice for $S=1/2$ and $T=0.2$, $0.4$ and $0.6$ from top (a) to bottom (c). Note that the horizontal axis is scaled by $T$.}
\label{fig:chi_distribution}
\end{figure}

The six cases mentioned above lead to six peaks in the local susceptibility
distribution $P(\chi_{\rm loc})$, as displayed in the middle panel of Fig~\ref{fig:chi_distribution} and ranked right to left (the free spin has obviously the largest local susceptibility). Given the very similar responses, it might be hard to distinguish the three left-most peaks which are likely to form a main broad peak and correspond to the pure case as well as few impurities non-orphan physics. Note that in theses cases, the impurity susceptibility is smaller (respectively larger) than the pure one at low (respectively high) temperature, in agreement with recent series expansion results~\cite{bernu2020}. 
The orphan physics is manifested through the presence of secondary peak (where the cases $P2$ and $P3$ might also be hard to resolve too). While this secondary peak is about only $2\%$ of the main peak in 
strength, it is well separated from it, which gives hope for a possible experimental detection.  Also shown in Fig.~\ref{fig:chi_distribution} are results for $T=0.2$ and 
$0.6$, which show the variation of the histogram with slight variations of
the temperature away from the temperature ($T=0.4$) which best fits the
orphan picture. For $T=0.2$, the quantum correlations may be too strong for the approximations $P1-6$
to strictly apply.

\section{Conclusions}\label{Sec5}

In this work, we have investigated whether and how the physics of half-orphans, originally described using a classical theory, survives quantum effects and leads to clearly identifiable signatures in the response of quantum spin-$S$ Heisenberg antiferromagnets on the kagom\'e lattice when nearest-neighbours non-magnetic impurities are present. We found that orphan-specific signatures are evident in local susceptibility data in an intermediate temperature range. We also found that the classical predictions for the screening cloud around an orphan spin are justified in the quantum case. Indeed the resulting spin texture is a more robust signature than the local susceptibility of the orphan spin itself, and appears to provide a window to orphan physics even in the $S=1/2$ case. 
Another striking result, for samples with a single orphan, is the strong dependence of the ground state spin quantum number on the spin-$S$ of the local moment itself. In complete agreement with the classical expectation, we found that the $S\geq 2$ kagom\'e antiferromagnet with an orphan impurity complex has a ground state spin which is the nearest (half-)integer to $S/2$ allowed in most cases. Finally, we also presented experimental predictions for NMR Knight shifts on $S=1/2$ kagom\'e systems doped with non-magnetic impurities, which point to the possibility of observing the enhanced orphan-induced magnetic response through a small, but resolvable, secondary peak.

In the appropriate temperature regime corresponding to spin liquid behaviour, these classical results on half-orphan spins are also expected to apply mutatis mutandis to other corner-sharing lattices such as the pyrochlore. Our results, taken in conjunction with this expectation, therefore provide strong motivation for other related studies. 
For instance, one follow-up suggested by our work is to consider whether the nonzero ground-state spin quantum number that we find in samples with an orphan impurity complex is also present for the Heisenberg model with sufficiently large spin-$S$ on other corner-sharing lattices, for which the original $T=0$ arguments go through unchanged. We hope to report on this in the near future.

A noteworthy feature of the orphan-induced spin texture studied here
is that it appears to be a very local feature that exists in an intermediate temperature window, and this feature seems largely independent of the nature of the ground state, whose character can vary with the spin value $S$. For example,
the Heisenberg kagom\'e antiferromagnet with spin-1/2 has a ground state which is non-magnetic but the precise nature of this spin liquid (gapless vs gapped) is still debated~\cite{Jiang2008,Evenbly2010,Yan2011,Depenbrock2012,Iqbal2013,Capponi2013,kagome_DMRG_U1,Liao2017}; 
in the spin-1 case, the ground state is also non-magnetic although whether it breaks lattice symmetries~\cite{Changlani2015,Liu2015} or not~\cite{Nishimoto2015} is still a matter of debate; for larger values of $S$, spin-wave analysis, coupled-cluster or series expansions methods point to an ordered phase~\cite{Chubukov1992,Goetze2011,Oitmaa2016} while a large-$N$ analysis, which generalizes the symmetry from SU(2) to a larger symmetry group,  leads to a non-magnetic state~\cite{Sachdev1992}.

It would be worth understanding how to incorporate the quantum orphan features found in our work in various effective field theories for these putative ground states~\cite{gregor_nonmagnetic_2008}. Indeed, preliminary results suggest that some signatures of orphan physics are already present when samples with orphan impurity complex are studied using relatively crude extensions of variational treatments inspired by such effective field theories. It would also be interesting to complement our study of quantum half-orphans by series expansion methods; this may be particularly fruitful since the textures induced by the orphans are relatively compact in the temperature range of their existence.

\begin{acknowledgements}

We thank Ph. Mendels and L. Messio for discussions and A.~Wietek for useful insights about the typicality method. KD would like to gratefully acknowledge earlier collaborations on closely related topics with R.~Moessner, A.~Sen, and S.~Sanyal, which provided part of the motivation for this work. This work benefited from the support of the project LINK ANR-18-CE30-0022-04 of the French National Research Agency (ANR). We acknowledge the use of HPC resources from CALMIP (grants 2019-P0677 and 2020-P0677) and GENCI (grant x2020050225). We use the libraries ITensor for DMRG simulations~\cite{itensor} (available at \url{http://itensor.org}), and PETSc~\cite{petsc-efficient,petsc-user-ref}, SLEPc~\cite{slepc-toms,slepc-manual} for the typicality computations. We gratefully thank J. Roman for improving the Krylov methods implementation in SLEPc to further ease typicality computations.

\end{acknowledgements}

\bibliography{orphan}

\begin{thebibliography}{78}%
\makeatletter
\providecommand \@ifxundefined [1]{%
 \@ifx{#1\undefined}
}%
\providecommand \@ifnum [1]{%
 \ifnum #1\expandafter \@firstoftwo
 \else \expandafter \@secondoftwo
 \fi
}%
\providecommand \@ifx [1]{%
 \ifx #1\expandafter \@firstoftwo
 \else \expandafter \@secondoftwo
 \fi
}%
\providecommand \natexlab [1]{#1}%
\providecommand \enquote  [1]{``#1''}%
\providecommand \bibnamefont  [1]{#1}%
\providecommand \bibfnamefont [1]{#1}%
\providecommand \citenamefont [1]{#1}%
\providecommand \href@noop [0]{\@secondoftwo}%
\providecommand \href [0]{\begingroup \@sanitize@url \@href}%
\providecommand \@href[1]{\@@startlink{#1}\@@href}%
\providecommand \@@href[1]{\endgroup#1\@@endlink}%
\providecommand \@sanitize@url [0]{\catcode `\\12\catcode `\$12\catcode
  `\&12\catcode `\#12\catcode `\^12\catcode `\_12\catcode `\%12\relax}%
\providecommand \@@startlink[1]{}%
\providecommand \@@endlink[0]{}%
\providecommand \url  [0]{\begingroup\@sanitize@url \@url }%
\providecommand \@url [1]{\endgroup\@href {#1}{\urlprefix }}%
\providecommand \urlprefix  [0]{URL }%
\providecommand \Eprint [0]{\href }%
\providecommand \doibase [0]{http://dx.doi.org/}%
\providecommand \selectlanguage [0]{\@gobble}%
\providecommand \bibinfo  [0]{\@secondoftwo}%
\providecommand \bibfield  [0]{\@secondoftwo}%
\providecommand \translation [1]{[#1]}%
\providecommand \BibitemOpen [0]{}%
\providecommand \bibitemStop [0]{}%
\providecommand \bibitemNoStop [0]{.\EOS\space}%
\providecommand \EOS [0]{\spacefactor3000\relax}%
\providecommand \BibitemShut  [1]{\csname bibitem#1\endcsname}%
\let\auto@bib@innerbib\@empty
\bibitem [{\citenamefont {Savary}\ and\ \citenamefont
  {Balents}(2016)}]{ExpQSL}%
  \BibitemOpen
  \bibfield  {author} {\bibinfo {author} {\bibfnamefont {Lucile}\ \bibnamefont
  {Savary}}\ and\ \bibinfo {author} {\bibfnamefont {Leon}\ \bibnamefont
  {Balents}},\ }\bibfield  {title} {\enquote {\bibinfo {title} {Quantum spin
  liquids: a review},}\ }\href@noop {} {\bibfield  {journal} {\bibinfo
  {journal} {Reports on Progress in Physics}\ }\textbf {\bibinfo {volume}
  {80}},\ \bibinfo {pages} {016502} (\bibinfo {year} {2016})}\BibitemShut
  {NoStop}%
\bibitem [{\citenamefont {Castelnovo}\ \emph {et~al.}(2008)\citenamefont
  {Castelnovo}, \citenamefont {Moessner},\ and\ \citenamefont
  {Sondhi}}]{castelnovo2008magnetic}%
  \BibitemOpen
  \bibfield  {author} {\bibinfo {author} {\bibfnamefont {Claudio}\ \bibnamefont
  {Castelnovo}}, \bibinfo {author} {\bibfnamefont {Roderich}\ \bibnamefont
  {Moessner}}, \ and\ \bibinfo {author} {\bibfnamefont {Shivaji~L}\
  \bibnamefont {Sondhi}},\ }\bibfield  {title} {\enquote {\bibinfo {title}
  {Magnetic monopoles in spin ice},}\ }\href@noop {} {\bibfield  {journal}
  {\bibinfo  {journal} {Nature}\ }\textbf {\bibinfo {volume} {451}},\ \bibinfo
  {pages} {42--45} (\bibinfo {year} {2008})}\BibitemShut {NoStop}%
\bibitem [{\citenamefont {Kitaev}(2006)}]{kitaev2006anyons}%
  \BibitemOpen
  \bibfield  {author} {\bibinfo {author} {\bibfnamefont {Alexei}\ \bibnamefont
  {Kitaev}},\ }\bibfield  {title} {\enquote {\bibinfo {title} {Anyons in an
  exactly solved model and beyond},}\ }\href@noop {} {\bibfield  {journal}
  {\bibinfo  {journal} {Annals of Physics}\ }\textbf {\bibinfo {volume}
  {321}},\ \bibinfo {pages} {2--111} (\bibinfo {year} {2006})}\BibitemShut
  {NoStop}%
\bibitem [{\citenamefont {Poilblanc}\ \emph {et~al.}(2006)\citenamefont
  {Poilblanc}, \citenamefont {L\"auchli}, \citenamefont {Mambrini},\ and\
  \citenamefont {Mila}}]{Poilblanc2006}%
  \BibitemOpen
  \bibfield  {author} {\bibinfo {author} {\bibfnamefont {Didier}\ \bibnamefont
  {Poilblanc}}, \bibinfo {author} {\bibfnamefont {Andreas}\ \bibnamefont
  {L\"auchli}}, \bibinfo {author} {\bibfnamefont {Matthieu}\ \bibnamefont
  {Mambrini}}, \ and\ \bibinfo {author} {\bibfnamefont {Fr\'ed\'eric}\
  \bibnamefont {Mila}},\ }\bibfield  {title} {\enquote {\bibinfo {title}
  {Spinon deconfinement in doped frustrated quantum antiferromagnets},}\ }\href
  {\doibase 10.1103/PhysRevB.73.100403} {\bibfield  {journal} {\bibinfo
  {journal} {Phys. Rev. B}\ }\textbf {\bibinfo {volume} {73}},\ \bibinfo
  {pages} {100403} (\bibinfo {year} {2006})}\BibitemShut {NoStop}%
\bibitem [{\citenamefont {L\"auchli}\ and\ \citenamefont
  {Poilblanc}(2004)}]{Laeuchli2004}%
  \BibitemOpen
  \bibfield  {author} {\bibinfo {author} {\bibfnamefont {Andreas}\ \bibnamefont
  {L\"auchli}}\ and\ \bibinfo {author} {\bibfnamefont {Didier}\ \bibnamefont
  {Poilblanc}},\ }\bibfield  {title} {\enquote {\bibinfo {title} {Spin-charge
  separation in two-dimensional frustrated quantum magnets},}\ }\href {\doibase
  10.1103/PhysRevLett.92.236404} {\bibfield  {journal} {\bibinfo  {journal}
  {Phys. Rev. Lett.}\ }\textbf {\bibinfo {volume} {92}},\ \bibinfo {pages}
  {236404} (\bibinfo {year} {2004})}\BibitemShut {NoStop}%
\bibitem [{\citenamefont {Alloul}\ \emph {et~al.}(2009)\citenamefont {Alloul},
  \citenamefont {Bobroff}, \citenamefont {Gabay},\ and\ \citenamefont
  {Hirschfeld}}]{Alloul2009}%
  \BibitemOpen
  \bibfield  {author} {\bibinfo {author} {\bibfnamefont {H.}~\bibnamefont
  {Alloul}}, \bibinfo {author} {\bibfnamefont {J.}~\bibnamefont {Bobroff}},
  \bibinfo {author} {\bibfnamefont {M.}~\bibnamefont {Gabay}}, \ and\ \bibinfo
  {author} {\bibfnamefont {P.~J.}\ \bibnamefont {Hirschfeld}},\ }\bibfield
  {title} {\enquote {\bibinfo {title} {Defects in correlated metals and
  superconductors},}\ }\href {\doibase 10.1103/RevModPhys.81.45} {\bibfield
  {journal} {\bibinfo  {journal} {Rev. Mod. Phys.}\ }\textbf {\bibinfo {volume}
  {81}},\ \bibinfo {pages} {45--108} (\bibinfo {year} {2009})}\BibitemShut
  {NoStop}%
\bibitem [{\citenamefont {Martins}\ \emph {et~al.}(1997)\citenamefont
  {Martins}, \citenamefont {Laukamp}, \citenamefont {Riera},\ and\
  \citenamefont {Dagotto}}]{Martins}%
  \BibitemOpen
  \bibfield  {author} {\bibinfo {author} {\bibfnamefont {George~Balster}\
  \bibnamefont {Martins}}, \bibinfo {author} {\bibfnamefont {Markus}\
  \bibnamefont {Laukamp}}, \bibinfo {author} {\bibfnamefont {Jos\'e}\
  \bibnamefont {Riera}}, \ and\ \bibinfo {author} {\bibfnamefont {Elbio}\
  \bibnamefont {Dagotto}},\ }\bibfield  {title} {\enquote {\bibinfo {title}
  {Local enhancement of antiferromagnetic correlations by nonmagnetic
  impurities},}\ }\href {\doibase 10.1103/PhysRevLett.78.3563} {\bibfield
  {journal} {\bibinfo  {journal} {Phys. Rev. Lett.}\ }\textbf {\bibinfo
  {volume} {78}},\ \bibinfo {pages} {3563--3566} (\bibinfo {year}
  {1997})}\BibitemShut {NoStop}%
\bibitem [{\citenamefont {Kaul}\ \emph {et~al.}(2008)\citenamefont {Kaul},
  \citenamefont {Melko}, \citenamefont {Metlitski},\ and\ \citenamefont
  {Sachdev}}]{Kaul}%
  \BibitemOpen
  \bibfield  {author} {\bibinfo {author} {\bibfnamefont {Ribhu~K.}\
  \bibnamefont {Kaul}}, \bibinfo {author} {\bibfnamefont {Roger~G.}\
  \bibnamefont {Melko}}, \bibinfo {author} {\bibfnamefont {Max~A.}\
  \bibnamefont {Metlitski}}, \ and\ \bibinfo {author} {\bibfnamefont {Subir}\
  \bibnamefont {Sachdev}},\ }\bibfield  {title} {\enquote {\bibinfo {title}
  {Imaging bond order near nonmagnetic impurities in square-lattice
  antiferromagnets},}\ }\href {\doibase 10.1103/PhysRevLett.101.187206}
  {\bibfield  {journal} {\bibinfo  {journal} {Phys. Rev. Lett.}\ }\textbf
  {\bibinfo {volume} {101}},\ \bibinfo {pages} {187206} (\bibinfo {year}
  {2008})}\BibitemShut {NoStop}%
\bibitem [{\citenamefont {Sachdev}\ and\ \citenamefont
  {Vojta}(2003)}]{sachdev}%
  \BibitemOpen
  \bibfield  {author} {\bibinfo {author} {\bibfnamefont {Subir}\ \bibnamefont
  {Sachdev}}\ and\ \bibinfo {author} {\bibfnamefont {Matthias}\ \bibnamefont
  {Vojta}},\ }\bibfield  {title} {\enquote {\bibinfo {title} {Quantum impurity
  in an antiferromagnet: Nonlinear sigma model theory},}\ }\href {\doibase
  10.1103/PhysRevB.68.064419} {\bibfield  {journal} {\bibinfo  {journal} {Phys.
  Rev. B}\ }\textbf {\bibinfo {volume} {68}},\ \bibinfo {pages} {064419}
  (\bibinfo {year} {2003})}\BibitemShut {NoStop}%
\bibitem [{\citenamefont {H\"oglund}\ \emph {et~al.}(2007)\citenamefont
  {H\"oglund}, \citenamefont {Sandvik},\ and\ \citenamefont
  {Sachdev}}]{hoglund}%
  \BibitemOpen
  \bibfield  {author} {\bibinfo {author} {\bibfnamefont {Kaj~H.}\ \bibnamefont
  {H\"oglund}}, \bibinfo {author} {\bibfnamefont {Anders~W.}\ \bibnamefont
  {Sandvik}}, \ and\ \bibinfo {author} {\bibfnamefont {Subir}\ \bibnamefont
  {Sachdev}},\ }\bibfield  {title} {\enquote {\bibinfo {title} {Impurity
  induced spin texture in quantum critical 2d antiferromagnets},}\ }\href
  {\doibase 10.1103/PhysRevLett.98.087203} {\bibfield  {journal} {\bibinfo
  {journal} {Phys. Rev. Lett.}\ }\textbf {\bibinfo {volume} {98}},\ \bibinfo
  {pages} {087203} (\bibinfo {year} {2007})}\BibitemShut {NoStop}%
\bibitem [{\citenamefont {Haravifard}\ \emph {et~al.}(2006)\citenamefont
  {Haravifard}, \citenamefont {Dunsiger}, \citenamefont {El~Shawish},
  \citenamefont {Gaulin}, \citenamefont {Dabkowska}, \citenamefont {Telling},
  \citenamefont {Perring},\ and\ \citenamefont {Bon\ifmmode~\check{c}\else
  \v{c}\fi{}a}}]{Haravifard2006}%
  \BibitemOpen
  \bibfield  {author} {\bibinfo {author} {\bibfnamefont {S.}~\bibnamefont
  {Haravifard}}, \bibinfo {author} {\bibfnamefont {S.~R.}\ \bibnamefont
  {Dunsiger}}, \bibinfo {author} {\bibfnamefont {S.}~\bibnamefont
  {El~Shawish}}, \bibinfo {author} {\bibfnamefont {B.~D.}\ \bibnamefont
  {Gaulin}}, \bibinfo {author} {\bibfnamefont {H.~A.}\ \bibnamefont
  {Dabkowska}}, \bibinfo {author} {\bibfnamefont {M.~T.~F.}\ \bibnamefont
  {Telling}}, \bibinfo {author} {\bibfnamefont {T.~G.}\ \bibnamefont
  {Perring}}, \ and\ \bibinfo {author} {\bibfnamefont {J.}~\bibnamefont
  {Bon\ifmmode~\check{c}\else \v{c}\fi{}a}},\ }\bibfield  {title} {\enquote
  {\bibinfo {title} {In-gap spin excitations and finite triplet lifetimes in
  the dilute singlet ground state system
  {SrCu}$_{2-x}${Mg}$_x$({BO}$_3$)$_2$},}\ }\href {\doibase
  10.1103/PhysRevLett.97.247206} {\bibfield  {journal} {\bibinfo  {journal}
  {Phys. Rev. Lett.}\ }\textbf {\bibinfo {volume} {97}},\ \bibinfo {pages}
  {247206} (\bibinfo {year} {2006})}\BibitemShut {NoStop}%
\bibitem [{\citenamefont {Yoshida}\ \emph {et~al.}(2015)\citenamefont
  {Yoshida}, \citenamefont {Kobayashi}, \citenamefont {Yamauchi}, \citenamefont
  {Takigawa}, \citenamefont {Capponi}, \citenamefont {Poilblanc}, \citenamefont
  {Mila}, \citenamefont {Kudo}, \citenamefont {Koike},\ and\ \citenamefont
  {Kobayashi}}]{Yoshida2015}%
  \BibitemOpen
  \bibfield  {author} {\bibinfo {author} {\bibfnamefont {M.}~\bibnamefont
  {Yoshida}}, \bibinfo {author} {\bibfnamefont {H.}~\bibnamefont {Kobayashi}},
  \bibinfo {author} {\bibfnamefont {I.}~\bibnamefont {Yamauchi}}, \bibinfo
  {author} {\bibfnamefont {M.}~\bibnamefont {Takigawa}}, \bibinfo {author}
  {\bibfnamefont {S.}~\bibnamefont {Capponi}}, \bibinfo {author} {\bibfnamefont
  {D.}~\bibnamefont {Poilblanc}}, \bibinfo {author} {\bibfnamefont
  {F.}~\bibnamefont {Mila}}, \bibinfo {author} {\bibfnamefont {K.}~\bibnamefont
  {Kudo}}, \bibinfo {author} {\bibfnamefont {Y.}~\bibnamefont {Koike}}, \ and\
  \bibinfo {author} {\bibfnamefont {N.}~\bibnamefont {Kobayashi}},\ }\bibfield
  {title} {\enquote {\bibinfo {title} {Real space imaging of spin polarons in
  {Zn}-doped {SrCu}$_2$({BO}$_3$)$_2$},}\ }\href {\doibase
  10.1103/PhysRevLett.114.056402} {\bibfield  {journal} {\bibinfo  {journal}
  {Phys. Rev. Lett.}\ }\textbf {\bibinfo {volume} {114}},\ \bibinfo {pages}
  {056402} (\bibinfo {year} {2015})}\BibitemShut {NoStop}%
\bibitem [{\citenamefont {Tedoldi}\ \emph {et~al.}(1999)\citenamefont
  {Tedoldi}, \citenamefont {Santachiara},\ and\ \citenamefont
  {Horvati\ifmmode~\acute{c}\else \'{c}\fi{}}}]{Tedoldi}%
  \BibitemOpen
  \bibfield  {author} {\bibinfo {author} {\bibfnamefont {F.}~\bibnamefont
  {Tedoldi}}, \bibinfo {author} {\bibfnamefont {R.}~\bibnamefont
  {Santachiara}}, \ and\ \bibinfo {author} {\bibfnamefont {M.}~\bibnamefont
  {Horvati\ifmmode~\acute{c}\else \'{c}\fi{}}},\ }\bibfield  {title} {\enquote
  {\bibinfo {title} {$^{89}$ {Y} {NMR} imaging of the staggered magnetization
  in the doped haldane chain {Y}$_{2}${BaNi}$_{1-x}${Mg}$_x${O}$_5$},}\ }\href
  {\doibase 10.1103/PhysRevLett.83.412} {\bibfield  {journal} {\bibinfo
  {journal} {Phys. Rev. Lett.}\ }\textbf {\bibinfo {volume} {83}},\ \bibinfo
  {pages} {412--415} (\bibinfo {year} {1999})}\BibitemShut {NoStop}%
\bibitem [{\citenamefont {Olariu}\ \emph {et~al.}(2008)\citenamefont {Olariu},
  \citenamefont {Mendels}, \citenamefont {Bert}, \citenamefont {Duc},
  \citenamefont {Trombe}, \citenamefont {de~Vries},\ and\ \citenamefont
  {Harrison}}]{Olariu2008}%
  \BibitemOpen
  \bibfield  {author} {\bibinfo {author} {\bibfnamefont {A.}~\bibnamefont
  {Olariu}}, \bibinfo {author} {\bibfnamefont {P.}~\bibnamefont {Mendels}},
  \bibinfo {author} {\bibfnamefont {F.}~\bibnamefont {Bert}}, \bibinfo {author}
  {\bibfnamefont {F.}~\bibnamefont {Duc}}, \bibinfo {author} {\bibfnamefont
  {J.~C.}\ \bibnamefont {Trombe}}, \bibinfo {author} {\bibfnamefont {M.~A.}\
  \bibnamefont {de~Vries}}, \ and\ \bibinfo {author} {\bibfnamefont
  {A.}~\bibnamefont {Harrison}},\ }\bibfield  {title} {\enquote {\bibinfo
  {title} {$^{17}${O} {NMR} study of the intrinsic magnetic susceptibility and
  spin dynamics of the quantum kagome antiferromagnet
  {ZnCu}$_3$({OH})$_6${Cl}$_2$},}\ }\href {\doibase
  10.1103/PhysRevLett.100.087202} {\bibfield  {journal} {\bibinfo  {journal}
  {Phys. Rev. Lett.}\ }\textbf {\bibinfo {volume} {100}},\ \bibinfo {pages}
  {087202} (\bibinfo {year} {2008})}\BibitemShut {NoStop}%
\bibitem [{\citenamefont {Yin}\ \emph {et~al.}(2020)\citenamefont {Yin},
  \citenamefont {Shumiya}, \citenamefont {Jiang}, \citenamefont {Zhou},
  \citenamefont {Macam}, \citenamefont {Sura}, \citenamefont {Zhang},
  \citenamefont {Cheng}, \citenamefont {Guguchia}, \citenamefont {Li} \emph
  {et~al.}}]{yin2020spin}%
  \BibitemOpen
  \bibfield  {author} {\bibinfo {author} {\bibfnamefont {Jia-Xin}\ \bibnamefont
  {Yin}}, \bibinfo {author} {\bibfnamefont {Nana}\ \bibnamefont {Shumiya}},
  \bibinfo {author} {\bibfnamefont {Yuxiao}\ \bibnamefont {Jiang}}, \bibinfo
  {author} {\bibfnamefont {Huibin}\ \bibnamefont {Zhou}}, \bibinfo {author}
  {\bibfnamefont {Gennevieve}\ \bibnamefont {Macam}}, \bibinfo {author}
  {\bibfnamefont {Hano Omar~Mohammad}\ \bibnamefont {Sura}}, \bibinfo {author}
  {\bibfnamefont {Songtian~S}\ \bibnamefont {Zhang}}, \bibinfo {author}
  {\bibfnamefont {Zi-Jia}\ \bibnamefont {Cheng}}, \bibinfo {author}
  {\bibfnamefont {Zurab}\ \bibnamefont {Guguchia}}, \bibinfo {author}
  {\bibfnamefont {Yangmu}\ \bibnamefont {Li}},  \emph {et~al.},\ }\bibfield
  {title} {\enquote {\bibinfo {title} {Spin-orbit quantum impurity in a
  topological magnet},}\ }\href@noop {} {\bibfield  {journal} {\bibinfo
  {journal} {Nature Communications}\ }\textbf {\bibinfo {volume} {11}},\
  \bibinfo {pages} {1--6} (\bibinfo {year} {2020})}\BibitemShut {NoStop}%
\bibitem [{\citenamefont {Obradors}\ \emph {et~al.}(1988)\citenamefont
  {Obradors}, \citenamefont {Labarta}, \citenamefont {Isalgué}, \citenamefont
  {Tejada}, \citenamefont {Rodriguez},\ and\ \citenamefont
  {Pernet}}]{SCGO1988}%
  \BibitemOpen
  \bibfield  {author} {\bibinfo {author} {\bibfnamefont {X.}~\bibnamefont
  {Obradors}}, \bibinfo {author} {\bibfnamefont {A.}~\bibnamefont {Labarta}},
  \bibinfo {author} {\bibfnamefont {A.}~\bibnamefont {Isalgué}}, \bibinfo
  {author} {\bibfnamefont {J.}~\bibnamefont {Tejada}}, \bibinfo {author}
  {\bibfnamefont {J.}~\bibnamefont {Rodriguez}}, \ and\ \bibinfo {author}
  {\bibfnamefont {M.}~\bibnamefont {Pernet}},\ }\bibfield  {title} {\enquote
  {\bibinfo {title} {Magnetic frustration and lattice dimensionality in
  srcr8ga4o19},}\ }\href {\doibase
  https://doi.org/10.1016/0038-1098(88)90885-X} {\bibfield  {journal} {\bibinfo
   {journal} {Solid State Communications}\ }\textbf {\bibinfo {volume} {65}},\
  \bibinfo {pages} {189 -- 192} (\bibinfo {year} {1988})}\BibitemShut {NoStop}%
\bibitem [{\citenamefont {Ramirez}\ \emph {et~al.}(1990)\citenamefont
  {Ramirez}, \citenamefont {Espinosa},\ and\ \citenamefont
  {Cooper}}]{Ramirez1990}%
  \BibitemOpen
  \bibfield  {author} {\bibinfo {author} {\bibfnamefont {A.~P.}\ \bibnamefont
  {Ramirez}}, \bibinfo {author} {\bibfnamefont {G.~P.}\ \bibnamefont
  {Espinosa}}, \ and\ \bibinfo {author} {\bibfnamefont {A.~S.}\ \bibnamefont
  {Cooper}},\ }\bibfield  {title} {\enquote {\bibinfo {title} {Strong
  frustration and dilution-enhanced order in a quasi-2d spin glass},}\ }\href
  {\doibase 10.1103/PhysRevLett.64.2070} {\bibfield  {journal} {\bibinfo
  {journal} {Phys. Rev. Lett.}\ }\textbf {\bibinfo {volume} {64}},\ \bibinfo
  {pages} {2070} (\bibinfo {year} {1990})}\BibitemShut {NoStop}%
\bibitem [{\citenamefont {Limot}\ \emph {et~al.}(2002)\citenamefont {Limot},
  \citenamefont {Mendels}, \citenamefont {Collin}, \citenamefont {Mondelli},
  \citenamefont {Ouladdiaf}, \citenamefont {Mutka}, \citenamefont {Blanchard},\
  and\ \citenamefont {Mekata}}]{Limot2002}%
  \BibitemOpen
  \bibfield  {author} {\bibinfo {author} {\bibfnamefont {L.}~\bibnamefont
  {Limot}}, \bibinfo {author} {\bibfnamefont {P.}~\bibnamefont {Mendels}},
  \bibinfo {author} {\bibfnamefont {G.}~\bibnamefont {Collin}}, \bibinfo
  {author} {\bibfnamefont {C.}~\bibnamefont {Mondelli}}, \bibinfo {author}
  {\bibfnamefont {B.}~\bibnamefont {Ouladdiaf}}, \bibinfo {author}
  {\bibfnamefont {H.}~\bibnamefont {Mutka}}, \bibinfo {author} {\bibfnamefont
  {N.}~\bibnamefont {Blanchard}}, \ and\ \bibinfo {author} {\bibfnamefont
  {M.}~\bibnamefont {Mekata}},\ }\bibfield  {title} {\enquote {\bibinfo {title}
  {Susceptibility and dilution effects of the kagom\'e bilayer geometrically
  frustrated network: A {Ga} {NMR} study of
  {SrCr}$_{9p}${Ga}$_{12-9p}${O}$_{19}$},}\ }\href {\doibase
  10.1103/PhysRevB.65.144447} {\bibfield  {journal} {\bibinfo  {journal} {Phys.
  Rev. B}\ }\textbf {\bibinfo {volume} {65}},\ \bibinfo {pages} {144447}
  (\bibinfo {year} {2002})}\BibitemShut {NoStop}%
\bibitem [{\citenamefont {Moessner}\ and\ \citenamefont
  {Berlinsky}(1999)}]{Moessner_Berlinsky}%
  \BibitemOpen
  \bibfield  {author} {\bibinfo {author} {\bibfnamefont {R}~\bibnamefont
  {Moessner}}\ and\ \bibinfo {author} {\bibfnamefont {AJ}~\bibnamefont
  {Berlinsky}},\ }\bibfield  {title} {\enquote {\bibinfo {title} {Magnetic
  susceptibility of diluted pyrochlore and
  {SrCr}$_{9-9x}${Ga}$_{3+9x}${O}$_{19}$ antiferromagnets},}\ }\href@noop {}
  {\bibfield  {journal} {\bibinfo  {journal} {Phys. Rev. Lett.}\ }\textbf
  {\bibinfo {volume} {83}},\ \bibinfo {pages} {3293} (\bibinfo {year}
  {1999})}\BibitemShut {NoStop}%
\bibitem [{\citenamefont {Moessner}\ and\ \citenamefont
  {Chalker}(1998)}]{moessner1998properties}%
  \BibitemOpen
  \bibfield  {author} {\bibinfo {author} {\bibfnamefont {Roderich}\
  \bibnamefont {Moessner}}\ and\ \bibinfo {author} {\bibfnamefont {John~T}\
  \bibnamefont {Chalker}},\ }\bibfield  {title} {\enquote {\bibinfo {title}
  {Properties of a classical spin liquid: the heisenberg pyrochlore
  antiferromagnet},}\ }\href@noop {} {\bibfield  {journal} {\bibinfo  {journal}
  {Phys. Rev. Lett.}\ }\textbf {\bibinfo {volume} {80}},\ \bibinfo {pages}
  {2929} (\bibinfo {year} {1998})}\BibitemShut {NoStop}%
\bibitem [{\citenamefont {Schiffer}\ and\ \citenamefont
  {Daruka}(1997)}]{schiffer_two_1997}%
  \BibitemOpen
  \bibfield  {author} {\bibinfo {author} {\bibfnamefont {P.}~\bibnamefont
  {Schiffer}}\ and\ \bibinfo {author} {\bibfnamefont {I.}~\bibnamefont
  {Daruka}},\ }\bibfield  {title} {\enquote {\bibinfo {title} {Two-population
  model for anomalous low-temperature magnetism in geometrically frustrated
  magnets},}\ }\href {\doibase 10.1103/PhysRevB.56.13712} {\bibfield  {journal}
  {\bibinfo  {journal} {Phys. Rev. B}\ }\textbf {\bibinfo {volume} {56}},\
  \bibinfo {pages} {13712--13715} (\bibinfo {year} {1997})}\BibitemShut
  {NoStop}%
\bibitem [{\citenamefont {Henley}(2001)}]{henley_effective_2001}%
  \BibitemOpen
  \bibfield  {author} {\bibinfo {author} {\bibfnamefont {C~L}\ \bibnamefont
  {Henley}},\ }\bibfield  {title} {\enquote {\bibinfo {title} {Effective
  hamiltonians and dilution effects in kagome and related anti-ferromagnets},}\
  }\href {\doibase 10.1139/p01-097} {\bibfield  {journal} {\bibinfo  {journal}
  {Canadian Journal of Physics}\ }\textbf {\bibinfo {volume} {79}},\ \bibinfo
  {pages} {1307--1321} (\bibinfo {year} {2001})}\BibitemShut {NoStop}%
\bibitem [{\citenamefont {Sen}\ \emph {et~al.}(2011)\citenamefont {Sen},
  \citenamefont {Damle},\ and\ \citenamefont {Moessner}}]{sen_fractional_2011}%
  \BibitemOpen
  \bibfield  {author} {\bibinfo {author} {\bibfnamefont {Arnab}\ \bibnamefont
  {Sen}}, \bibinfo {author} {\bibfnamefont {Kedar}\ \bibnamefont {Damle}}, \
  and\ \bibinfo {author} {\bibfnamefont {Roderich}\ \bibnamefont {Moessner}},\
  }\bibfield  {title} {\enquote {\bibinfo {title} {Fractional spin textures in
  the frustrated magnet {SrCr}$_{9p}${Ga}$_{12-9p}${O}$_{19}$},}\ }\href
  {\doibase 10.1103/PhysRevLett.106.127203} {\bibfield  {journal} {\bibinfo
  {journal} {Phys. Rev. Lett.}\ }\textbf {\bibinfo {volume} {106}},\ \bibinfo
  {pages} {127203} (\bibinfo {year} {2011})}\BibitemShut {NoStop}%
\bibitem [{\citenamefont {Sen}\ \emph {et~al.}(2012)\citenamefont {Sen},
  \citenamefont {Damle},\ and\ \citenamefont {Moessner}}]{sen_vacancy_2012}%
  \BibitemOpen
  \bibfield  {author} {\bibinfo {author} {\bibfnamefont {Arnab}\ \bibnamefont
  {Sen}}, \bibinfo {author} {\bibfnamefont {Kedar}\ \bibnamefont {Damle}}, \
  and\ \bibinfo {author} {\bibfnamefont {R.}~\bibnamefont {Moessner}},\
  }\bibfield  {title} {\enquote {\bibinfo {title} {Vacancy-induced spin
  textures and their interactions in a classical spin liquid},}\ }\href
  {\doibase 10.1103/PhysRevB.86.205134} {\bibfield  {journal} {\bibinfo
  {journal} {Phys. Rev. B}\ }\textbf {\bibinfo {volume} {86}},\ \bibinfo
  {pages} {205134} (\bibinfo {year} {2012})}\BibitemShut {NoStop}%
\bibitem [{\citenamefont {Bono}\ \emph {et~al.}(2004)\citenamefont {Bono},
  \citenamefont {Mendels}, \citenamefont {Collin}, \citenamefont {Blanchard},
  \citenamefont {Bert}, \citenamefont {Amato}, \citenamefont {Baines},\ and\
  \citenamefont {Hillier}}]{Bono2004}%
  \BibitemOpen
  \bibfield  {author} {\bibinfo {author} {\bibfnamefont {D.}~\bibnamefont
  {Bono}}, \bibinfo {author} {\bibfnamefont {P.}~\bibnamefont {Mendels}},
  \bibinfo {author} {\bibfnamefont {G.}~\bibnamefont {Collin}}, \bibinfo
  {author} {\bibfnamefont {N.}~\bibnamefont {Blanchard}}, \bibinfo {author}
  {\bibfnamefont {F.}~\bibnamefont {Bert}}, \bibinfo {author} {\bibfnamefont
  {A.}~\bibnamefont {Amato}}, \bibinfo {author} {\bibfnamefont
  {C.}~\bibnamefont {Baines}}, \ and\ \bibinfo {author} {\bibfnamefont {A.~D.}\
  \bibnamefont {Hillier}},\ }\bibfield  {title} {\enquote {\bibinfo {title}
  {$\ensuremath{\mu}\mathrm{S}\mathrm{R}$ study of the quantum dynamics in the
  frustrated $s=\frac{3}{2}$ kagom\'e bilayers},}\ }\href {\doibase
  10.1103/PhysRevLett.93.187201} {\bibfield  {journal} {\bibinfo  {journal}
  {Phys. Rev. Lett.}\ }\textbf {\bibinfo {volume} {93}},\ \bibinfo {pages}
  {187201} (\bibinfo {year} {2004})}\BibitemShut {NoStop}%
\bibitem [{\citenamefont {Mendels}\ and\ \citenamefont
  {Bert}(2011)}]{Mendels_2011}%
  \BibitemOpen
  \bibfield  {author} {\bibinfo {author} {\bibfnamefont {Philippe}\
  \bibnamefont {Mendels}}\ and\ \bibinfo {author} {\bibfnamefont {Fabrice}\
  \bibnamefont {Bert}},\ }\bibfield  {title} {\enquote {\bibinfo {title}
  {Quantum kagome antiferromagnet : {ZnCu}$_3$({OH})$_6$cl$_2$},}\ }\href
  {\doibase 10.1088/1742-6596/320/1/012004} {\bibfield  {journal} {\bibinfo
  {journal} {Journal of Physics: Conference Series}\ }\textbf {\bibinfo
  {volume} {320}},\ \bibinfo {pages} {012004} (\bibinfo {year}
  {2011})}\BibitemShut {NoStop}%
\bibitem [{\citenamefont {Khuntia}\ \emph {et~al.}(2020)\citenamefont
  {Khuntia}, \citenamefont {Velazquez}, \citenamefont {Barth{\'e}lemy},
  \citenamefont {Bert}, \citenamefont {Kermarrec}, \citenamefont {Legros},
  \citenamefont {Bernu}, \citenamefont {Messio}, \citenamefont {Zorko},\ and\
  \citenamefont {Mendels}}]{Khuntia2020}%
  \BibitemOpen
  \bibfield  {author} {\bibinfo {author} {\bibfnamefont {P}~\bibnamefont
  {Khuntia}}, \bibinfo {author} {\bibfnamefont {Matias}\ \bibnamefont
  {Velazquez}}, \bibinfo {author} {\bibfnamefont {Quentin}\ \bibnamefont
  {Barth{\'e}lemy}}, \bibinfo {author} {\bibfnamefont {Fabrice}\ \bibnamefont
  {Bert}}, \bibinfo {author} {\bibfnamefont {Edwin}\ \bibnamefont {Kermarrec}},
  \bibinfo {author} {\bibfnamefont {A}~\bibnamefont {Legros}}, \bibinfo
  {author} {\bibfnamefont {Bernard}\ \bibnamefont {Bernu}}, \bibinfo {author}
  {\bibfnamefont {L}~\bibnamefont {Messio}}, \bibinfo {author} {\bibfnamefont
  {Andrej}\ \bibnamefont {Zorko}}, \ and\ \bibinfo {author} {\bibfnamefont
  {P}~\bibnamefont {Mendels}},\ }\bibfield  {title} {\enquote {\bibinfo {title}
  {Gapless ground state in the archetypal quantum kagome antiferromagnet
  {ZnCu}$_3$({OH})$_6${Cl}$_2$},}\ }\href@noop {} {\bibfield  {journal}
  {\bibinfo  {journal} {Nature Physics}\ }\textbf {\bibinfo {volume} {16}},\
  \bibinfo {pages} {469--474} (\bibinfo {year} {2020})}\BibitemShut {NoStop}%
\bibitem [{\citenamefont {Chern}\ and\ \citenamefont
  {Moessner}(2013)}]{Chern_Moessner}%
  \BibitemOpen
  \bibfield  {author} {\bibinfo {author} {\bibfnamefont {Gia-Wei}\ \bibnamefont
  {Chern}}\ and\ \bibinfo {author} {\bibfnamefont {Roderich}\ \bibnamefont
  {Moessner}},\ }\bibfield  {title} {\enquote {\bibinfo {title} {Dipolar order
  by disorder in the classical heisenberg antiferromagnet on the kagome
  lattice},}\ }\href@noop {} {\bibfield  {journal} {\bibinfo  {journal} {Phys.
  Rev. Lett.}\ }\textbf {\bibinfo {volume} {110}},\ \bibinfo {pages} {077201}
  (\bibinfo {year} {2013})}\BibitemShut {NoStop}%
\bibitem [{\citenamefont {Zhitomirsky}(2008)}]{Zhitomirsky}%
  \BibitemOpen
  \bibfield  {author} {\bibinfo {author} {\bibfnamefont {Michael~E}\
  \bibnamefont {Zhitomirsky}},\ }\bibfield  {title} {\enquote {\bibinfo {title}
  {Octupolar ordering of classical kagome antiferromagnets in two and three
  dimensions},}\ }\href@noop {} {\bibfield  {journal} {\bibinfo  {journal}
  {Phys. Rev. B}\ }\textbf {\bibinfo {volume} {78}},\ \bibinfo {pages} {094423}
  (\bibinfo {year} {2008})}\BibitemShut {NoStop}%
\bibitem [{\citenamefont {Huse}\ and\ \citenamefont
  {Rutenberg}(1992)}]{Rutenberg_Huse}%
  \BibitemOpen
  \bibfield  {author} {\bibinfo {author} {\bibfnamefont {David~A}\ \bibnamefont
  {Huse}}\ and\ \bibinfo {author} {\bibfnamefont {Andrew~D}\ \bibnamefont
  {Rutenberg}},\ }\bibfield  {title} {\enquote {\bibinfo {title} {Classical
  antiferromagnets on the kagom{\'e} lattice},}\ }\href@noop {} {\bibfield
  {journal} {\bibinfo  {journal} {Physical Review B}\ }\textbf {\bibinfo
  {volume} {45}},\ \bibinfo {pages} {7536} (\bibinfo {year}
  {1992})}\BibitemShut {NoStop}%
\bibitem [{\citenamefont {Dommange}\ \emph {et~al.}(2003)\citenamefont
  {Dommange}, \citenamefont {Mambrini}, \citenamefont {Normand},\ and\
  \citenamefont {Mila}}]{Dommange2003}%
  \BibitemOpen
  \bibfield  {author} {\bibinfo {author} {\bibfnamefont {S.}~\bibnamefont
  {Dommange}}, \bibinfo {author} {\bibfnamefont {M.}~\bibnamefont {Mambrini}},
  \bibinfo {author} {\bibfnamefont {B.}~\bibnamefont {Normand}}, \ and\
  \bibinfo {author} {\bibfnamefont {F.}~\bibnamefont {Mila}},\ }\bibfield
  {title} {\enquote {\bibinfo {title} {Static impurities in the $s=1/2$ kagome
  lattice: Dimer freezing and mutual repulsion},}\ }\href {\doibase
  10.1103/PhysRevB.68.224416} {\bibfield  {journal} {\bibinfo  {journal} {Phys.
  Rev. B}\ }\textbf {\bibinfo {volume} {68}},\ \bibinfo {pages} {224416}
  (\bibinfo {year} {2003})}\BibitemShut {NoStop}%
\bibitem [{\citenamefont {L\"auchli}\ \emph {et~al.}(2007)\citenamefont
  {L\"auchli}, \citenamefont {Dommange}, \citenamefont {Normand},\ and\
  \citenamefont {Mila}}]{Laeuchli2007}%
  \BibitemOpen
  \bibfield  {author} {\bibinfo {author} {\bibfnamefont {A.}~\bibnamefont
  {L\"auchli}}, \bibinfo {author} {\bibfnamefont {S.}~\bibnamefont {Dommange}},
  \bibinfo {author} {\bibfnamefont {B.}~\bibnamefont {Normand}}, \ and\
  \bibinfo {author} {\bibfnamefont {F.}~\bibnamefont {Mila}},\ }\bibfield
  {title} {\enquote {\bibinfo {title} {Static impurities in the $s=\frac{3}{2}$
  kagome lattice: Exact diagonalization calculations on small clusters},}\
  }\href {\doibase 10.1103/PhysRevB.76.144413} {\bibfield  {journal} {\bibinfo
  {journal} {Phys. Rev. B}\ }\textbf {\bibinfo {volume} {76}},\ \bibinfo
  {pages} {144413} (\bibinfo {year} {2007})}\BibitemShut {NoStop}%
\bibitem [{\citenamefont {Singh}(2010)}]{Singh2010}%
  \BibitemOpen
  \bibfield  {author} {\bibinfo {author} {\bibfnamefont {R.~R.~P.}\
  \bibnamefont {Singh}},\ }\bibfield  {title} {\enquote {\bibinfo {title}
  {Valence bond glass phase in dilute kagome antiferromagnets},}\ }\href
  {\doibase 10.1103/PhysRevLett.104.177203} {\bibfield  {journal} {\bibinfo
  {journal} {Phys. Rev. Lett.}\ }\textbf {\bibinfo {volume} {104}},\ \bibinfo
  {pages} {177203} (\bibinfo {year} {2010})}\BibitemShut {NoStop}%
\bibitem [{\citenamefont {Gregor}\ and\ \citenamefont
  {Motrunich}(2008)}]{gregor_nonmagnetic_2008}%
  \BibitemOpen
  \bibfield  {author} {\bibinfo {author} {\bibfnamefont {Karol}\ \bibnamefont
  {Gregor}}\ and\ \bibinfo {author} {\bibfnamefont {Olexei~I.}\ \bibnamefont
  {Motrunich}},\ }\bibfield  {title} {\enquote {\bibinfo {title} {Nonmagnetic
  impurities in the spin-1/2 kagome antiferromagnet},}\ }\href {\doibase
  10.1103/PhysRevB.77.184423} {\bibfield  {journal} {\bibinfo  {journal} {Phys.
  Rev. B}\ }\textbf {\bibinfo {volume} {77}},\ \bibinfo {pages} {184423}
  (\bibinfo {year} {2008})}\BibitemShut {NoStop}%
\bibitem [{\citenamefont {White}(1992)}]{White1992}%
  \BibitemOpen
  \bibfield  {author} {\bibinfo {author} {\bibfnamefont {S.~R.}\ \bibnamefont
  {White}},\ }\bibfield  {title} {\enquote {\bibinfo {title} {{Density matrix
  formulation for quantum renormalization groups}},}\ }\href
  {http://link.aps.org/doi/10.1103/PhysRevLett.69.2863} {\bibfield  {journal}
  {\bibinfo  {journal} {Phys. Rev. Lett.}\ }\textbf {\bibinfo {volume} {69}},\
  \bibinfo {pages} {2863--2866} (\bibinfo {year} {1992})}\BibitemShut {NoStop}%
\bibitem [{\citenamefont {Schollw\"{o}ck}(2011)}]{Schollwoeck2011}%
  \BibitemOpen
  \bibfield  {author} {\bibinfo {author} {\bibfnamefont {Ulrich}\ \bibnamefont
  {Schollw\"{o}ck}},\ }\bibfield  {title} {\enquote {\bibinfo {title} {The
  density-matrix renormalization group in the age of matrix product states},}\
  }\href {\doibase 10.1016/j.aop.2010.09.012} {\bibfield  {journal} {\bibinfo
  {journal} {Annals of Physics}\ }\textbf {\bibinfo {volume} {326}},\ \bibinfo
  {pages} {96 -- 192} (\bibinfo {year} {2011})}\BibitemShut {NoStop}%
\bibitem [{\citenamefont {Fishman}\ \emph {et~al.}(2020)\citenamefont
  {Fishman}, \citenamefont {White},\ and\ \citenamefont
  {Stoudenmire}}]{itensor}%
  \BibitemOpen
  \bibfield  {author} {\bibinfo {author} {\bibfnamefont {Matthew}\ \bibnamefont
  {Fishman}}, \bibinfo {author} {\bibfnamefont {Steven~R.}\ \bibnamefont
  {White}}, \ and\ \bibinfo {author} {\bibfnamefont {E.~Miles}\ \bibnamefont
  {Stoudenmire}},\ }\bibfield  {title} {\enquote {\bibinfo {title} {The
  \mbox{ITensor} software library for tensor network calculations},}\ }\href
  {https://arxiv.org/abs/2007.14822} {\bibfield  {journal} {\bibinfo  {journal}
  {arXiv:2007.14822}\ } (\bibinfo {year} {2020})}\BibitemShut {NoStop}%
\bibitem [{\citenamefont {Jakli\ifmmode~\check{c}\else \v{c}\fi{}}\ and\
  \citenamefont {Prelov\ifmmode~\check{s}\else
  \v{s}\fi{}ek}(1994)}]{Prelovsek1994}%
  \BibitemOpen
  \bibfield  {author} {\bibinfo {author} {\bibfnamefont {J.}~\bibnamefont
  {Jakli\ifmmode~\check{c}\else \v{c}\fi{}}}\ and\ \bibinfo {author}
  {\bibfnamefont {P.}~\bibnamefont {Prelov\ifmmode~\check{s}\else
  \v{s}\fi{}ek}},\ }\bibfield  {title} {\enquote {\bibinfo {title} {Lanczos
  method for the calculation of finite-temperature quantities in correlated
  systems},}\ }\href {\doibase 10.1103/PhysRevB.49.5065} {\bibfield  {journal}
  {\bibinfo  {journal} {Phys. Rev. B}\ }\textbf {\bibinfo {volume} {49}},\
  \bibinfo {pages} {5065} (\bibinfo {year} {1994})}\BibitemShut {NoStop}%
\bibitem [{\citenamefont {Hams}\ and\ \citenamefont
  {De~Raedt}(2000)}]{Hams2000}%
  \BibitemOpen
  \bibfield  {author} {\bibinfo {author} {\bibfnamefont {Anthony}\ \bibnamefont
  {Hams}}\ and\ \bibinfo {author} {\bibfnamefont {Hans}\ \bibnamefont
  {De~Raedt}},\ }\bibfield  {title} {\enquote {\bibinfo {title} {Fast algorithm
  for finding the eigenvalue distribution of very large matrices},}\ }\href
  {\doibase 10.1103/PhysRevE.62.4365} {\bibfield  {journal} {\bibinfo
  {journal} {Phys. Rev. E}\ }\textbf {\bibinfo {volume} {62}},\ \bibinfo
  {pages} {4365--4377} (\bibinfo {year} {2000})}\BibitemShut {NoStop}%
\bibitem [{\citenamefont {Goldstein}\ \emph {et~al.}(2006)\citenamefont
  {Goldstein}, \citenamefont {Lebowitz}, \citenamefont {Tumulka},\ and\
  \citenamefont {Zangh\`{\i}}}]{goldstein_canonical_2006}%
  \BibitemOpen
  \bibfield  {author} {\bibinfo {author} {\bibfnamefont {Sheldon}\ \bibnamefont
  {Goldstein}}, \bibinfo {author} {\bibfnamefont {Joel~L.}\ \bibnamefont
  {Lebowitz}}, \bibinfo {author} {\bibfnamefont {Roderich}\ \bibnamefont
  {Tumulka}}, \ and\ \bibinfo {author} {\bibfnamefont {Nino}\ \bibnamefont
  {Zangh\`{\i}}},\ }\bibfield  {title} {\enquote {\bibinfo {title} {Canonical
  typicality},}\ }\href {\doibase 10.1103/PhysRevLett.96.050403} {\bibfield
  {journal} {\bibinfo  {journal} {Phys. Rev. Lett.}\ }\textbf {\bibinfo
  {volume} {96}},\ \bibinfo {pages} {050403} (\bibinfo {year}
  {2006})}\BibitemShut {NoStop}%
\bibitem [{\citenamefont {Popescu}\ \emph {et~al.}(2006)\citenamefont
  {Popescu}, \citenamefont {Short},\ and\ \citenamefont
  {Winter}}]{Popescu2006}%
  \BibitemOpen
  \bibfield  {author} {\bibinfo {author} {\bibfnamefont {S.}~\bibnamefont
  {Popescu}}, \bibinfo {author} {\bibfnamefont {A.}~\bibnamefont {Short}}, \
  and\ \bibinfo {author} {\bibfnamefont {A.}~\bibnamefont {Winter}},\
  }\bibfield  {title} {\enquote {\bibinfo {title} {Entanglement and the
  foundations of statistical mechanics},}\ }\href {\doibase 10.1038/nphys444}
  {\bibfield  {journal} {\bibinfo  {journal} {Nat. Phys.}\ }\textbf {\bibinfo
  {volume} {2}},\ \bibinfo {pages} {754} (\bibinfo {year} {2006})}\BibitemShut
  {NoStop}%
\bibitem [{\citenamefont {Reimann}(2007)}]{Reimann2007}%
  \BibitemOpen
  \bibfield  {author} {\bibinfo {author} {\bibfnamefont {Peter}\ \bibnamefont
  {Reimann}},\ }\bibfield  {title} {\enquote {\bibinfo {title} {Typicality for
  generalized microcanonical ensembles},}\ }\href {\doibase
  10.1103/PhysRevLett.99.160404} {\bibfield  {journal} {\bibinfo  {journal}
  {Phys. Rev. Lett.}\ }\textbf {\bibinfo {volume} {99}},\ \bibinfo {pages}
  {160404} (\bibinfo {year} {2007})}\BibitemShut {NoStop}%
\bibitem [{\citenamefont {Sugiura}\ and\ \citenamefont
  {Shimizu}(2013)}]{sugiura_canonical_2013}%
  \BibitemOpen
  \bibfield  {author} {\bibinfo {author} {\bibfnamefont {Sho}\ \bibnamefont
  {Sugiura}}\ and\ \bibinfo {author} {\bibfnamefont {Akira}\ \bibnamefont
  {Shimizu}},\ }\bibfield  {title} {\enquote {\bibinfo {title} {Canonical
  thermal pure quantum state},}\ }\href {\doibase
  10.1103/PhysRevLett.111.010401} {\bibfield  {journal} {\bibinfo  {journal}
  {Phys. Rev. Lett.}\ }\textbf {\bibinfo {volume} {111}},\ \bibinfo {pages}
  {010401} (\bibinfo {year} {2013})}\BibitemShut {NoStop}%
\bibitem [{\citenamefont {Sugiura}\ and\ \citenamefont
  {Shimizu}(2012)}]{sugiura_thermal_2012}%
  \BibitemOpen
  \bibfield  {author} {\bibinfo {author} {\bibfnamefont {Sho}\ \bibnamefont
  {Sugiura}}\ and\ \bibinfo {author} {\bibfnamefont {Akira}\ \bibnamefont
  {Shimizu}},\ }\bibfield  {title} {\enquote {\bibinfo {title} {Thermal pure
  quantum states at finite temperature},}\ }\href {\doibase
  10.1103/PhysRevLett.108.240401} {\bibfield  {journal} {\bibinfo  {journal}
  {Phys. Rev. Lett.}\ }\textbf {\bibinfo {volume} {108}},\ \bibinfo {pages}
  {240401} (\bibinfo {year} {2012})}\BibitemShut {NoStop}%
\bibitem [{\citenamefont {Shimokawa}\ and\ \citenamefont
  {Kawamura}(2016)}]{Shimokawa2016}%
  \BibitemOpen
  \bibfield  {author} {\bibinfo {author} {\bibfnamefont {Tokuro}\ \bibnamefont
  {Shimokawa}}\ and\ \bibinfo {author} {\bibfnamefont {Hikaru}\ \bibnamefont
  {Kawamura}},\ }\bibfield  {title} {\enquote {\bibinfo {title}
  {Finite-temperature crossover phenomenon in the s = 1/2 antiferromagnetic
  heisenberg model on the kagome lattice},}\ }\href {\doibase
  10.7566/JPSJ.85.113702} {\bibfield  {journal} {\bibinfo  {journal} {J. Phys.
  Soc. Jpn}\ }\textbf {\bibinfo {volume} {85}},\ \bibinfo {pages} {113702}
  (\bibinfo {year} {2016})}\BibitemShut {NoStop}%
\bibitem [{\citenamefont {Schnack}\ \emph {et~al.}(2018)\citenamefont
  {Schnack}, \citenamefont {Schulenburg},\ and\ \citenamefont
  {Richter}}]{Schnack2018}%
  \BibitemOpen
  \bibfield  {author} {\bibinfo {author} {\bibfnamefont {J\"urgen}\
  \bibnamefont {Schnack}}, \bibinfo {author} {\bibfnamefont {J\"org}\
  \bibnamefont {Schulenburg}}, \ and\ \bibinfo {author} {\bibfnamefont
  {Johannes}\ \bibnamefont {Richter}},\ }\bibfield  {title} {\enquote {\bibinfo
  {title} {Magnetism of the $n=42$ kagome lattice antiferromagnet},}\ }\href
  {\doibase 10.1103/PhysRevB.98.094423} {\bibfield  {journal} {\bibinfo
  {journal} {Phys. Rev. B}\ }\textbf {\bibinfo {volume} {98}},\ \bibinfo
  {pages} {094423} (\bibinfo {year} {2018})}\BibitemShut {NoStop}%
\bibitem [{\citenamefont {Wietek}\ \emph {et~al.}(2019)\citenamefont {Wietek},
  \citenamefont {Corboz}, \citenamefont {Wessel}, \citenamefont {Normand},
  \citenamefont {Mila},\ and\ \citenamefont
  {Honecker}}]{wietek_thermodynamic_2019}%
  \BibitemOpen
  \bibfield  {author} {\bibinfo {author} {\bibfnamefont {Alexander}\
  \bibnamefont {Wietek}}, \bibinfo {author} {\bibfnamefont {Philippe}\
  \bibnamefont {Corboz}}, \bibinfo {author} {\bibfnamefont {Stefan}\
  \bibnamefont {Wessel}}, \bibinfo {author} {\bibfnamefont {B.}~\bibnamefont
  {Normand}}, \bibinfo {author} {\bibfnamefont {Fr\'ed\'eric}\ \bibnamefont
  {Mila}}, \ and\ \bibinfo {author} {\bibfnamefont {Andreas}\ \bibnamefont
  {Honecker}},\ }\bibfield  {title} {\enquote {\bibinfo {title} {Thermodynamic
  properties of the shastry-sutherland model throughout the dimer-product
  phase},}\ }\href {\doibase 10.1103/PhysRevResearch.1.033038} {\bibfield
  {journal} {\bibinfo  {journal} {Phys. Rev. Research}\ }\textbf {\bibinfo
  {volume} {1}},\ \bibinfo {pages} {033038} (\bibinfo {year}
  {2019})}\BibitemShut {NoStop}%
\bibitem [{\citenamefont {Prelov\ifmmode~\check{s}\else \v{s}\fi{}ek}\ and\
  \citenamefont {Kokalj}(2020)}]{Prelovsek2020}%
  \BibitemOpen
  \bibfield  {author} {\bibinfo {author} {\bibfnamefont {P.}~\bibnamefont
  {Prelov\ifmmode~\check{s}\else \v{s}\fi{}ek}}\ and\ \bibinfo {author}
  {\bibfnamefont {J.}~\bibnamefont {Kokalj}},\ }\bibfield  {title} {\enquote
  {\bibinfo {title} {Similarity of thermodynamic properties of the heisenberg
  model on triangular and kagome lattices},}\ }\href {\doibase
  10.1103/PhysRevB.101.075105} {\bibfield  {journal} {\bibinfo  {journal}
  {Phys. Rev. B}\ }\textbf {\bibinfo {volume} {101}},\ \bibinfo {pages}
  {075105} (\bibinfo {year} {2020})}\BibitemShut {NoStop}%
\bibitem [{\citenamefont {Schnack}\ \emph {et~al.}(2020)\citenamefont
  {Schnack}, \citenamefont {Richter},\ and\ \citenamefont
  {Steinigeweg}}]{Schnack2020}%
  \BibitemOpen
  \bibfield  {author} {\bibinfo {author} {\bibfnamefont {J\"urgen}\
  \bibnamefont {Schnack}}, \bibinfo {author} {\bibfnamefont {Johannes}\
  \bibnamefont {Richter}}, \ and\ \bibinfo {author} {\bibfnamefont {Robin}\
  \bibnamefont {Steinigeweg}},\ }\bibfield  {title} {\enquote {\bibinfo {title}
  {Accuracy of the finite-temperature lanczos method compared to simple
  typicality-based estimates},}\ }\href {\doibase
  10.1103/PhysRevResearch.2.013186} {\bibfield  {journal} {\bibinfo  {journal}
  {Phys. Rev. Research}\ }\textbf {\bibinfo {volume} {2}},\ \bibinfo {pages}
  {013186} (\bibinfo {year} {2020})}\BibitemShut {NoStop}%
\bibitem [{\citenamefont {Aichhorn}\ \emph {et~al.}(2003)\citenamefont
  {Aichhorn}, \citenamefont {Daghofer}, \citenamefont {Evertz},\ and\
  \citenamefont {von~der Linden}}]{Aichhorn2003}%
  \BibitemOpen
  \bibfield  {author} {\bibinfo {author} {\bibfnamefont {Markus}\ \bibnamefont
  {Aichhorn}}, \bibinfo {author} {\bibfnamefont {Maria}\ \bibnamefont
  {Daghofer}}, \bibinfo {author} {\bibfnamefont {Hans~Gerd}\ \bibnamefont
  {Evertz}}, \ and\ \bibinfo {author} {\bibfnamefont {Wolfgang}\ \bibnamefont
  {von~der Linden}},\ }\bibfield  {title} {\enquote {\bibinfo {title}
  {Low-temperature lanczos method for strongly correlated systems},}\ }\href
  {\doibase 10.1103/PhysRevB.67.161103} {\bibfield  {journal} {\bibinfo
  {journal} {Phys. Rev. B}\ }\textbf {\bibinfo {volume} {67}},\ \bibinfo
  {pages} {161103} (\bibinfo {year} {2003})}\BibitemShut {NoStop}%
\bibitem [{\citenamefont {Hiroi}\ \emph {et~al.}(2001)\citenamefont {Hiroi},
  \citenamefont {Hanawa}, \citenamefont {Kobayashi}, \citenamefont {Nohara},
  \citenamefont {Takagi}, \citenamefont {Kato},\ and\ \citenamefont
  {Takigawa}}]{Hiroi2001}%
  \BibitemOpen
  \bibfield  {author} {\bibinfo {author} {\bibfnamefont {Zenji}\ \bibnamefont
  {Hiroi}}, \bibinfo {author} {\bibfnamefont {Masafumi}\ \bibnamefont
  {Hanawa}}, \bibinfo {author} {\bibfnamefont {Naoya}\ \bibnamefont
  {Kobayashi}}, \bibinfo {author} {\bibfnamefont {Minoru}\ \bibnamefont
  {Nohara}}, \bibinfo {author} {\bibfnamefont {Hidenori}\ \bibnamefont
  {Takagi}}, \bibinfo {author} {\bibfnamefont {Yoshitomo}\ \bibnamefont
  {Kato}}, \ and\ \bibinfo {author} {\bibfnamefont {Masashi}\ \bibnamefont
  {Takigawa}},\ }\bibfield  {title} {\enquote {\bibinfo {title} {Spin-1/2
  kagom\'e -like lattice in volborthite
  {Cu}$_3${V}$_2${O}$_7$({OH}).2{H}$_2${O}},}\ }\href {\doibase
  10.1143/JPSJ.70.3377} {\bibfield  {journal} {\bibinfo  {journal} {J. Phys.
  Soc. Jpn.}\ }\textbf {\bibinfo {volume} {70}},\ \bibinfo {pages} {3377}
  (\bibinfo {year} {2001})}\BibitemShut {NoStop}%
\bibitem [{\citenamefont {Mendels}\ \emph {et~al.}(2007)\citenamefont
  {Mendels}, \citenamefont {Bert}, \citenamefont {de~Vries}, \citenamefont
  {Olariu}, \citenamefont {Harrison}, \citenamefont {Duc}, \citenamefont
  {Trombe}, \citenamefont {Lord}, \citenamefont {Amato},\ and\ \citenamefont
  {Baines}}]{Mendels2007}%
  \BibitemOpen
  \bibfield  {author} {\bibinfo {author} {\bibfnamefont {P.}~\bibnamefont
  {Mendels}}, \bibinfo {author} {\bibfnamefont {F.}~\bibnamefont {Bert}},
  \bibinfo {author} {\bibfnamefont {M.~A.}\ \bibnamefont {de~Vries}}, \bibinfo
  {author} {\bibfnamefont {A.}~\bibnamefont {Olariu}}, \bibinfo {author}
  {\bibfnamefont {A.}~\bibnamefont {Harrison}}, \bibinfo {author}
  {\bibfnamefont {F.}~\bibnamefont {Duc}}, \bibinfo {author} {\bibfnamefont
  {J.~C.}\ \bibnamefont {Trombe}}, \bibinfo {author} {\bibfnamefont {J.~S.}\
  \bibnamefont {Lord}}, \bibinfo {author} {\bibfnamefont {A.}~\bibnamefont
  {Amato}}, \ and\ \bibinfo {author} {\bibfnamefont {C.}~\bibnamefont
  {Baines}},\ }\bibfield  {title} {\enquote {\bibinfo {title} {Quantum
  magnetism in the paratacamite family: Towards an ideal kagom\'e lattice},}\
  }\href {\doibase 10.1103/PhysRevLett.98.077204} {\bibfield  {journal}
  {\bibinfo  {journal} {Phys. Rev. Lett.}\ }\textbf {\bibinfo {volume} {98}},\
  \bibinfo {pages} {077204} (\bibinfo {year} {2007})}\BibitemShut {NoStop}%
\bibitem [{\citenamefont {Arh}\ \emph {et~al.}(2020)\citenamefont {Arh},
  \citenamefont {Gomil\ifmmode~\check{s}\else \v{s}\fi{}ek}, \citenamefont
  {Prelov\ifmmode~\check{s}\else \v{s}\fi{}ek}, \citenamefont {Pregelj},
  \citenamefont {Klanj\ifmmode~\check{s}\else \v{s}\fi{}ek}, \citenamefont
  {Ozarowski}, \citenamefont {Clark}, \citenamefont {Lancaster}, \citenamefont
  {Sun}, \citenamefont {Mi},\ and\ \citenamefont {Zorko}}]{Arh2020}%
  \BibitemOpen
  \bibfield  {author} {\bibinfo {author} {\bibfnamefont {T.}~\bibnamefont
  {Arh}}, \bibinfo {author} {\bibfnamefont {M.}~\bibnamefont
  {Gomil\ifmmode~\check{s}\else \v{s}\fi{}ek}}, \bibinfo {author}
  {\bibfnamefont {P.}~\bibnamefont {Prelov\ifmmode~\check{s}\else
  \v{s}\fi{}ek}}, \bibinfo {author} {\bibfnamefont {M.}~\bibnamefont
  {Pregelj}}, \bibinfo {author} {\bibfnamefont {M.}~\bibnamefont
  {Klanj\ifmmode~\check{s}\else \v{s}\fi{}ek}}, \bibinfo {author}
  {\bibfnamefont {A.}~\bibnamefont {Ozarowski}}, \bibinfo {author}
  {\bibfnamefont {S.~J.}\ \bibnamefont {Clark}}, \bibinfo {author}
  {\bibfnamefont {T.}~\bibnamefont {Lancaster}}, \bibinfo {author}
  {\bibfnamefont {W.}~\bibnamefont {Sun}}, \bibinfo {author} {\bibfnamefont
  {J.-X.}\ \bibnamefont {Mi}}, \ and\ \bibinfo {author} {\bibfnamefont
  {A.}~\bibnamefont {Zorko}},\ }\bibfield  {title} {\enquote {\bibinfo {title}
  {Origin of magnetic ordering in a structurally perfect quantum kagome
  antiferromagnet},}\ }\href {\doibase 10.1103/PhysRevLett.125.027203}
  {\bibfield  {journal} {\bibinfo  {journal} {Phys. Rev. Lett.}\ }\textbf
  {\bibinfo {volume} {125}},\ \bibinfo {pages} {027203} (\bibinfo {year}
  {2020})}\BibitemShut {NoStop}%
\bibitem [{\citenamefont {Schulenburg}\ \emph {et~al.}(2002)\citenamefont
  {Schulenburg}, \citenamefont {Honecker}, \citenamefont {Schnack},
  \citenamefont {Richter},\ and\ \citenamefont {Schmidt}}]{Schulenburg2002}%
  \BibitemOpen
  \bibfield  {author} {\bibinfo {author} {\bibfnamefont {J.}~\bibnamefont
  {Schulenburg}}, \bibinfo {author} {\bibfnamefont {A.}~\bibnamefont
  {Honecker}}, \bibinfo {author} {\bibfnamefont {J.}~\bibnamefont {Schnack}},
  \bibinfo {author} {\bibfnamefont {J.}~\bibnamefont {Richter}}, \ and\
  \bibinfo {author} {\bibfnamefont {H.-J.}\ \bibnamefont {Schmidt}},\
  }\bibfield  {title} {\enquote {\bibinfo {title} {Macroscopic magnetization
  jumps due to independent magnons in frustrated quantum spin lattices},}\
  }\href {\doibase 10.1103/PhysRevLett.88.167207} {\bibfield  {journal}
  {\bibinfo  {journal} {Phys. Rev. Lett.}\ }\textbf {\bibinfo {volume} {88}},\
  \bibinfo {pages} {167207} (\bibinfo {year} {2002})}\BibitemShut {NoStop}%
\bibitem [{\citenamefont {Cabra}\ \emph {et~al.}(2005)\citenamefont {Cabra},
  \citenamefont {Grynberg}, \citenamefont {Holdsworth}, \citenamefont
  {Honecker}, \citenamefont {Pujol}, \citenamefont {Richter}, \citenamefont
  {Schmalfu\ss{}},\ and\ \citenamefont {Schulenburg}}]{Cabra2005}%
  \BibitemOpen
  \bibfield  {author} {\bibinfo {author} {\bibfnamefont {D.~C.}\ \bibnamefont
  {Cabra}}, \bibinfo {author} {\bibfnamefont {M.~D.}\ \bibnamefont {Grynberg}},
  \bibinfo {author} {\bibfnamefont {P.~C.~W.}\ \bibnamefont {Holdsworth}},
  \bibinfo {author} {\bibfnamefont {A.}~\bibnamefont {Honecker}}, \bibinfo
  {author} {\bibfnamefont {P.}~\bibnamefont {Pujol}}, \bibinfo {author}
  {\bibfnamefont {J.}~\bibnamefont {Richter}}, \bibinfo {author} {\bibfnamefont
  {D.}~\bibnamefont {Schmalfu\ss{}}}, \ and\ \bibinfo {author} {\bibfnamefont
  {J.}~\bibnamefont {Schulenburg}},\ }\bibfield  {title} {\enquote {\bibinfo
  {title} {Quantum kagom\'e antiferromagnet in a magnetic field: Low-lying
  nonmagnetic excitations versus valence-bond crystal order},}\ }\href
  {\doibase 10.1103/PhysRevB.71.144420} {\bibfield  {journal} {\bibinfo
  {journal} {Phys. Rev. B}\ }\textbf {\bibinfo {volume} {71}},\ \bibinfo
  {pages} {144420} (\bibinfo {year} {2005})}\BibitemShut {NoStop}%
\bibitem [{\citenamefont {Nishimoto}\ \emph {et~al.}(2013)\citenamefont
  {Nishimoto}, \citenamefont {Shibata},\ and\ \citenamefont
  {Hotta}}]{Nishimoto2013}%
  \BibitemOpen
  \bibfield  {author} {\bibinfo {author} {\bibfnamefont {S.}~\bibnamefont
  {Nishimoto}}, \bibinfo {author} {\bibfnamefont {N.}~\bibnamefont {Shibata}},
  \ and\ \bibinfo {author} {\bibfnamefont {C.}~\bibnamefont {Hotta}},\
  }\bibfield  {title} {\enquote {\bibinfo {title} {Controlling frustrated
  liquids and solids with an applied field in a kagome heisenberg
  antiferromagnet},}\ }\href {\doibase 10.1038/ncomms3287} {\bibfield
  {journal} {\bibinfo  {journal} {Nat. Comm.}\ }\textbf {\bibinfo {volume}
  {4}},\ \bibinfo {pages} {2287} (\bibinfo {year} {2013})}\BibitemShut
  {NoStop}%
\bibitem [{\citenamefont {Capponi}\ \emph
  {et~al.}(2013{\natexlab{a}})\citenamefont {Capponi}, \citenamefont {Derzhko},
  \citenamefont {Honecker}, \citenamefont {L\"auchli},\ and\ \citenamefont
  {Richter}}]{Capponi2013b}%
  \BibitemOpen
  \bibfield  {author} {\bibinfo {author} {\bibfnamefont {Sylvain}\ \bibnamefont
  {Capponi}}, \bibinfo {author} {\bibfnamefont {Oleg}\ \bibnamefont {Derzhko}},
  \bibinfo {author} {\bibfnamefont {Andreas}\ \bibnamefont {Honecker}},
  \bibinfo {author} {\bibfnamefont {Andreas~M.}\ \bibnamefont {L\"auchli}}, \
  and\ \bibinfo {author} {\bibfnamefont {Johannes}\ \bibnamefont {Richter}},\
  }\bibfield  {title} {\enquote {\bibinfo {title} {Numerical study of
  magnetization plateaus in the spin-$\frac{1}{2}$ kagome heisenberg
  antiferromagnet},}\ }\href {\doibase 10.1103/PhysRevB.88.144416} {\bibfield
  {journal} {\bibinfo  {journal} {Phys. Rev. B}\ }\textbf {\bibinfo {volume}
  {88}},\ \bibinfo {pages} {144416} (\bibinfo {year}
  {2013}{\natexlab{a}})}\BibitemShut {NoStop}%
\bibitem [{\citenamefont {Bernu}\ \emph {et~al.}(2020)\citenamefont {Bernu},
  \citenamefont {Pierre}, \citenamefont {Essafi},\ and\ \citenamefont
  {Messio}}]{bernu2020}%
  \BibitemOpen
  \bibfield  {author} {\bibinfo {author} {\bibfnamefont {Bernard}\ \bibnamefont
  {Bernu}}, \bibinfo {author} {\bibfnamefont {Laurent}\ \bibnamefont {Pierre}},
  \bibinfo {author} {\bibfnamefont {Karim}\ \bibnamefont {Essafi}}, \ and\
  \bibinfo {author} {\bibfnamefont {Laura}\ \bibnamefont {Messio}},\ }\bibfield
   {title} {\enquote {\bibinfo {title} {Effect of perturbations on the kagome
  $s=1/2$ antiferromagnet at all temperatures},}\ }\href@noop {} {\bibfield
  {journal} {\bibinfo  {journal} {Physical Review B}\ }\textbf {\bibinfo
  {volume} {101}},\ \bibinfo {pages} {140403} (\bibinfo {year}
  {2020})}\BibitemShut {NoStop}%
\bibitem [{\citenamefont {Jiang}\ \emph {et~al.}(2008)\citenamefont {Jiang},
  \citenamefont {Weng},\ and\ \citenamefont {Sheng}}]{Jiang2008}%
  \BibitemOpen
  \bibfield  {author} {\bibinfo {author} {\bibfnamefont {H.~C.}\ \bibnamefont
  {Jiang}}, \bibinfo {author} {\bibfnamefont {Z.~Y.}\ \bibnamefont {Weng}}, \
  and\ \bibinfo {author} {\bibfnamefont {D.~N.}\ \bibnamefont {Sheng}},\
  }\bibfield  {title} {\enquote {\bibinfo {title} {Density matrix
  renormalization group numerical study of the kagome antiferromagnet},}\
  }\href {\doibase 10.1103/PhysRevLett.101.117203} {\bibfield  {journal}
  {\bibinfo  {journal} {Phys. Rev. Lett.}\ }\textbf {\bibinfo {volume} {101}},\
  \bibinfo {pages} {117203} (\bibinfo {year} {2008})}\BibitemShut {NoStop}%
\bibitem [{\citenamefont {Evenbly}\ and\ \citenamefont
  {Vidal}(2010)}]{Evenbly2010}%
  \BibitemOpen
  \bibfield  {author} {\bibinfo {author} {\bibfnamefont {G.}~\bibnamefont
  {Evenbly}}\ and\ \bibinfo {author} {\bibfnamefont {G.}~\bibnamefont
  {Vidal}},\ }\bibfield  {title} {\enquote {\bibinfo {title} {Frustrated
  antiferromagnets with entanglement renormalization: Ground state of the
  spin-$\frac{1}{2}$ heisenberg model on a kagome lattice},}\ }\href {\doibase
  10.1103/PhysRevLett.104.187203} {\bibfield  {journal} {\bibinfo  {journal}
  {Phys. Rev. Lett.}\ }\textbf {\bibinfo {volume} {104}},\ \bibinfo {pages}
  {187203} (\bibinfo {year} {2010})}\BibitemShut {NoStop}%
\bibitem [{\citenamefont {Yan}\ \emph {et~al.}(2011)\citenamefont {Yan},
  \citenamefont {Huse},\ and\ \citenamefont {White}}]{Yan2011}%
  \BibitemOpen
  \bibfield  {author} {\bibinfo {author} {\bibfnamefont {Simeng}\ \bibnamefont
  {Yan}}, \bibinfo {author} {\bibfnamefont {David~A.}\ \bibnamefont {Huse}}, \
  and\ \bibinfo {author} {\bibfnamefont {Steven~R.}\ \bibnamefont {White}},\
  }\bibfield  {title} {\enquote {\bibinfo {title} {Spin-liquid ground state of
  the $s=1/2$ kagome heisenberg antiferromagnet},}\ }\href {\doibase
  10.1126/science.1201080} {\bibfield  {journal} {\bibinfo  {journal}
  {Science}\ }\textbf {\bibinfo {volume} {332}},\ \bibinfo {pages} {1173}
  (\bibinfo {year} {2011})}\BibitemShut {NoStop}%
\bibitem [{\citenamefont {Depenbrock}\ \emph {et~al.}(2012)\citenamefont
  {Depenbrock}, \citenamefont {McCulloch},\ and\ \citenamefont
  {Schollw\"ock}}]{Depenbrock2012}%
  \BibitemOpen
  \bibfield  {author} {\bibinfo {author} {\bibfnamefont {Stefan}\ \bibnamefont
  {Depenbrock}}, \bibinfo {author} {\bibfnamefont {Ian~P.}\ \bibnamefont
  {McCulloch}}, \ and\ \bibinfo {author} {\bibfnamefont {Ulrich}\ \bibnamefont
  {Schollw\"ock}},\ }\bibfield  {title} {\enquote {\bibinfo {title} {Nature of
  the spin-liquid ground state of the $s=1/2$ heisenberg model on the kagome
  lattice},}\ }\href {\doibase 10.1103/PhysRevLett.109.067201} {\bibfield
  {journal} {\bibinfo  {journal} {Phys. Rev. Lett.}\ }\textbf {\bibinfo
  {volume} {109}},\ \bibinfo {pages} {067201} (\bibinfo {year}
  {2012})}\BibitemShut {NoStop}%
\bibitem [{\citenamefont {Iqbal}\ \emph {et~al.}(2013)\citenamefont {Iqbal},
  \citenamefont {Becca}, \citenamefont {Sorella},\ and\ \citenamefont
  {Poilblanc}}]{Iqbal2013}%
  \BibitemOpen
  \bibfield  {author} {\bibinfo {author} {\bibfnamefont {Yasir}\ \bibnamefont
  {Iqbal}}, \bibinfo {author} {\bibfnamefont {Federico}\ \bibnamefont {Becca}},
  \bibinfo {author} {\bibfnamefont {Sandro}\ \bibnamefont {Sorella}}, \ and\
  \bibinfo {author} {\bibfnamefont {Didier}\ \bibnamefont {Poilblanc}},\
  }\bibfield  {title} {\enquote {\bibinfo {title} {Gapless spin-liquid phase in
  the kagome spin-$\frac{1}{2}$ heisenberg antiferromagnet},}\ }\href {\doibase
  10.1103/PhysRevB.87.060405} {\bibfield  {journal} {\bibinfo  {journal} {Phys.
  Rev. B}\ }\textbf {\bibinfo {volume} {87}},\ \bibinfo {pages} {060405}
  (\bibinfo {year} {2013})}\BibitemShut {NoStop}%
\bibitem [{\citenamefont {Capponi}\ \emph
  {et~al.}(2013{\natexlab{b}})\citenamefont {Capponi}, \citenamefont {Chandra},
  \citenamefont {Auerbach},\ and\ \citenamefont {Weinstein}}]{Capponi2013}%
  \BibitemOpen
  \bibfield  {author} {\bibinfo {author} {\bibfnamefont {Sylvain}\ \bibnamefont
  {Capponi}}, \bibinfo {author} {\bibfnamefont {V.~Ravi}\ \bibnamefont
  {Chandra}}, \bibinfo {author} {\bibfnamefont {Assa}\ \bibnamefont
  {Auerbach}}, \ and\ \bibinfo {author} {\bibfnamefont {Marvin}\ \bibnamefont
  {Weinstein}},\ }\bibfield  {title} {\enquote {\bibinfo {title} {$p6$ chiral
  resonating valence bonds in the kagome antiferromagnet},}\ }\href {\doibase
  10.1103/PhysRevB.87.161118} {\bibfield  {journal} {\bibinfo  {journal} {Phys.
  Rev. B}\ }\textbf {\bibinfo {volume} {87}},\ \bibinfo {pages} {161118}
  (\bibinfo {year} {2013}{\natexlab{b}})}\BibitemShut {NoStop}%
\bibitem [{\citenamefont {He}\ \emph {et~al.}(2017)\citenamefont {He},
  \citenamefont {Zaletel}, \citenamefont {Oshikawa},\ and\ \citenamefont
  {Pollmann}}]{kagome_DMRG_U1}%
  \BibitemOpen
  \bibfield  {author} {\bibinfo {author} {\bibfnamefont {Yin-Chen}\
  \bibnamefont {He}}, \bibinfo {author} {\bibfnamefont {Michael~P.}\
  \bibnamefont {Zaletel}}, \bibinfo {author} {\bibfnamefont {Masaki}\
  \bibnamefont {Oshikawa}}, \ and\ \bibinfo {author} {\bibfnamefont {Frank}\
  \bibnamefont {Pollmann}},\ }\bibfield  {title} {\enquote {\bibinfo {title}
  {Signatures of dirac cones in a dmrg study of the kagome heisenberg model},}\
  }\href {\doibase 10.1103/PhysRevX.7.031020} {\bibfield  {journal} {\bibinfo
  {journal} {Phys. Rev. X}\ }\textbf {\bibinfo {volume} {7}},\ \bibinfo {pages}
  {031020} (\bibinfo {year} {2017})}\BibitemShut {NoStop}%
\bibitem [{\citenamefont {Liao}\ \emph {et~al.}(2017)\citenamefont {Liao},
  \citenamefont {Xie}, \citenamefont {Chen}, \citenamefont {Liu}, \citenamefont
  {Xie}, \citenamefont {Huang}, \citenamefont {Normand},\ and\ \citenamefont
  {Xiang}}]{Liao2017}%
  \BibitemOpen
  \bibfield  {author} {\bibinfo {author} {\bibfnamefont {H.~J.}\ \bibnamefont
  {Liao}}, \bibinfo {author} {\bibfnamefont {Z.~Y.}\ \bibnamefont {Xie}},
  \bibinfo {author} {\bibfnamefont {J.}~\bibnamefont {Chen}}, \bibinfo {author}
  {\bibfnamefont {Z.~Y.}\ \bibnamefont {Liu}}, \bibinfo {author} {\bibfnamefont
  {H.~D.}\ \bibnamefont {Xie}}, \bibinfo {author} {\bibfnamefont {R.~Z.}\
  \bibnamefont {Huang}}, \bibinfo {author} {\bibfnamefont {B.}~\bibnamefont
  {Normand}}, \ and\ \bibinfo {author} {\bibfnamefont {T.}~\bibnamefont
  {Xiang}},\ }\bibfield  {title} {\enquote {\bibinfo {title} {Gapless
  spin-liquid ground state in the $s=1/2$ kagome antiferromagnet},}\ }\href
  {\doibase 10.1103/PhysRevLett.118.137202} {\bibfield  {journal} {\bibinfo
  {journal} {Phys. Rev. Lett.}\ }\textbf {\bibinfo {volume} {118}},\ \bibinfo
  {pages} {137202} (\bibinfo {year} {2017})}\BibitemShut {NoStop}%
\bibitem [{\citenamefont {Changlani}\ and\ \citenamefont
  {L\"auchli}(2015)}]{Changlani2015}%
  \BibitemOpen
  \bibfield  {author} {\bibinfo {author} {\bibfnamefont {Hitesh~J.}\
  \bibnamefont {Changlani}}\ and\ \bibinfo {author} {\bibfnamefont
  {Andreas~M.}\ \bibnamefont {L\"auchli}},\ }\bibfield  {title} {\enquote
  {\bibinfo {title} {Trimerized ground state of the spin-1 heisenberg
  antiferromagnet on the kagome lattice},}\ }\href {\doibase
  10.1103/PhysRevB.91.100407} {\bibfield  {journal} {\bibinfo  {journal} {Phys.
  Rev. B}\ }\textbf {\bibinfo {volume} {91}},\ \bibinfo {pages} {100407}
  (\bibinfo {year} {2015})}\BibitemShut {NoStop}%
\bibitem [{\citenamefont {Liu}\ \emph {et~al.}(2015)\citenamefont {Liu},
  \citenamefont {Li}, \citenamefont {Weichselbaum}, \citenamefont {von Delft},\
  and\ \citenamefont {Su}}]{Liu2015}%
  \BibitemOpen
  \bibfield  {author} {\bibinfo {author} {\bibfnamefont {Tao}\ \bibnamefont
  {Liu}}, \bibinfo {author} {\bibfnamefont {Wei}\ \bibnamefont {Li}}, \bibinfo
  {author} {\bibfnamefont {Andreas}\ \bibnamefont {Weichselbaum}}, \bibinfo
  {author} {\bibfnamefont {Jan}\ \bibnamefont {von Delft}}, \ and\ \bibinfo
  {author} {\bibfnamefont {Gang}\ \bibnamefont {Su}},\ }\bibfield  {title}
  {\enquote {\bibinfo {title} {Simplex valence-bond crystal in the spin-1
  kagome heisenberg antiferromagnet},}\ }\href {\doibase
  10.1103/PhysRevB.91.060403} {\bibfield  {journal} {\bibinfo  {journal} {Phys.
  Rev. B}\ }\textbf {\bibinfo {volume} {91}},\ \bibinfo {pages} {060403}
  (\bibinfo {year} {2015})}\BibitemShut {NoStop}%
\bibitem [{\citenamefont {Nishimoto}\ and\ \citenamefont
  {Nakamura}(2015)}]{Nishimoto2015}%
  \BibitemOpen
  \bibfield  {author} {\bibinfo {author} {\bibfnamefont {Satoshi}\ \bibnamefont
  {Nishimoto}}\ and\ \bibinfo {author} {\bibfnamefont {Masaaki}\ \bibnamefont
  {Nakamura}},\ }\bibfield  {title} {\enquote {\bibinfo {title}
  {Non-symmetry-breaking ground state of the $s=1$ heisenberg model on the
  kagome lattice},}\ }\href {\doibase 10.1103/PhysRevB.92.140412} {\bibfield
  {journal} {\bibinfo  {journal} {Phys. Rev. B}\ }\textbf {\bibinfo {volume}
  {92}},\ \bibinfo {pages} {140412} (\bibinfo {year} {2015})}\BibitemShut
  {NoStop}%
\bibitem [{\citenamefont {Chubukov}(1992)}]{Chubukov1992}%
  \BibitemOpen
  \bibfield  {author} {\bibinfo {author} {\bibfnamefont {Andrey}\ \bibnamefont
  {Chubukov}},\ }\bibfield  {title} {\enquote {\bibinfo {title} {Order from
  disorder in a kagom\'e antiferromagnet},}\ }\href {\doibase
  10.1103/PhysRevLett.69.832} {\bibfield  {journal} {\bibinfo  {journal} {Phys.
  Rev. Lett.}\ }\textbf {\bibinfo {volume} {69}},\ \bibinfo {pages} {832}
  (\bibinfo {year} {1992})}\BibitemShut {NoStop}%
\bibitem [{\citenamefont {G\"otze}\ \emph {et~al.}(2011)\citenamefont
  {G\"otze}, \citenamefont {Farnell}, \citenamefont {Bishop}, \citenamefont
  {Li},\ and\ \citenamefont {Richter}}]{Goetze2011}%
  \BibitemOpen
  \bibfield  {author} {\bibinfo {author} {\bibfnamefont {O.}~\bibnamefont
  {G\"otze}}, \bibinfo {author} {\bibfnamefont {D.~J.~J.}\ \bibnamefont
  {Farnell}}, \bibinfo {author} {\bibfnamefont {R.~F.}\ \bibnamefont {Bishop}},
  \bibinfo {author} {\bibfnamefont {P.~H.~Y.}\ \bibnamefont {Li}}, \ and\
  \bibinfo {author} {\bibfnamefont {J.}~\bibnamefont {Richter}},\ }\bibfield
  {title} {\enquote {\bibinfo {title} {Heisenberg antiferromagnet on the kagome
  lattice with arbitrary spin: A higher-order coupled cluster treatment},}\
  }\href {\doibase 10.1103/PhysRevB.84.224428} {\bibfield  {journal} {\bibinfo
  {journal} {Phys. Rev. B}\ }\textbf {\bibinfo {volume} {84}},\ \bibinfo
  {pages} {224428} (\bibinfo {year} {2011})}\BibitemShut {NoStop}%
\bibitem [{\citenamefont {Oitmaa}\ and\ \citenamefont
  {Singh}(2016)}]{Oitmaa2016}%
  \BibitemOpen
  \bibfield  {author} {\bibinfo {author} {\bibfnamefont {J.}~\bibnamefont
  {Oitmaa}}\ and\ \bibinfo {author} {\bibfnamefont {R.~R.~P.}\ \bibnamefont
  {Singh}},\ }\bibfield  {title} {\enquote {\bibinfo {title} {Competing orders
  in spin-1 and $\text{spin}\ensuremath{-}\frac{3}{2}$ {XXZ} kagome
  antiferromagnets: A series expansion study},}\ }\href {\doibase
  10.1103/PhysRevB.93.014424} {\bibfield  {journal} {\bibinfo  {journal} {Phys.
  Rev. B}\ }\textbf {\bibinfo {volume} {93}},\ \bibinfo {pages} {014424}
  (\bibinfo {year} {2016})}\BibitemShut {NoStop}%
\bibitem [{\citenamefont {Sachdev}(1992)}]{Sachdev1992}%
  \BibitemOpen
  \bibfield  {author} {\bibinfo {author} {\bibfnamefont {Subir}\ \bibnamefont
  {Sachdev}},\ }\bibfield  {title} {\enquote {\bibinfo {title} {Kagome- and
  triangular-lattice heisenberg antiferromagnets: Ordering from quantum
  fluctuations and quantum-disordered ground states with unconfined bosonic
  spinons},}\ }\href {\doibase 10.1103/PhysRevB.45.12377} {\bibfield  {journal}
  {\bibinfo  {journal} {Phys. Rev. B}\ }\textbf {\bibinfo {volume} {45}},\
  \bibinfo {pages} {12377} (\bibinfo {year} {1992})}\BibitemShut {NoStop}%
\bibitem [{\citenamefont {Balay}\ \emph {et~al.}(1997)\citenamefont {Balay},
  \citenamefont {Gropp}, \citenamefont {McInnes},\ and\ \citenamefont
  {Smith}}]{petsc-efficient}%
  \BibitemOpen
  \bibfield  {author} {\bibinfo {author} {\bibfnamefont {Satish}\ \bibnamefont
  {Balay}}, \bibinfo {author} {\bibfnamefont {William~D.}\ \bibnamefont
  {Gropp}}, \bibinfo {author} {\bibfnamefont {Lois~Curfman}\ \bibnamefont
  {McInnes}}, \ and\ \bibinfo {author} {\bibfnamefont {Barry~F.}\ \bibnamefont
  {Smith}},\ }\bibfield  {title} {\enquote {\bibinfo {title} {Efficient
  management of parallelism in object oriented numerical software libraries},}\
  }in\ \href {\doibase 10.1007/978-1-4612-1986-6_8} {\emph {\bibinfo
  {booktitle} {Modern Software Tools in Scientific Computing}}},\ \bibinfo
  {editor} {edited by\ \bibinfo {editor} {\bibfnamefont {E.}~\bibnamefont
  {Arge}}, \bibinfo {editor} {\bibfnamefont {A.~M.}\ \bibnamefont {Bruaset}}, \
  and\ \bibinfo {editor} {\bibfnamefont {H.~P.}\ \bibnamefont {Langtangen}}}\
  (\bibinfo  {publisher} {Birkh{\"{a}}user Press},\ \bibinfo {year} {1997})\
  pp.\ \bibinfo {pages} {163--202}\BibitemShut {NoStop}%
\bibitem [{\citenamefont {Balay}\ \emph {et~al.}(2017)\citenamefont {Balay},
  \citenamefont {Abhyankar}, \citenamefont {Adams}, \citenamefont {Brown},
  \citenamefont {Brune}, \citenamefont {Buschelman}, \citenamefont {Dalcin},
  \citenamefont {Eijkhout}, \citenamefont {Gropp}, \citenamefont {Kaushik},
  \citenamefont {Knepley}, \citenamefont {McInnes}, \citenamefont {Rupp},
  \citenamefont {Smith}, \citenamefont {Zampini}, \citenamefont {Zhang},\ and\
  \citenamefont {Zhang}}]{petsc-user-ref}%
  \BibitemOpen
  \bibfield  {author} {\bibinfo {author} {\bibfnamefont {Satish}\ \bibnamefont
  {Balay}}, \bibinfo {author} {\bibfnamefont {Shrirang}\ \bibnamefont
  {Abhyankar}}, \bibinfo {author} {\bibfnamefont {Mark~F.}\ \bibnamefont
  {Adams}}, \bibinfo {author} {\bibfnamefont {Jed}\ \bibnamefont {Brown}},
  \bibinfo {author} {\bibfnamefont {Peter}\ \bibnamefont {Brune}}, \bibinfo
  {author} {\bibfnamefont {Kris}\ \bibnamefont {Buschelman}}, \bibinfo {author}
  {\bibfnamefont {Lisandro}\ \bibnamefont {Dalcin}}, \bibinfo {author}
  {\bibfnamefont {Victor}\ \bibnamefont {Eijkhout}}, \bibinfo {author}
  {\bibfnamefont {William~D.}\ \bibnamefont {Gropp}}, \bibinfo {author}
  {\bibfnamefont {Dinesh}\ \bibnamefont {Kaushik}}, \bibinfo {author}
  {\bibfnamefont {Matthew~G.}\ \bibnamefont {Knepley}}, \bibinfo {author}
  {\bibfnamefont {Lois~Curfman}\ \bibnamefont {McInnes}}, \bibinfo {author}
  {\bibfnamefont {Karl}\ \bibnamefont {Rupp}}, \bibinfo {author} {\bibfnamefont
  {Barry~F.}\ \bibnamefont {Smith}}, \bibinfo {author} {\bibfnamefont
  {Stefano}\ \bibnamefont {Zampini}}, \bibinfo {author} {\bibfnamefont {Hong}\
  \bibnamefont {Zhang}}, \ and\ \bibinfo {author} {\bibfnamefont {Hong}\
  \bibnamefont {Zhang}},\ }\href
  {https://www.mcs.anl.gov/petsc/petsc-current/docs/manual.pdf} {\emph
  {\bibinfo {title} {{PETS}c Users Manual}}},\ \bibinfo {type} {Tech. Rep.}\
  \bibinfo {number} {ANL-95/11 - Revision 3.8}\ (\bibinfo  {institution}
  {Argonne National Laboratory},\ \bibinfo {year} {2017})\BibitemShut {NoStop}%
\bibitem [{\citenamefont {Hernandez}\ \emph {et~al.}(2005)\citenamefont
  {Hernandez}, \citenamefont {Roman},\ and\ \citenamefont
  {Vidal}}]{slepc-toms}%
  \BibitemOpen
  \bibfield  {author} {\bibinfo {author} {\bibfnamefont {Vicente}\ \bibnamefont
  {Hernandez}}, \bibinfo {author} {\bibfnamefont {Jose~E.}\ \bibnamefont
  {Roman}}, \ and\ \bibinfo {author} {\bibfnamefont {Vicente}\ \bibnamefont
  {Vidal}},\ }\bibfield  {title} {\enquote {\bibinfo {title} {{SLEPc}: A
  scalable and flexible toolkit for the solution of eigenvalue problems},}\
  }\href {\doibase 10.1145/1089014.1089019} {\bibfield  {journal} {\bibinfo
  {journal} {{ACM} Trans. Math. Software}\ }\textbf {\bibinfo {volume} {31}},\
  \bibinfo {pages} {351--362} (\bibinfo {year} {2005})}\BibitemShut {NoStop}%
\bibitem [{\citenamefont {Roman}\ \emph {et~al.}(2017)\citenamefont {Roman},
  \citenamefont {Campos}, \citenamefont {Romero},\ and\ \citenamefont
  {Tomas}}]{slepc-manual}%
  \BibitemOpen
  \bibfield  {author} {\bibinfo {author} {\bibfnamefont {J.~E.}\ \bibnamefont
  {Roman}}, \bibinfo {author} {\bibfnamefont {C.}~\bibnamefont {Campos}},
  \bibinfo {author} {\bibfnamefont {E.}~\bibnamefont {Romero}}, \ and\ \bibinfo
  {author} {\bibfnamefont {A.}~\bibnamefont {Tomas}},\ }\href
  {https://slepc.upv.es/documentation/slepc.pdf} {\emph {\bibinfo {title}
  {{SLEPc} Users Manual}}},\ \bibinfo {type} {Tech. Rep.}\ \bibinfo {number}
  {DSIC-II/24/02 - Revision 3.8}\ (\bibinfo  {institution} {D. Sistemes
  Inform\`atics i Computaci\'o, Universitat Polit\`ecnica de Val\`encia},\
  \bibinfo {year} {2017})\BibitemShut {NoStop}%
\bibitem [{\citenamefont {Lee}\ and\ \citenamefont
  {Young}(2007)}]{lee2007large}%
  \BibitemOpen
  \bibfield  {author} {\bibinfo {author} {\bibfnamefont {LW}~\bibnamefont
  {Lee}}\ and\ \bibinfo {author} {\bibfnamefont {AP}~\bibnamefont {Young}},\
  }\bibfield  {title} {\enquote {\bibinfo {title} {Large-scale monte carlo
  simulations of the isotropic three-dimensional heisenberg spin glass},}\
  }\href@noop {} {\bibfield  {journal} {\bibinfo  {journal} {Phys. Rev. B}\
  }\textbf {\bibinfo {volume} {76}},\ \bibinfo {pages} {024405} (\bibinfo
  {year} {2007})}\BibitemShut {NoStop}%
\end{thebibliography}%
	
\appendix*
\section{Classical orphan physics on the kagom\'e lattice}\label{app}

In this section, we expand on our introductory remarks on orphan spins on corner-sharing lattices and their main signatures in the finite-temperature properties of {\it classical} frustrated magnets, with particular emphasis on the kagom\'e antiferromagnet of interest to us here. To this end, we consider in this section classical spins $S$ (vectors of magnitude $|S|$), coupled by a bilinear Heisenberg interaction on a network of corner-sharing simplices. For nearest-neighbour interactions, the Hamiltonian can be rewritten in terms of frustrated plaquette terms~\cite{Moessner_Berlinsky,henley_effective_2001}. We have:
\begin{equation}
H=\sum_{\langle i,j\rangle} \vec{S_i}\cdot\vec{S_j}=
\frac{1}{2}\sum_{p} \left(\sum_{i\in p} \vec{S_i}\right)^2 + c,
\label{eq:Hinsumofsquaresform}
\end{equation}
where $p$ labels plaquettes in the lattice. This is true for kagom\'e and
pyrochlore and pyrochlore-slab structures, and other structures where all pairs of spin within a
plaquette are interacting. 

The Hamiltonian written in plaquette language is minimized by setting $\sum_{i\in p} \vec{S_i}=0$ for all 
plaquettes~\cite{henley_effective_2001}. Replacing spins by non-magnetic impurities
reduces the number of spins participating in a plaquette term, but the energetic constraint can still be satisfied as long as there is more than one spin in a plaquette.
The orphan spin situation illustrated in Fig~\ref{figlat}a, corresponds to the case where only one spin is left after substitution
with non-magnetic impurities.

\begin{figure}[t]
\includegraphics[width=\hsize]{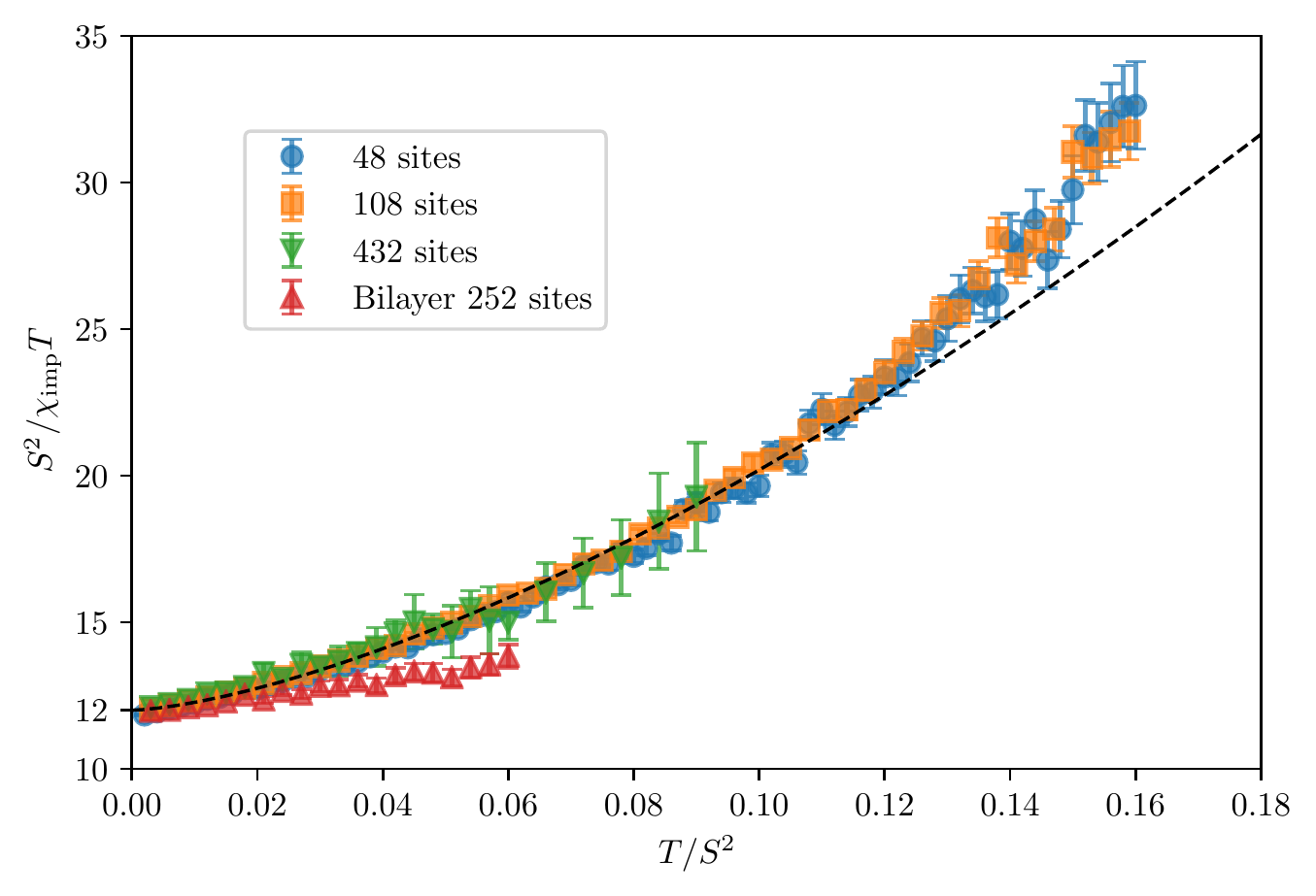}
\caption{Difference in the susceptibility between a sample with a single orphan complex and a sample with no impurities, $\chi_{\rm imp}=\chi_{\rm tot}^{\rm orphan}-\chi_{\rm tot}^{\rm pure}$, as a function of temperature for the classical Heisenberg model on a kagom\'e lattices of different sizes. A fit to the form $12+aT^b$ at low temperature is also displayed. A comparison with the bilayer system shows that the orphan signature is stronger in the latter.}
\label{fig:chi_imp_classical}
\end{figure}

As the orphan spin is released from the constraints coming from one plaquette (triangle for the kagom\'e lattice), one expects that this spin remains ``partially free'' in some sense. As summarized in the Introduction, this expectation found a precise expression in the idea of a half-orphan degree of freedom in the work of Henley, and of Moessner and Berlinsky~\cite{henley_effective_2001,Moessner_Berlinsky}. The detailed characterization of the response~\cite{sen_fractional_2011} indicates that the orphan behaves as a spin $S$ object that sees an external field $h/2$ when an external field of $h$ is applied. This paramagnetic response of the lone spin is partially screened by the net diamagnetic response of the surrounding spin texture, leading to a net low temperature susceptibility identical to that of a spin $S/2$, as befits the response of an emergent half-orphan degree of freedom which behaves like a net spin-$S/2$ particle smeared across a few lattice spacings around the orphan~\cite{sen_fractional_2011}. 

For the specific case of the kagom\'e lattice, there is a caveat: The classical spin liquid at intermediate temperatures gives way at low enough temperature to a weakly ordered coplanar state of the classical Heisenberg model, where the coplanarity is itself a symmetry-breaking crossover phenomenon that occurs at $T^{*} \sim 10^{-3} JS^2$, and the subsequent entropically driven choice of coplanar ordered state by anharmonic fluctuations only becomes apparent at even lower temperatures~\cite{Chern_Moessner,Zhitomirsky,Rutenberg_Huse}.  The smallness of $T^{*}$ is fortuitous from the point of view of orphan physics: In the very broad temperature regime $T^{*} \ll T \ll JS^2$ in which there is spin liquid behaviour, we expect a pair of vacancies on the same triangle to again nucleate an emergent half-orphan degree of freedom. Below, we confirm via classical Monte Carlo simulations that this is indeed true, although the temperature window in which half-orphan physics is well-established classically is seen to be smaller compared to the pyrochlore-slab lattice~\cite{sen_fractional_2011}. 

Ref.~\onlinecite{sen_fractional_2011} showed in the pyrochlore-slab magnet that the difference, denoted $\chi_{\rm imp}$ in Ref.~\onlinecite{sen_fractional_2011}, of the total magnetic susceptibility $\chi_{\rm tot}$ between a pure sample (with no impurities) and an otherwise pure sample with two nearest-neighbour non-magnetic impurities (thus creating a single orphan) indeed corresponds to the response of a free spin of length $S/2$. As $\chi=S^2/(3T)$ for a single free spin, we should analogously expect to see at low temperatures a
difference of the form $\chi_{\rm imp}=\chi_{\rm tot}^{\rm orphan}-\chi_{\rm tot}^{\rm pure}=S^2/(12T)$ on the kagom\'e lattice at intermediate temperatures well above the extremely low-temperature crossover to coplanar ordering, but well below the temperature scale set by $JS^2$. 

We have computed this difference $\chi_{\rm imp}$ for several kagom\'e samples using classical Monte Carlo simulations and standard Metropolis, Wolff cluster and over-relaxation~\cite{lee2007large} updates to ensure ergodicity. Our results are displayed in Fig.~\ref{fig:chi_imp_classical} where we plot $S^2/(\chi_{\rm imp} T)$ as function of $T$ to check the dependence with temperature in a clear manner. Data at the lower end of our temperature range converge to the expected constant $12$, and we find a size-independent and smooth deviation away from this result. For sake of completeness, we also reproduced the results of Ref.~\onlinecite{sen_fractional_2011} for the bilayer with 252 sites, where we find the approach to the $T=0$ limit to be even flatter.

\begin{figure}[t]
\includegraphics[width=\hsize]{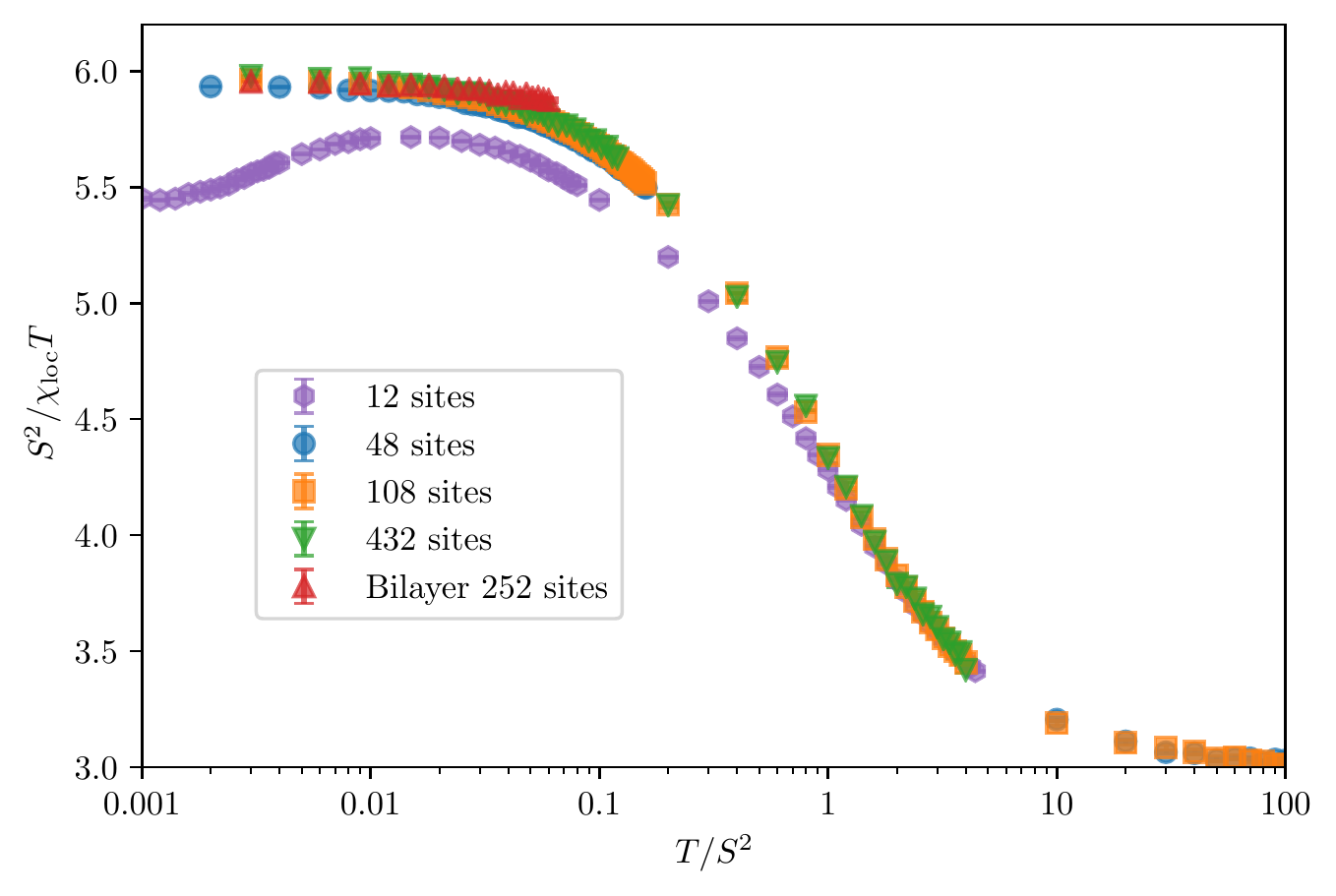}
\caption{Local susceptibility of the classical Heisenberg model on various samples
sizes of the kagom\'e lattice and bilayer, as a function of temperature.}
\label{fig:chi_local_classical}
\end{figure}

The local susceptibility $\chi_{loc}=\langle S^z_{\mathrm{orphan}}S^z_{\mathrm{total}}\rangle/T$, should reflect that the
orphan spin behaves as a free spin of length $S$ in a magnetic field of
strength $h/2$ (where $h$ is the applied external field). This is expected
due to the screening of the magnetic field by the neighbors of the orphan
spin~\cite{sen_vacancy_2012}. In the low-field linear regime, we thus expect $S^2/\chi_{loc} T=6$ at sufficiently low but not-too-low temperature. Our Monte-Carlo simulations (Fig.~\ref{fig:chi_local_classical}) are in good agreement with this expectation at low 
temperatures, with an approximately size-independent peel-off towards the
high temperature result of a free spin-$S$ at temperature $T$: the data once again displayed as $S^2/(\chi_{\rm loc} T)$ clearly exhibits a crossover from the value $6$ (obtained for spin-$S/2$) at low-temperature to the high-temperature spin-$S$ value $3$. Fig.~\ref{fig:chi_local_classical} also represents data for the small size cluster of 12 sites which is shown to be already close to the thermodynamic limit (with at most a difference less than 10\%). This particular 12-site sample (which is represented in Fig.~\ref{figlat}a, along with the location of impurities and the orphan spin) is useful to compare with the quantum simulations reported in the main text, as we are limited to small sizes in our studies of the quantum problem.

\end{document}